\documentclass[fleqn,usenatbib]{mnras}

\usepackage[T1]{fontenc}

\DeclareRobustCommand{\VAN}[3]{#2}
\let\VANthebibliography\thebibliography
\def\thebibliography{\DeclareRobustCommand{\VAN}[3]{##3}\VANthebibliography}

\usepackage{graphicx}	
\usepackage{amsmath}	
\usepackage{xcolor}
\newcommand{\CHEOPS}{\mbox{\it CHEOPS}}
\newcommand{\pycheops}{\mbox{\sc pycheops}}
\newcommand{\software}[1]{\mbox{\sc {#1}}}
\newcommand{\var}[1]{\mbox{\sf {#1}}}
\newcommand{\code}[1]{\mbox{\sf {#1}}}
\newcommand{\Rearth}{$R_\oplus$}

\title[CHEOPS Early Science and pycheops]{Analysis of Early Science observations with the CHaracterising ExOPlanets Satellite (\textit{CHEOPS}) using \pycheops}

\author[P. F. L. Maxted et al.]{P. F. L. Maxted,$^{1}$\thanks{E-mail: p.maxted@keele.ac.uk}
D. Ehrenreich,$^{2}$
T. G. Wilson,$^{3}$
Y. Alibert,$^{4}$
A. Collier Cameron,$^{3}$ \newauthor
S. Hoyer,$^{5}$
S. G. Sousa,$^{6}$
G. Olofsson,$^{7}$
A. Bekkelien,$^{2}$
A. Deline,$^{2}$
L. Delrez,$^{8,9}$ \newauthor
A. Bonfanti,$^{10}$
L. Borsato,$^{11}$
R. Alonso,$^{12,13}$
G. Anglada Escudé,$^{14,15}$
D. Barrado,$^{16}$ \newauthor
S. C. C. Barros,$^{6,17}$
W. Baumjohann,$^{10}$
M. Beck,$^{2}$
T. Beck,$^{4}$
W. Benz,$^{4,18}$
N. Billot,$^{2}$ \newauthor
F. Biondi,$^{11,19}$
X. Bonfils,$^{20}$
A. Brandeker,$^{7}$
C. Broeg,$^{4,18}$
T. Bárczy,$^{21}$
J. Cabrera,$^{22}$ \newauthor
S. Charnoz,$^{23}$
C. Corral Van Damme,$^{24}$
Sz. Csizmadia,$^{22}$
M. B. Davies,$^{25}$
M. Deleuil,$^{5}$ \newauthor
O. D. S. Demangeon,$^{6,17}$
B.-O. Demory,$^{18}$
A. Erikson,$^{22}$
H.G. Flor{\'e}n,$^{7}$
A. Fortier,$^{4,18}$ \newauthor
L. Fossati,$^{10}$
M. Fridlund,$^{26,27}$
D. Futyan,$^{2}$
D. Gandolfi,$^{28}$
M. Gillon,$^{8}$
M. Guedel,$^{29}$ \newauthor
P. Guterman,$^{5,30}$
K. Heng,$^{18,31}$
K. G. Isaak,$^{32}$
L. Kiss,$^{33}$
J. Laskar,$^{34}$ \newauthor
A. Lecavelier des Etangs,$^{35}$
M. Lendl,$^{2}$
C. Lovis,$^{2}$
D. Magrin,$^{11}$
V. Nascimbeni,$^{11}$ \newauthor
R. Ottensamer,$^{36}$
I. Pagano,$^{37}$
E. Pallé,$^{12,13}$
G. Peter,$^{38}$
G. Piotto,$^{11,39}$
D. Pollacco,$^{31}$ \newauthor
F. J. Pozuelos,$^{8,9}$
D. Queloz,$^{2,40}$
R. Ragazzoni,$^{11,39}$
N. Rando,$^{24}$
H. Rauer,$^{22,41,42}$ \newauthor
C. Reimers,$^{36}$
I. Ribas,$^{14,15}$
S. Salmon, $^{2}$,
N. C. Santos,$^{6,17}$
G. Scandariato,$^{37}$ \newauthor
A. E. Simon,$^{4}$
A. M. S. Smith,$^{22}$
M. Steller,$^{10}$
M. I. Swayne,$^{1}$
Gy. M. Szab{\'o},$^{43,44}$\newauthor
D. Ségransan,$^{2}$ 
N. Thomas,$^{4}$
S. Udry,$^{2}$
V. Van Grootel,$^{9}$
and N. A. Walton,$^{45}$
\\
(Affiliations are listed at the end of the paper)
}

\date{Accepted XXX. Received YYY; in original form ZZZ}

\pubyear{2015}

\begin{document}
\label{firstpage}
\pagerange{\pageref{firstpage}--\pageref{lastpage}}
\maketitle

\begin{abstract}
\CHEOPS\ (CHaracterising ExOPlanet Satellite) is an ESA S-class mission that observes bright stars at high cadence from low-Earth orbit. The main aim of the mission is to characterize exoplanets that transit nearby stars using ultrahigh precision photometry. Here we report the analysis of transits observed by \CHEOPS\ during its Early Science observing programme for four well-known exoplanets: GJ~436\,b, HD~106315\,b, HD~97658\,b and GJ~1132\,b. The analysis is done using \pycheops, an open-source software package we have developed to easily and efficiently analyse \CHEOPS\ light curve data using state-of-the-art techniques that are fully described herein.  We show that the precision of the transit parameters measured using \CHEOPS\ is comparable to that from larger space telescopes such as  Spitzer Space Telescope and {\it Kepler}. We use the updated planet parameters from our analysis to derive new constraints on the internal structure of these four exoplanets. 
 \end{abstract}




\section{Introduction}
\label{sec:intro}

The CHaracterising ExOPlanet Satellite (\CHEOPS) was selected as the first S-class mission in the European Space Agency (ESA) science programme and was successfully launched on December 18, 2019 \citep{2021ExA....51..109B}. Nominal science operations started on April 18, 2020 after a period of in-orbit commissioning \citep{2020SPIE11443E..14R}.  \CHEOPS\ is a follow-up mission that generates ultrahigh precision photometry for bright stars already known to host exoplanets \citep{2018haex.bookE..84B}. It has the flexibility to observe stars at specified times over a large fraction of the sky.\footnote{\url{https://www.cosmos.esa.int/web/cheops/the-cheops-sky}} The observing time is split between the Guaranteed Time Observation (GTO) programme (72\%), the Guest Observers (GO) programme (18\%) and the Monitoring and Characterisation (M\&C) programme (10\%). The \CHEOPS\ GTO programme includes observations to search for transits of planets detected in radial velocity surveys \citep{delrez2021}, to provide precise radius measurements for known transiting exoplanets \citep{2021A+A...646A.157B, 2021arXiv210109260L}, to characterize exoplanet atmospheres from measurements of their eclipses \citep{2020A+A...643A..94L},  to study the dynamics of exoplanet systems using transit time variations \citep[TTVs,][]{borsato2021}, to search for moons and rings in exoplanets systems \citep{2018EPSC...12..401A}, to measure the tidal deformation of planets \citep{2019A+A...621A.117A}, and some stellar science that is relevant to exoplanet studies, e.g. characterisation of very low-mass stars in eclipsing binary star systems.

The {\it CoRoT} \citep{Baglin06} and {\it Kepler} \citep{Borucki10} surveys  have provided valuable information on the Galactic exoplanet population based on intensive monitoring of small areas of the sky. However, most of the exoplanets identified from their transits by those surveys are too faint to allow for detailed characterisation. The best-characterised exoplanets are typically those  discovered by radial velocity surveys orbiting bright stars that were subsequently found to be transiting, e.g.  HD~209458\,b ($V=7.8$), HD~189733\,b ($V=7.8$), GJ~436\,b ($V=10.2$)  or 55~Cancri\,e ($V=6.0$). Detailed characterisation has also been possible for  gas- and ice-giant planets transiting bright stars discovered by ground-based transit surveys such as WASP \citep{Pollacco06},  HAT \citep{Bakos07},  KELT \citep{Pepper18} and MASCARA \citep{Snellen12}. Surveys such as Mearth \citep{Charbonneau09} and SPECULOOS \citep{2018SPIE10700E..1ID} are able to discover Earth-size planets by looking for transits around M-dwarf host stars. The {\it Kepler K2} mission surveyed a larger area of the sky around the ecliptic than the original mission and so increased the number of planets discovered orbiting bright stars with this instrument, e.g. HD~106315 \citep{2017A+A...608A..25B, 2017AJ....153..255C, Rodriguez17}. NASA's Transiting Exoplanet Survey Satellite \citep[{\it TESS};][]{Ricker14} is an all-sky survey with the aim to discover exoplanets orbiting stars bright enough for detailed  characterisation with NASA’s James Webb Space Telescope \citep[{\it JWST};][]{2006SSRv..123..485G}. The focus of the \CHEOPS\ mission is the characterisation of a set of most promising objects for constraining planet formation and evolution theories, and to support spectroscopic studies of these planets' atmospheres with {\it JWST},  {\it Ariel} \citep{2018ExA....46..135T}, and instrumentation on 30-m class telescopes \citep{2021Msngr.182...27M}.  With its unique characteristics, \CHEOPS\ is complementary to all other transit survey missions as it provides the agility and the photometric precision necessary to re-visit sufficiently interesting targets for which further measurements are deemed valuable. The \CHEOPS\ mission is also providing valuable experience for the European space science community that is feeding into the development of the PLATO mission, an ESA M-class mission with the challenging goal to detect and characterise Earth-size planets with orbital periods up to one year that transit bright stars \citep{2014ExA....38..249R}.

During the first 8 months of science operations, the \CHEOPS\ guaranteed-time observing programme (GTO) scheduled and observed over 300 transits and eclipses of known transiting exoplanet and eclipsing binary star systems. Another 24 long-duration observations were obtained for 12 bright stars to search for transits due to exoplanets discovered by radial velocity surveys. In addition, over 600 observations with a duration of 1-3 orbits\footnote{The duration of \CHEOPS\ observations are measured in orbits of 98.725 minutes each.} each were obtained. These ``filler'' observations ensure that \CHEOPS\ continues to collect useful science observations during short intervals between time-critical observations of transits and eclipses. The filler programmes within the GTO are being used to study the variability of low mass stars on short time scales, and to search for remnants of planetary systems around hot subdwarf stars \citep{2021arXiv210410462V}.

These large data rates, the peculiarities of observing from a nadir-locked orbit with a rotating field-of-view,  and the very high precision of the \CHEOPS\ data require specialised software to enable efficient and accurate analysis of the light curves, and timely publication of the results. Very accurate models are needed to precisely match the features visible in these ultrahigh precision light curves. The software should be easy to run and efficient so that everyone on the science team members has the opportunity to contribute to the data analysis effort without requiring access to large computing resources or extensive training. These requirements led us to develop the \pycheops\ software package, building on previous work to test the power-2 limb-darkening law \citep{2018A+A...616A..39M} and the development of the qpower2 algorithm \citep{2018A+A...616A..39M}.

The \pycheops\ software package is described fully in Section~\ref{sec:pycheops} of this paper. The analysis of the \CHEOPS\ light curves for four transiting exoplanets observed during the Early Science observing programme is described in Section~\ref{sec:analysis}. Section~\ref{sec:structure} describes the method we have used to place constraints on the internal structure of these planets. These results are discussed in Section~\ref{sec:discussion} and conclusions are briefly given in Section~\ref{sec:conclusion}.

\section{The pycheops software package} \label{sec:pycheops}
\subsection{Implementation and dependencies}
 \pycheops\ is written in \software{python} version 3.7 and makes extensive use of the packages \software{numpy} \citep{numpy} and \software{scipy} \citep{scipy}. \software{matplotlib} \citep{matplotlib} is used for data visualisation and plotting. Tabular data, celestial coordinates and time scales are handled using routines from the \software{astropy}\footnote{\url{https://www.astropy.org/}} software package \citep{astropy}. The package \software{lmfit}\footnote{\url{https://lmfit.github.io/lmfit-py/}} \citep{lmfit} is used for non-linear least-squares minimization and parameter handling. For Bayesian data analysis techniques, we use the affine-invariant Markov chain Monte Carlo ensemble sampler by \citet{2010CAMCS...5...65G} implemented in \software{emcee}\footnote{\url{https://github.com/dfm/emcee}} \citep{emcee} to generate samples from the posterior probability distribution. Correlated noise is modelled using Gaussian process (GP) regression in the form of the {\it celerite} algorithm implemented in software package \software{celerite2}\footnote{\url{https://github.com/dfm/celerite2}}  \citep{celerite1, celerite2}. We use run-time compilation with \software{numba}\footnote{\url{https://numba.pydata.org/}} \citep{numba} to reduce the execution time for a few key subroutines that are called frequently by \software{emcee}. 
 Parameter correlation plots are generated using the \software{python} module \software{corner}\footnote{\url{https://corner.readthedocs.io}} \citep{corner}. \CHEOPS\ data are archived at the Data \& Analysis Center for Exoplanets (DACE) hosted by the University of Geneva. These data can be accessed directly from \pycheops\ using the \software{python-dace-client} \software{python} module available from the DACE website.\footnote{\url{https://dace.unige.ch}} {\bf This client handles access to both proprietary
data for science team members and public data for general users.}

 We have successfully installed and tested \pycheops\ on machines running macOS, Windows 10 and Linux operating systems.
 
\subsection{Package structure}
 Almost all the functionality of \pycheops\ is implemented as a single \software{python} module of the same name that contains the following sub-modules.
 
\begin{description}
\item [\tt core]{-- handles the software configuration, e.g. data locations and user options.}
\item [\tt constants]{-- contains fundamental constants and nominal values for selected solar and planetary quantities defined by IAU 2015 Resolution B3 \citep{2015arXiv151007674M}. The Newtonian constant is taken to be $G=6.67408\times 10^{-11}\,{\rm m}^3\,{\rm kg}^{-1}\,{\rm s}^{-2}$ (2014 CODATA value). The radius of the Earth is defined to be $R_{\oplus}=6371$\,km so that the volume of a sphere with this radius equals the nominal volume of the Earth defined in IAU 2015 Resolution B3. Similarly, the radius of Jupiter is defined to be $R_{\rm jup} = 69911$\,km.}
\item [\tt funcs]{-- provides functions related to orbits and eclipses of stars and planets in Keplerian orbits, e.g. the solution of Kepler's equation \citep{1995CeMDA..63..101M} and the time of mid-eclipse in an eccentric orbit using Lacy's method \citep{1992AJ....104.2213L}. This sub-module also includes a function to calculate the mass and radius of a planet from the observed parameters of its transit, and to plot the planet in the mass-radius plane compared to various models and/or the parameters of other known exoplanets taken from TEPCat \citep{2011MNRAS.417.2166S}.}
\item [\tt instrument]{-- contains data specific to the \CHEOPS\ instrument, e.g. the instrument response function.}
\item[\tt utils]{-- provides utility functions, e.g. formatting of values with errors for output and light curve binning.}
\item [\tt ld]{-- provides the parameters of the power-2 limb-darkening as a function of stellar effective temperature (T$_{\rm eff}$), surface gravity ($\log g$) and metallicity ([Fe/H]).  This sub-module also contains functions to convert between different parametrisations of the power-2 limb-darkening law. Data are included for the \CHEOPS\, {\it TESS}, {\it Kepler}, NGTS and {\it CoRoT} passbands as well as various filters within the SDSS and Johnson/Cousins photometric systems.
The parameters are interpolated from tables generated from synthetic 3D-LTE spectra from
the {\sc stagger}-grid calculated by \citet{2015A+A...573A..90M}. For stars outside the range covered by the {\sc stagger}-grid we use the coefficients for a 4-parameter limb-darkening law provided by \citet{2019RNAAS...3...17C} for the {\it Gaia} G band, which gives a close approximation to the \CHEOPS\ passband. The transformation from the coefficients $a_1, \dots a_4$ from Table~10 of \citet{2019RNAAS...3...17C} to the parameters $h_1$ and $h_2$ of the power-2 limb-darkening law was done using a least-squares fit to the intensity profile as a function of  $r=\sqrt{1-\mu^2}$ in the region $r<0.99$.}
\item[\tt models]{-- provides models for photometric effects observed in transiting exoplanet and eclipsing binary star systems, e.g. transits, eclipses, ellipsoidal effect, etc., and a 2-body Keplerian radial velocity model. These models are provided in the form of \software{lmfit} \code{Model} classes and so can be easily combined using arithmetic operators. Trends in the data correlated with parameters such as spacecraft roll angle, sky background level, telescope tube temperature, etc. can be modelled using the \code{FactorModel} class provided by this sub-module. }
\item[\software{dataset}]{-- provides the \code{Dataset} class that is used to download, inspect and analyse a single eclipse or transit observation obtained with \CHEOPS. The light curve plots in this paper for observations consisting of a single visit were generated using this class.}
\item[\software{multivisit}]{-- provides the  \code{MultiVisit} class for the combined analysis of multiple \code{Dataset} objects. For the analysis of multiple transits it is possible to include parameters \var{ttv\_01}, \var{ttv\_02}, etc. in the model to allow for transit timing variations around a linear ephemeris. Similarly, the depths of the eclipses \var{L\_01}, \var{L\_02}, etc. can be included as free parameters in the analysis of visits obtained during different occultations. \code{MultiVisit} can be used for the analysis of a single visit. The light curve plots in this paper for observations composed of multiple visits were generated using this class.}
\item[\software{starproperties}]{-- provides the  \code{StarProperties} class for convenient handling of information about the target star. This class will automatically download and extract stellar atmospheric parameters for the target star from the SWEET-Cat catalogue \citep{Santos2013, Sousa2018}, if available.}
\item[\software{planetproperties}]{-- provides the  \code{PlanetProperties} class for convenient handling of information about planets orbiting the target star. This class will automatically download and extract the properties of the transiting planet from the TEPCat catalogue \citep{2011MNRAS.417.2166S}, if available.}
\end{description}

In addition to these sub-modules, the package distribution includes a script \software{make\_xml\_files} as an aid to planning and execution of observing requests, and the script \software{combine} to calculate the weighted mean of values with error estimates accounting for possible systematic errors using the algorithm described in appendix~\ref{sec:combine}.

Distribution of \pycheops\ is done via the python package index website.\footnote{\url{https://pypi.org/project/pycheops/}} Bug reports and software development are coordinated using \software{github}.\footnote{\url{https://github.com/pmaxted/pycheops}} Several examples that demonstrate and test the capabilities of \pycheops\ are included with the software distribution package in the form of Jupyter Notebooks.\footnote{\url{https://jupyter.org/}} These include an analysis of the \CHEOPS\ data for 4 eclipses of the transiting hot Jupiter WASP-189\,b first presented by \citet{2020A+A...643A..94L} and a tutorial based on the observation of a single transit of KELT-11\,b using the same data analysed by \citet{2021ExA....51..109B}. The ``pycheops cookbook'' included in the distribution provides installation instructions and data analysis recipes.

\subsection{Transit and eclipse models}

Transit light curves are calculated using the qpower2 algorithm \citep{2019A+A...622A..33M}. This algorithm uses an analytic approximation to efficiently calculate the flux blocked by a spherical planet of radius $R_{\rm p}$ orbiting a spherical star of radius $R_{\star}$ with an intensity profile described by the power-2 limb darkening law $I_{\lambda}(\mu) = 1 - c\left(1-\mu^{\alpha}\right)$, where $\mu$ is the cosine of the angle between the surface normal and the line of sight. The algorithm is accurate to about 100 ppm for broad-band optical light curves of systems with a star-planet radius ratio $k = R_{\rm p}/R_{\star} = 0.1$. This is sufficient to recover transit parameters accurate to $\pm 0.5$\% or better for planets with $k<0.15$ \citep{2019A+A...622A..33M}.

The parameters of the transit model for a planet with orbital semi-major axis $a$ and orbital inclination $i$ are as follows.
\begin{description}
\item[$T_0$]{ = time of mid-transit}
\item[$P$]{= orbital period in days}
\item[$b$]{= $a\cos(i)/R_{\star}$}
\item[$D$]{= $(R_{\rm p}/R_{\star})^2 = k^2$}
\item[$W$]{=$(R_{\star}/a)\sqrt{(1+k)^2 - b^2}/\pi$}
\item[$f_c$]{= $\sqrt{e}\,\cos(\omega)$}
\item[$f_s$]{= $\sqrt{e}\,\sin(\omega)$}
\item[$h_1$]{= $I_{\lambda}(\frac{1}{2}) = 1 - c(1-2^{-\alpha})$}
\item[$h_2$]{= $I_{\lambda}(\frac{1}{2}) - I_{\lambda}(0) = c2^{-\alpha}$}
\end{description}

For planets in circular orbits (eccentricity $e=0$), the parameter $W$ is the width of the transit in phase units and $b$ is the transit impact parameter. $D$ is the depth of the transit in the absence of limb darkening. The parameters $f_c$ and $f_s$ are used because they have a uniform prior probability distribution assuming that the eccentricity, $e$, and the longitude of periastron, $\omega$, both have uniform prior probability distributions \citep{2013PASP..125...83E, 2011ApJ...726L..19A}. The parameters $h_1$ and $h_2$ are used because suitable priors can be applied to these parameters independently based on the results from \citet{2018A+A...616A..39M}, at least for inactive solar-type stars -- see also \citet{2019RNAAS...3..117S} for the correct calculation of the physical limits on these parameters. 

The secondary eclipse model uses the same parametrisation for the geometry of the star-planet system. The additional parameters for this model are the planet-star flux ratio, $L$, and the correction for the light travel time across the orbit $a_c$. The eclipse models assumes that the flux distribution across the visible hemisphere of the planet is uniform. 

For the sampling of the posterior probability distribution of the model parameters within \code{Dataset} and \code{MultiVisit} we assume that $\cos i$, $\log k$ and $\log a/R_{\star}$ have uniform prior probability distributions. The logarithm of the prior probability distribution for the parameters of the transit model is then  \[\log(P(D, W, b)) = \log(2kW) - \log(k) - \log(a/R_{\star}),\] where the factor $2kW$ is the absolute value of the determinant of the Jacobian matrix \mbox{$ \bmath{J} = d(D, W, b)/d(\cos i, k, a/R_{\star})$} \citep{2008ApJ...689..499C}.

\subsection{Parameter decorrelation}
Trends in a data set due to instrumental noise are often correlated with parameters such as the instrument temperature, the position of the star on the detector, background count rate, etc. Removing these trends is known as {\it decorrelation} or {\it detrending} and the coefficient that relates the change in a parameter to the change in count rate is known as a {\it decorrelation parameter} or {\it detrending parameter}. Several decorrelation parameters are available for use within \pycheops. These decorrelation parameters can be included as free parameters in the analysis of transits and eclipses so that the covariance between the parameters of interest (transit depth, eclipse depth, etc.) and these ``nuisance parameters'' can be quantified. Of particular relevance to \CHEOPS\ are trends in count rate, $f$, that depend on spacecraft roll-angle, $\phi$. \CHEOPS\ is nadir-locked, which results in the rotation of the stellar field around the line-of-sight once per orbit. Stray light from the Earth (an important background contamination in the images) is highly dependent on the roll angle. The decorrelation parameters available to model these trends are ${df}/{d\sin(j\phi)}$ and ${df}/{d\cos(j\phi)}$. Within the module \code{Dataset} the decorrelation can be done for these roll-angle decorrelation parameters up to the 2$^{\rm nd}$ harmonic of the roll angle, i.e. $j=1,2$ or $3$. Within the module \code{MultiVisit} the decorrelation against roll angle is done implicitly, i.e. without explicit calculation of the decorrelation parameters, and there is no limit to the number of harmonics that can be used -- see Section \ref{sec:unroll} for details.  

Since sine and cosine functions have a range from $-1$ to $+1$, the magnitude of the decorrelation parameters ${df}/{d\sin(j\phi)}$ and ${df}/{d\cos(j\phi)}$ are approximately equal to the amplitude of the instrumental noise in the light curve due to correlations with each harmonic of $\phi$. In a similar way, we shift and scale the variables used for decorrelation  so that all the decorrelation parameters are approximately equal to the amplitude of the instrumental noise in the light curve that is correlated with the parameter. For example, decorrelation against the $x$ position of the star on the detector due to the pointing jitter of the spacecraft uses the variable \[\Delta x = \frac{x-(x_{\rm max}+x_{\rm min})/2}{\left(x_{\rm max} - x_{\rm min}\right)/2},\] where $x_{\rm min}$ and $x_{\rm max}$ are the minimum and maximum values of $x$, and similarly for $\Delta y$.
The metadata provided with \CHEOPS\ light curves includes estimates of the count rate in the photometric aperture due to three effects -- the background level in the images, photo-electrons from nearby stars accumulated during the CCD frame-transfer,\footnote{\CHEOPS\ has no shutter so pixels remain exposed during the readout process. During the 25\,ms of the frame transfer, each charge well collects light from each pixel crossed on its way to the storage area. This produces vertical ``smear'' trails on the image from nearby stars.} and extra counts accumulated during the exposure due to contamination of the photometric aperture by nearby stars. These are all positive quantities so we scale them between their minimum and maximum values so that the decorrelation is done against the variables \var{bg}, \var{smear} and \var{contam}, respectively, that range from 0 to 1.

Linear and quadratic trends with time, e.g. due to intrinsic stellar variability, can be accounted from using the decorrelation parameters \var{dfdt} and \var{d2fdt2}, respectively. The decorrelation is done against the variable $t-t_{\rm med}$, where $t_{\rm med}$ in the median observation time of observations for the visit in days.

\subsection{Internal reflections (glint)}
Bright objects within 24\degr\ from the target can cause internal reflections that appear as small peaks in the light curve once per spacecraft rotation cycle. We refer to this phenomenon as ``glint''. Glint due to moonlight does not occur at exactly the same spacecraft roll angle  every cycle because of the motion of the Moon on the sky during the observation. The module \code{Dataset} includes a function \code{add\_glint} that can be used to create a periodic cubic spline function to model this effect. The cubic spline is calculated using a least-squares fit to the residuals from the previous transit or eclipse fit to the light curve, or to the data either side of the transit or eclipse. The independent variable for this cubic spline is either the spacecraft roll angle, or the position angle of the Moon relative to the spacecraft roll angle on the sky. Once the glint function, $f_{\rm glint}(t)$ has been created, the light curve model will include a term \var{glint\_scale}$\times f_{\rm glint}(t)$. The factor $\var{glint\_scale}\approx 1$ can be included in the analysis as a free parameter so that the impact of the  uncertainty in correcting for glint can be quantified. The function added to the model to correct for glint also accounts for much of the instrumental noise due to spacecraft roll angle, so this feature can also be used as an alternative to linear decorrelation against $\sin(\phi)$, $\cos(\phi), \sin(2\phi)$, etc.

\subsection{Ramp effect}
Long-duration observations of bright stars with \CHEOPS\ sometimes show changes in the count rate at the start of a visit with an amplitude up to a few hundred parts-per-million (ppm) that decays smoothly over several hours. This is an instrumental effect caused by changes in the instrument point spread function (PSF). The changes in the PSF are correlated with temperature changes recorded at various points on the telescope tube, particularly the value of \var{thermFront\_2} provided in the metadata for each visit. Based on this correlation, the following equation has been developed to correct for the ``ramp'' effect:
\[ {\rm flux}_{\rm corrected} = {\rm flux}_{\rm measured}\times\left(1 + \beta_r\times(\var{thermFront\_2} + 12\degr {\rm C})\right).\]
The value of the coefficient $\beta_r$ varies from $\beta_r = 140\,{\rm ppm}\,\degr {\rm C}^{-1}$ to 
$\beta_r = 330\,{\rm ppm}\,\degr {\rm C}^{-1}$ for photometric aperture radii from 22.5\,pixels to 30\,pixels, respectively. This ramp correction is not implemented by default in \pycheops, but can be easily applied using the function \code{Dataset.correct\_ramp}. This empirical approach to correcting the ramp effect is sufficient for most purposes, but investigations are continuing into more complex methods that may provide a more accurate correction for this effect (Wilson et al. 2021, in prep.)

\subsection{Model selection}
\label{sec:bayesfac}
\subsubsection{Akaike and Bayesian information criteria}
For a model with $k$ free parameters and maximum likelihood $\hat{\cal{L}}$ for a fit to $n$ observations, the Akaike and Bayesian information criteria  have the following definitions:
 \[{\rm AIC} = 2k-2\ln(\hat{\cal L});\]
 \[{\rm BIC} = k\ln(n)-2\ln(\hat{\cal L}).\]
Models with a lower AIC and/or BIC have a better balance between the complexity of the model and the quality of the fit. For a least-squares fit to observations $o_i$ with independent Gaussian standard errors, $\sigma_i$, the log-likelihood for a model that predicts values $c_i$ is 
\begin{equation} 
\label{eqn:lnlike}
\ln({\cal L}) = -\frac{\chi^2}{2} -\frac{1}{2}\sum_{i=1}^{n}\ln(\sigma_i^2) -\frac{n}{2}\ln(2\pi), 
\end{equation}
where 
\[\chi^2 = \sum_{i=1}^{n}\frac{(o_i-c_i)^2}{\sigma_i^2} .\]
The constant $-\frac{n}{2}\ln(2\pi)$ is sometimes dropped from this definition. This is the case for the values of the AIC and BIC returned by functions in \software{lmfit}, but not for the log-likelihood values returned by \software{celerite2}. For consistency, and to enable like-for-like comparison, we overwrite the values of AIC and BIC returned by \software{lmfit} with values calculated using equation (\ref{eqn:lnlike}) before the values are reported in the output from routines in \code{Dataset} and \code{MultiVisit}.
\subsubsection{Bayes Factors}
The question of which decorrelation parameters to include in the analysis of a given light curve is a model selection problem. For nested models $M_0$ and $M_1$ with parameters $\theta_0 = \{p_1, p_2, \dots, p_n, 0\}$ and $\theta_1 = \{p_1, p_2, \dots, p_n, \psi\}$, given the data ${\cal D}$, the Bayes factor $B_{01}$ is defined by  
\[ \frac{P(M_0\,|\,{\cal D})}{P(M_1\,|\,{\cal D})} = \frac{P(M_0)}{P(M_1)} \frac{P({\cal D}\,|\,M_0)}{P({\cal D}\,|\,M_1)} = \frac{P(M_0)}{P(M_1)} B_{01},\]
where
$ P({\cal D}\,|\,M_0) = \int P({\cal D}\,|\,\theta_0)P(\theta_0)d^n\theta$ and similarly for $P({\cal D}\,|\,M_1)$. $P(\theta_0)$ is the prior probability distribution for the parameters of model $M_0$. The prior on the extra parameter $\psi$ is the same for both models so we can use the Savage-Dickey density ratio \citep{dickey1970weighted, 2007MNRAS.378...72T} to calculate the Bayes factor
\[ B_{01} = \frac{P(\psi=0\,|\,{\cal D})}{P(\psi=0)}. \]
For a parameter assumed to have  a normal prior with standard deviation \mbox{$\sigma_0$, $P(\psi=0) = 1/\sigma_0\sqrt{2\pi}$}.

 For the specific case where ${\cal D}$ is a \CHEOPS\ light curve, we find that the posterior probability distributions for the decorrelation parameters are usually well-behaved and close to Gaussian, as expected for a linear model. Assuming that they are normally distributed and that the standard deviation is given accurately by the error on the parameter given by \software{lmfit}, and that {\it a priori} the two models are equally likely, we can calculate the Bayes factor for models with/without a parameter with value $p\pm\sigma_p$ using
\[B_p = e^{-(p/\sigma_p)^2/2}\,\sigma_0/\sigma_p.\]
These Bayes factors are listed in the output from the \code{lmfit\_report} method for \code{Dataset} objects. Parameters with Bayes factors $\goa 1$  are not supported by the data and can be removed from the model. This statistic is only valid for comparison of the models with/without one parameter, so parameters should be added or removed one-by-one and the test repeated for every new pair of models.

\subsection{Noise models}
\label{sec:noise}
 The standard error estimates provided with \CHEOPS\ light curves account for the known sources of noise in the data, e.g. photon-counting statistics, detector read-out noise, errors in background subtraction, etc. There will be additional sources of noise that are not accounted for in these error estimates, e.g. undetected cosmic ray hits to the detector, variability of stars that contaminate the photometric aperture, thermal effects, scattered light, intrinsic variability of the target star, etc. The fitting routines in \code{Dataset} and \code{MultiVisit} include a parameter $\sigma_{\rm w}$ that accounts for this additional noise assuming that it is Gaussian white noise, i.e. a process that perturbs each measurement independently by some amount that has a normal distribution. The log-likelihood for the model using the same notation as above is then
 \[ \ln({\cal L}) = -\frac{\chi^2}{2} -\frac{1}{2}\sum_{i=1}^{n}\ln(\sigma_i^2+\sigma_w^2) -\frac{n}{2}\ln(2\pi), \]
where 
\[\chi^2 = \sum_{i=1}^{n}\frac{(o_i-c_i)^2}{\sigma_i^2+\sigma_w^2}.\]

The fitting routines in \code{Dataset} and \code{MultiVisit} can use a more sophisticated noise model that accounts for correlated noise assuming that this is described by a Gaussian process. The kernel that describes the correlations between observations obtained at times $t_n$ and $t_m$ is the \var{SHOTerm} kernel implemented in \software{celerite2}, i.e.
\begin{multline*}
k_\mathrm{SHO}(\tau;\,S_0,\,Q,\,\omega_0) = 
S_0\,\omega_0\,Q\,e^{-\frac{\omega_0\,\tau}{2Q}}\times \\
\begin{cases}
    \cosh{(\eta\,\omega_0\,\tau)} +
        \frac{1}{2\,\eta\,Q}\,\sinh{(\eta\,\omega_0\,\tau)}, & 0 < Q < 1/2;\\
    2\,(1+\omega_0\,\tau), & Q = 1/2;\\
    \cos{(\eta\,\omega_0\,\tau)} +
        \frac{1}{2\,\eta\,Q} \sin{(\eta\,\omega_0\,\tau)},& 1/2 < Q; \\
\end{cases}
\end{multline*}
where $\eta = \vert 1-(4\,Q^2)^{-1}\vert^{1/2}$ and  $\tau_{nm} = |t_n-t_m|$. This kernel represents a stochastically-driven, damped harmonic oscillator, and is commonly used with  $Q=1/\sqrt{2}$ to model granulation noise in stars \citep{celerite1}. The software package \software{celerite2} is used  to calculate the log-likelihood to observe a light curve for a given model and choice of the hyper-parameters $Q$, $\omega_0$ and $S_0$. The damping time scale for this process is $\tau =
2\,Q / \omega_0$ and the standard deviation of the process is $\sigma_{\rm GP} = \sqrt{S_0\,\omega_0\,Q}$. The \code{Dataset} module includes a function to plot the fast Fourier transform (FFT) of the residuals from the best-fit transit or eclipse model in log-log space so that the user can look for a slope or peaks in the power spectrum due to stellar granulation or oscillations \citep{2020A+A...636A..70S}.

\subsection{Implicit correction for trends correlated with spacecraft roll angle}
\label{sec:unroll}
The field of view of the \CHEOPS\ instrument rotates at an angular frequency $\Omega\approx 2\pi/98.725$ radians/minute. This rotation introduces instrumental noise at this frequency and its harmonics. The \CHEOPS\ point spread function (PSF) is approximately triangular in shape so to account for instrumental noise not removed by the data reduction pipeline (DRP) we typically  use a linear model of the form  $\sum_{j=1}^3 \alpha_j\sin(j\cdot\Omega t) + \beta_j\cos(j\cdot\Omega t)$. Adding the 6 extra coefficients $\alpha_j, \beta_j$ as free parameters in the analysis of a single observing sequence (``visit'') is not generally a problem, but this becomes inconvenient for the analysis of larger data sets because different coefficients are needed for each visit. Instead of explicitly including the nuisance parameters $\alpha_1, \beta_1, \dots$ in our analysis, we can marginalise over them using the trick described by \citet{2017RNAAS...1....7L}. This trick ({\it implicit decorrelation}) requires that we assume Gaussian priors on these nuisance parameters, in which case the likelihood to obtain the observed data $\bmath{y}$ from a mean model $\bmath{\mu}({\bmath \theta})$ with parameters $\bmath{\theta}$ is a multivariate normal distribution of the form  
\begin{equation}
p(\bmath{y} | \bmath{\theta}) = \mathcal{N}(\bmath{y};\bmath{\mu}, \mathbfit{C} + \bmath{A}\bmath{\Lambda}\bmath{A}^T ), 
\end{equation} 
 where the columns of the matrix $\bmath{A}$ are the basis functions of our instrumental linear model, i.e. $\sin(\Omega t), \cos(\Omega t)$, etc., and $\mathbfit{C}$ is the covariance matrix that describes the measurement errors on $\bmath{y}$. If we assume independent Gaussian priors on the nuisance parameters all with the same standard deviation $\sigma_\Omega$ then $\bmath{\Lambda} = \sigma_{\Omega}\bmath{I}$. The term $\sigma_{\Omega}\bmath{A}\bmath{A}^T$ is of the form \[\sum_{j=1}^{N_{\rm roll}} a_j\,e^{-c_j\,\tau_{nm}}\,\cos\left(d_j\,\tau_{nm}\right) + 
    b_j\,e^{-c_j\,\tau_{nm}}\,\sin\left(d_j\,\tau_{nm}\right),\]
where $\tau_{nm} = |t_n-t_m|$ for observations obtained at times $t_n$ and $t_m$. This means we can easily calculate the likelihood $p(\bmath{y}\,|\,\bmath{\theta}, \bmath{\alpha} )$ using the \software{celerite2} algorithm developed by \citet{celerite2}. Some simple trigonometry is sufficient to show that \mbox{$b_j=c_j=0$} and  \mbox{$\bmath{\alpha} = \left\{a_j = \sigma_\Omega, d_j= j\Omega, j=1, 2, \dots N_{\rm roll}\right\}$}. This instrumental noise model can be combined with both the white noise and the correlated noise models described in Section~\ref{sec:noise}.
  
 This implicit roll-angle decorrelation method is implemented in the sub-module \code{MultiVisit}. The number of harmonic terms $N_{\rm roll}$ can be selected with the keyword option \var{nroll}. \CHEOPS' roll-angle rotation rate  is not exactly constant, particularly for stars far from the celestial equator, so implicit decorrelation may not be as effective as explicit decorrelation using the parameters ${df}/{d\sin{\phi}}$, etc. This issue can be ignored if the trends with roll angle are weak, or mitigated by using a larger value of  $N_{\rm roll}$. A third option is to use the \var{unwrap} keyword option to remove the best-fit roll-angle trend from each data set prior to analysis with \code{MultiVisit} using implicit roll-angle decorrelation.
This is done by dividing the light curve data from each visit by the values generated by the following function: 
\[
1 + \sum_j \sin(j\phi(t_i)) {df}/{d\sin(j\phi)} + \cos(j\phi(t_i)){df}/{d\cos(j\phi)} ,\]
where $\phi(t_i)$ is the spacecraft roll angle at observation time $t_i$. The decorrelation parameters ${df}/{d\sin(j\phi)}$ and ${df}/{d\cos(j\phi)}$ are the best-fit values taken from the last fit to the light curve. These best-fit parameter values are stored together with other details of the fit when the data set is saved to an output file. For trends correlated with parameters other than roll angle,  \code{MultiVisit} automatically selects the same decorrelation parameters that were used in the last fit to the light curve from each visit.

\subsection{Analytic maximum-likelihood transit fit}
 A key part of the science case for the \CHEOPS\ mission is to have a facility that can be used to search for transits of small exoplanets orbiting bright stars discovered in radial velocity surveys.  The analysis of the long visits used to search for transits benefits from a method to inject and recover synthetic transits in the light curve. Transit injection and recovery can also be used to characterise the noise in the light curve on different time scales. The method we have developed for this task, described below, is implemented in the  \pycheops\ function \code{scaled\_transit\_fit}.

We can use a factor $s$ to modify the transit depth in a nominal model $\bmath{m_0}$ calculated with approximately the correct depth that is scaled as follows:
\[ \bmath{m}(s) = 1 + s\times(\bmath{m_0}- 1).\]
The data are normalised fluxes $\bmath{f} = f_1, \dots, f_N$ with nominal errors $\bmath{\sigma} = \sigma_1,\dots, \sigma_N$. 
Assume that the actual standard errors are underestimated by some factor $\beta$, and that these are normally distributed and independent, so that the log-likelihood is
\[\ln {\cal L}  = -\frac{1}{2b^2}\chi^2 - \frac{1}{2}\sum_{i=1}^N \ln \sigma_i^2  - N\ln \beta - \frac{N}{2}\ln(2\pi)\]
where
\[\chi^2 = \sum_i^N \frac{\left((f_i - 1) - s(m_{0,i}-1)\right)^2}{\sigma_i^2}. \]

The maximum likelihood occurs for parameter values $\hat{s}$, and $\hat{\beta}$ such that
$\left. \frac{\partial  \ln {\cal L}}{\partial s}\right|_{\hat{s},\hat{\beta}} = 0$
and 
$\left. \frac{\partial  \ln {\cal L}}{\partial \beta}\right|_{\hat{s},\hat{\beta}} = 0,$
from which we obtain 
\[\hat{s} = \sum_{i=1}^N \frac{(f_i - 1)(m_{0,i}-1) }{\sigma_i^2} \left[ \sum_{i=1}^N \frac{(m_{0,i}-1)^2}{\sigma_i^2}\right]^{-1},\]
 and 
\[\hat{\beta} = \sqrt{\chi^2/N}.\]
 
 For the standard errors on these parameters we use 
$\sigma_s^{-2}  = -\frac{\partial^2\ln{\cal L}}{\partial^2 s^2} \left.\right|_{\hat{s},\hat{\beta}}$
 and
$\sigma_\beta^{-2}  = -\frac{\partial^2\ln{\cal L}}{\partial^2 \beta^2} \left.\right|_{\hat{s},\hat{\beta}}$ 
to derive
\[\sigma_s = \beta\left[\sum_{i=1}^N \frac{(m_{0,i}-1)^2}{\sigma_i^2}\right]^{-1/2}\] 
 and
\[\sigma_\beta  = \left[3\chi^2/\beta^4 -N/\beta^2\right]^{-1/2} .\]
Whether or how much of the data outside transit to include depends on whether these data can be assumed to have the same noise characteristics as the data in transit. Note that including these data has no effect on $s$ or $\sigma_s$, because of the factors $(m_{0,i}-1)$ in their calculation, but will affect the estimates of $\beta$ and $\sigma_\beta$.

If the noise scaling factor $\beta$ is large ($\goa 2$) then it may be more appropriate to assume that the nominal errors provided with the data are a lower bound to the true standard errors, e.g. if there is an additional noise source that is not well quantified such as poor cosmic-ray rejection. We can assume that actual standard error on observation number $k$ is $\sigma_k$ with probability distribution 
\[ P(\sigma_k\,|\,\sigma_{0,k}) = \left\{
\begin{array}{ll}
0 & \sigma_k < \sigma_{0,k}\\
\frac{\sigma_{0,k}}{\sigma_k^2}  & \sigma_k \ge \sigma_{0,k}
\end{array}
\right.
\]
This is a less informative prior on the standard error distribution than the ``error scaling'' method and so the results tend to be more pessimistic. Assuming independent measurements and uniform priors, the posterior probability distribution is then
\[\ln {\cal L}  = C + \sum_{k=1}^N \ln \left[ \frac{1 - \exp(-R_k^2/2)}{R_k^2} \right], \]
where $C$ is a normalising constant and $R_k = (m_k - f_k)/\sigma_{0,k}$ \citep[][section 8.3.1]{Sivia2006}. This is a function of one parameter only so the minimum can be found efficiently using any suitable numerical algorithm. The standard error on $s$ is then found from the values of $s$ that give a log-likelihood that is 0.5 less than the maximum log-likelihood, i.e. one standard deviation (1-$\sigma$) assuming a Gaussian distribution.

\subsection{Mass and radius calculations for the star and planet}
The analysis of the light curve for a transiting exoplanet in a circular orbit provides constraints on three geometrical parameters -- the scaled semi-major axis, $a/R_{\star}$, the planet-star radius ratio, $k=R_{\rm p}/R_{\star}$, and the impact parameter, $b=a\cos i/R_{\star}$ \citep{2003ApJ...585.1038S}. Kepler's law can be used to convert the parameter $a/R_{\star}$ to a direct constraint on the mean stellar density 
\begin{equation}
\label{eqn:aR}
  \mbox{$\rho_{\star}$} =
   \frac{3\mbox{M$_{\star}$}}{4\pi R_{\star}^3} =
   \frac{3\pi}{GP^2(1+q)}
    \left(\frac{a}{\mbox{R$_{\star}$}}\right)^3. 
  \end{equation}
In general, the mass ratio $q=M_{\rm p}/M_{\star}$ is negligible for transiting exoplanets.
The same information is available from the analysis of transits for planets in non-circular orbits provided that independent constraints are available for both the eccentricity, $e$, and the longitude of periastron, $\omega$ \citep{2014MNRAS.440.2164K}. These parameters combined with the semi-amplitude of the star's spectroscopic orbit due to the planet, $K$, lead directly to a measurement of the planet's surface gravity, 
\begin{equation}
\label{eqn:g_p}
g_{\rm p} = \frac{2\pi}{P}\frac{(1-e^2)^{1/2}K}{(R_{\rm p}/a)^2\sin i}  
\end{equation} 
\citep{2007MNRAS.379L..11S}. One more constraint is needed to obtain the mass and radius of the planet. This is typically an estimate for either the mass or radius of the host star. Estimates for both mass and radius will be needed in cases where the stellar density is poorly constrained by the light curve, e.g. if the transits are shallow compared to the noise. 

 The function \code{funcs.massradius} within pycheops implements these calculations using the nominal solar and planetary constants defined in the module \code{constants}. Confidence limits and standard errors on parameters are calculated using a Monte Carlo approach with a sample of 100\,000 values per parameter. For parameters specified as a mean with standard error the sample of values is generated assuming a normal distribution. For parameters provided as a sample of points from the posterior probability distribution (PPD), e.g. using the output from \software{emcee}, we select 100\,000 values from the sample, with re-selection if required. Where multiple input samples with the same length are provided, e.g. samples generated from \software{emcee}, values are sampled in a way that preserves correlations between these parameters. Output statistics generated from the Monte Carlo sample include: mean, median and half-sample mode, standard error, and asymmetric error bars calculated from the 15.9\%, median and 84.1\% percentile points of the sample. The function \code{funcs.massradius} accepts input of the parameters $M_{\star}$, $R_{\star}$ and $a/R_{\star}$ independently, so it is possible to calculate a value of $\rho_{\star}$ from $a/R_{\star}$ that is inconsistent with the input values of $M_{\star}$ and $R_{\star}$. This leads to an ambiguity over which values of $M_{\star}$ and $R_{\star}$ to use in the calculation of the planet mass and radius. To resolve this ambiguity, $R_{\rm p}$ is calculated from $k$ and $R_{\star}$, and is only calculated if both of these values are provided. Similarly, $g_{\rm p}$ is only calculated from equation (\ref{eqn:g_p}). The mean planet density, $\rho_{\rm p}$, is calculated from $g_{\rm p}$ and $m_{\rm p}$, i.e. the input value of  $R_{\star}$ is not used in the calculation of $\rho_{\rm p}$. The sub-modules \code{MultiVisit} and \code{Dataset} both provide \code{massradius} class methods that use the output from the last fit to the light curve(s) as input to \code{funcs.massradius}. For these class methods, if only one of the parameters  $M_{\star}$ or $R_{\star}$ is provided by the user then the other is calculated from $a/R_{\star}$ using equation (\ref{eqn:aR}).
 
 If the width of the transit is not well defined by the light curve itself, e.g. due to gaps in the light curve or if the transit is shallow, then it is very useful to place a prior on the mean stellar density. As can be seen from equation (3), this stellar property is directly related to the parameter $R_{\star}/a$ and this parameter is itself directly related to the transit width, e.g. for circular orbits the transit width in phase units is $W=(R_{\star}/a)\sqrt{(1+k)^2 - b^2}/\pi$. The \code{StarProperties} class can be used to estimate the mean stellar density, $\rho_{\star}$, for stars with surface gravities $3.697<\log g < 4.65$ using a linear relation between $\log(\rho_{\star})$ and $\log g$ derived using the method and data described in \citet{2018ApJS..237...21M}.


 \begin{table*}
    \centering
    \caption{Log of \CHEOPS\ observations. Data sets are labelled by the sequence number given in the first column throughout this paper. Effic. is the fraction of the observing interval covered by valid observations of the target. $R_{\rm ap}$ is the aperture radius used to compute the light curve analysed in this paper. The column $T_\mathrm{exp}$  gives the exposure time in terms of the integration time per image multiplied by the number of images stacked on-board prior to download.} 
    \label{tab:obslog}
    \begin{tabular}{@{}llrrrrrrrr}
\hline\hline
\# & Target    & \multicolumn{1}{l}{G}  & 
\multicolumn{1}{l}{Start date}          & 
\multicolumn{1}{l}{Duration}            & 
\multicolumn{1}{l}{$T_\mathrm{exp}$}  & 
\multicolumn{1}{l}{$N_{\rm obs}$} & 
\multicolumn{1}{l}{Effic.} & 
\multicolumn{1}{l}{File key} & 
\multicolumn{1}{l}{$R_{\rm ap}$} \\
      &  &[mag]     & \multicolumn{1}{c}{[UTC]}   &     \multicolumn{1}{c}{[s]}         &     &        & [\%] &      & [pixels]    \\
\hline
1 & GJ~436    & 9.57 & 2020-03-27T23:56:16 & 27433 &  $1\times 60$s & 340 &   74 & CH\_PR100041\_TG000302\_V0102 & 25.0 \\
2 &           &      & 2020-04-02T06:53:35 & 27433 &  $1\times 60$s & 334 &   73 & CH\_PR100041\_TG000303\_V0102 & 25.0 \\
3 &           &      & 2020-04-23T11:05:36 & 28153 &  $1\times 60$s & 300 &   64 & CH\_PR100041\_TG001301\_V0102 & 25.0 \\
\noalign{\smallskip}
1 & HD~106315 & 8.89 & 2020-04-02T22:43:57 & 87305 &  $1\times 41$s & 1954     & 92  & CH\_PR100041\_TG000802\_V0102 & 25.0\\
2 &           &      & 2020-05-01T14:59:19 & 85992 &  $1\times 41$s & 1510     & 72  & CH\_PR100041\_TG001401\_V0102 & 25.0\\
\noalign{\smallskip}
1 & HD~97658  & 7.51 & 2020-04-22T04:59:16 & 27650 &  $3\times 11$s & 607 &  72  & CH\_PR100041\_TG001201\_V0102 & 25.0\\
\noalign{\smallskip}
1 & GJ~1132   &12.14 & 2020-03-26T23:52:36 & 26052 & $1\times 60$s & 301 &  70     & CH\_PR100041\_TG000401\_V0102 & 15.5 \\
2 &           &      & 2020-03-28T14:27:57 & 27613 & $1\times 60$s & 269 &  58     &  CH\_PR100041\_TG000402\_V0102 & 15.0 \\
3 &           &      & 2020-04-04T02:48:40 & 30674 & $1\times 60$s & 314 &  61     &  CH\_PR100041\_TG000403\_V0102 & 15.0 \\
\hline
    \end{tabular}
\end{table*}
 

\begin{table*}
    \centering
    \caption{Summary of the initial analysis for individual visits for targets with more than one visit using \code{Dataset.lmfit\_transit}. $T_{\rm c}$ is the time of mid-transit and RMS is the standard deviation of the residuals from the best fit. The numbering of the visits is the same as in Table~\ref{tab:obslog}. Note that the standard errors quoted here are based on the estimated covariance matrix, so may be underestimated. Values preceded by $=$ were held fixed in the analaysis. Data from the individual visits to GJ~1132 provide no useful constraint on the impact parameter, $b$. The variables in final column are as follows: time, $t$; spacecraft roll angle, $\phi$,  PSF centroid position, ($x$, $y$); smear correction, \var{smear}; aperture contamination, \var{contam}; image background level, \var{bg}. } 
    \label{tab:individual}
    \begin{tabular}{@{}llrrrrrl}
\hline\hline
\# & Target    & 
\multicolumn{1}{c}{BJD$_{\rm TDB}$ T$_{\rm c}$}  & 
\multicolumn{1}{c}{D}  & 
\multicolumn{1}{c}{W}  & 
\multicolumn{1}{c}{b}  & 
\multicolumn{1}{c}{RMS}          & 
Decorrelation parameters\\
 & &  \multicolumn{1}{c}{$-2458900$} & \multicolumn{1}{c}{[\%]} &   &  & \multicolumn{1}{l}{[ppm]}      &     \\
\hline
1 & GJ~436    & $36.6865 \pm 0.0001$ & $0.49 \pm 0.05$ & $0.0156 \pm 0.0004$ & $0.74 \pm 0.03$ & 262 & $t$ \\
2 &           & $41.9746 \pm 0.0010$ & $0.63 \pm 0.03$ & $0.0160 \pm 0.0005$ & $0.77 \pm 0.02$ & 265 & $t$, \var{contam}, \var{bg} \\
3 &           & $63.1321 \pm 0.0003$ & $0.65 \pm 0.01$ & $0.0196 \pm 0.0002$ & $0.67 \pm 0.02$ & 266 & $t$, $\sin(\phi)$ \\
\noalign{\smallskip}
1 & HD~106315 & $42.9442 \pm 0.0011$ &$0.031 \pm 0.002$& $0.0161 \pm {\bmath 0.0002}$ & $0.63 \pm 0.04$ & 238 & $t$, $x$, \var{bg}, \var{smear}, $x$\\  
2 &           & $\bmath{71.592 \pm 0.013}$ &$\bmath{0.027} \pm 0.002$& $\bmath{0.0160 \pm 0.0005}$ & ${\bmath =0.63}$ & 250 & $\sin(\phi)$, $x$ \\  
\noalign{\smallskip}

1 & GJ~1132   & $=35.6559$           & $0.30 \pm 0.03$ & $0.0193 \pm 0.0003$ &  =0.77 & 1262 & \var{contam}, \var{smear}, $t$, $\cos(\phi)$, $\sin(2\phi)$, $\cos(2\phi)$, \\
2 &           & $=37.2849$           & $0.22 \pm 0.04$ & $0.0118 \pm 0.0018 $&  =0.77 & 1125 & \var{contam}, \var{bg}, $t$, $x$, $y$, $\sin(\phi)$, $\cos(2\phi)$, $\sin(2\phi)$ \\ 
3 &           & $=43.8006$           & $0.27 \pm 0.13$ & $0.0138 \pm 0.0012 $&  =0.77 & 1408 & \var{contam}, \var{bg}, $t$, $\sin(\phi)$, $\cos(2\phi)$ \\ 
\hline
    \end{tabular}
\end{table*}

\section{Early Science programme}
\label{sec:analysis}
In this section we report the results from the first exoplanet transits observed by \CHEOPS\ during its Early Science programme for four well-known exoplanets: GJ~436\,b, HD~106315\,b, HD~97658\,b and GJ~1132\,b. These observations are used to assess the in-flight performances of \CHEOPS\ for measuring transit parameters, and to compare this performance with the results obtained by reanalysing  transit light curves from the {\it {\it Kepler} K2} mission, {\it TESS}, and Spitzer Space Telescope ({\it Spitzer}, hereafter). The targets were selected from a list of well-known transiting exoplanets based on their visibility around the dates when \CHEOPS\ nominal science operations were due to start. Several targets were selected in order to demonstrate the capabilities of \CHEOPS\ for transiting planets over a range of stellar and planetary properties. The Early Science programme also includes observations of the eclipses of WASP-189\,b, the orbital phase curve of 55~Cnc\,b and the transits of $\nu^2$~Lupi\,b. The results from these observations are reported elsewhere \citep{2020A+A...643A..94L,delrez2021,morris2021}.

\subsection{Observations}
The log of \CHEOPS\ observations is presented in Table~\ref{tab:obslog}. The data set comprises three transits each for GJ~436\,b and GJ~1132\,b, two transits of HD~106315\,b and one transit of HD~97658\,b. \CHEOPS\ observes from low-Earth orbit so observations are often  interrupted because the line of sight to the target is blocked by the Earth or because the satellite is passing through the South Atlantic Anomaly (SAA). The ratio between the uninterrupted observation time and the total duration of the observation sequence (``visit'') is also noted in Table~\ref{tab:obslog} and is at least 58\% for all of the visits analysed here.

\subsection{Photometric extraction}
\label{sec:drp}
All \CHEOPS\ data are automatically processed at the \CHEOPS\ science operations centre (SOC). The data reduction pipeline (DRP) calibrates the raw images, e.g. it applies bias, gain and non-linearity corrections, subtracts the dark current and scattered light, and applies a flat-field correction. The \CHEOPS\ field of view rotates continuously so the photometric aperture used to measure the flux from the target star is periodically contaminated by the read-out trail from other stars {\bf on the CCD}. This ``smear'' effect is also corrected for by the DRP. The DRP also simulates the field of view based on the positions and magnitudes of the target and nearby stars as listed in the {\it Gaia} DR2 catalogue \citep{2018A+A...616A...1G}. The contamination of the photometric aperture by nearby stars is reported in the DRP data products so that the user has the option to apply or ignore this contamination correction. Light curves are calculated using three pre-defined aperture radii with radii of 22.5, 25 and 30 pixels\footnote{The image scale for \CHEOPS\ is 1 arc~second per pixel.} labelled RINF, DEFAULT and RSUP, respectively. Light curves labelled OPTIMAL are also provided for a fourth aperture radius calculated to maximise the signal-to-noise ratio for the target while minimising contamination from other stars in the image. The data files generated by the DRP include a data reduction report that summarizes each data processing step and that provides various data quality metrics. Full details can be found in \citet{Hoyer20}. All light curves in this paper were processed using \CHEOPS\ DRP version cn03-20200703T111359.
 
 \subsection{Host star characterisation}
 \label{sec:irfm}
For all targets we determined the stellar radii utilising a modified version of the infrared flux method (IRFM; \citealt{Blackwell1977}). The method allows for derivation of angular diameters of stars using known relationships between this parameter, stellar effective temperature, and an estimate of the apparent bolometric flux. The angular diameter combined with the parallax can then be used to calculate the stellar radius. In this study we used a Markov-chain Monte Carlo (MCMC) method to compare the synthetic fluxes, determined by attenuating stellar atmospheric models with a galactic extinction law parameterised by the reddening $E(B-V)$. The reddened spectra were convolved with the broadband response functions for the chosen bandpasses. These were compared to the observed {\it Gaia} G, G$_{\rm BP}$, and G$_{\rm RP}$, 2MASS J, H, and K, and {\it WISE} W1 and W2 fluxes and relative uncertainties retrieved from the most recent data releases \citep{GaiaCollaboration2020,Skrutskie2006,Wright2010} in order to obtain the apparent bolometric fluxes. The resulting angular diameters are combined with the offset-corrected {\it Gaia} EDR3 parallaxes \citep{Lindegren2020} to derive stellar radii.

In this study, we used the \software{ATLAS} stellar atmospheric models \citep{Castelli2003} for HD\,106315 and HD\,97635, however for the cooler stars in the sample (GJ\,436 and GJ\,1132) we adopted the radii derived using \software{Phoenix} models \citep{Allard2011} as these spectral energy distributions contain molecular band absorption that can be important in the characterisation of M-dwarfs. Atmospheric models for calculation of the synthetic photometry were built from stellar parameters measured from the analysis of the star's spectrum, as described in the individual subsections on each star below.

For each star the effective temperature, $T_{\mathrm{eff}}$, the metallicity, [Fe/H], and the radius, $R_{\star}$, were used as input parameters to infer the mass $M_{\star}$ and age $t_{\star}$ from two different sets of stellar evolutionary models, namely PARSEC v1.2S \citep{marigo17} and CLES \citep{scuflaire08}. The isochronal $M_{\star}$ and $t_{\star}$ from PARSEC v1.2S were derived by applying the grid-based interpolation method known as isochrone placement and described in \citet{bonfanti15,bonfanti16}. In the case of CLES, instead, a direct computation of the evolutionary track based on the set of input parameters was performed. The consistency of the two pairs was successfully checked following the validation procedure based on the $\chi^2$ test presented in details in \citet{2021A+A...646A.157B}, so that we finally merge the two probability distributions of both $M_{\star}$ and $t_{\star}$ and computed their respective medians and standard deviation.

The results and additional details of the analysis are presented separately for each target in the subsection below. Photospheric abundance ratios are quoted relative to the solar composition from \citet{2009ARA+A..47..481A}.


\subsection{Light curve analysis}
\label{sec:lcanal}

We used \pycheops\ version 1.0.0 to analyse the data.  The photometric aperture was selected based on the lowest point-to-point root mean square (RMS) reported in the data reduction reports. The correction for contamination calculated by the DRP was applied to all light curves. We applied a correction for the ramp effect to all data sets apart from the observations of GJ~1132. This correction is generally very small ($\loa 100$ ppm). Observations with high background levels due to observing close to the Earth's limb ($>5$\% above the median background level) were excluded from the analysis. We also excluded data points more than 5 standard deviations from a median-smoothed version of each light curve. Typically, fewer than 5 data points are rejected from the analysis using this criterion. 

To select decorrelation parameters we did an initial fit to each light curve with no decorrelation and used the RMS of the residuals from this fit, $\sigma_p$, to set the prior on the decorrelation parameters, $\mathcal{N}(0, \sigma_p)$ or, for ${df}/{dt}$,  $\mathcal{N}(0, \sigma_p/\Delta t)$ where $\Delta t$ is the duration of the visit. We then added decorrelation parameters to the fit one-by-one, selecting the parameter with the lowest Bayes factor at each step and stopping when $B_p>1$ for all remaining parameters. This process sometimes leads to a set of parameters including some that are strongly correlated with one another and so are therefore not well determined, i.e. they have large Bayes factors. We therefore go through a process of repeatedly removing the parameter with the largest Bayes factor if any of the parameters have a Bayes factors $B_p>1$. The second step of this process typically removes no more than 1 or 2 parameters.

Gaussian-process (GP) regression is an effective way to account for the additional uncertainty in the parameters derived from observational data in cases where the time-correlated noise sources (``systematics'') are present. The use of GP regression is common practice within the exoplanet research community, partly because much of the research into exoplanets for the first two decades of this relatively new branch of astrophysics had to use instrumentation that was never designed to observe the weak signals from exoplanet systems. Time-correlated noise sources may arise within the instrument, the environment (particularly for ground-based observations) or from astrophysical noise sources, e.g. intrinsic variability of the host star. By design, \CHEOPS\ has very low levels of instrumental noise. Analysis of long-duration observations of bright stars with \CHEOPS\ have demonstrated that instrumental noise is between 15 and 80 ppm on timescales of a few hours for isolated stars in the magnitude range covered here. These observations also show that the standard error estimates on the count rates provided with the DRP data files are reliable but slightly under-estimate the true noise in the light curves by a factor $\approx 1.3$.   This may be due to small errors in the calibration of the data, e.g. flat-fielding errors, or weak cosmic ray events that are difficult to identify if they affect pixels near the peaks in the image of the star. To account for this small amount of extra noise we assume that it is Gaussian white noise with standard deviation $\sigma_w$. The amplitude of the noise due to stellar granulation and stochastically-driven oscillations for late-type star has been characterised in detail using data from the {\it Kepler} mission \citep{2014A+A...570A..41K}. For dwarf stars ($\log g \loa 4$), the amplitude of this noise on time scales relevant to the observations presented here ($\sim 10^2$ -- 10$^3$ $\mu$Hz) is typically no more than 100 ppm.  Therefore, there is little justification {\it a priori} to include a GP in the analysis of a \CHEOPS\ light curve for moderately bright dwarf stars. For all the light curves analysed here, we checked that the power spectrum plotted in log-log space is flat, i.e. consistent with white noise, as expected. Consequently, we do not include GPs in the analysis of the light curves analysed here. Note that the same argument does not apply to subgiant stars, e.g. we observed granulation noise in the \CHEOPS\ light curve of KELT-11 ($\log g \approx 3.7$) and included a GP in the analysis of that system using \pycheops\ \citep{2021ExA....51..109B}. Similarly, \CHEOPS\ is able to detect and characterise granulation noise and solar-like oscillations for very bright Sun-like stars such as $\nu^2$~Lupi \citep[V=5.65,][]{delrez2021}.

For all of the visits analysed here, we repeated the analysis using different photometric apertures, or without rejecting data with high background levels, or without the correction for the ramp effect, or (except for GJ 1132) excluding the correction for contaminating background stars. For the analysis with \code{MultiVisit} we also experimented with different values $N_{\rm roll}$. In all these cases, the results are negligibly different to the results reported here. 

Sampling of the PPD for the model parameters is done with \software{emcee} using 256 walkers and 512 steps following a ``burn-in'' phase of 1024 steps to ensure that the sampler has converged. Convergence of the sampler was checked using visual inspection of the parameters values from all the walkers plotted versus step number. These ``trail plots'' show no trends in mean value or width and all the walkers appeared to be randomly sampling the parameter values in very similar way.

For convenience, the light curves are normalized to their median value prior to analysis. We store the original light curve prior to normalisation and use this {\it post hoc} to convert the parameter $c_i$ used to model the out-of-transit level for data set $i$ to an observed out-of-transit count rate in photo-electrons per second [e$^-$/s].

\begin{table}
\centering
\caption{Results from our analysis of GJ~436\,b. Gaussian priors on parameters with mean $\mu$ and standard deviation $\sigma$ are noted using the notation $\mathcal{N}(\mu,\sigma)$. For each data set $i= 1, 2, 3$, $c_i$ is the mean count rate out of eclipse, $df_i/dt$ is the linear trend with time,  $df_i/d\var{contam}$ is the correlation of flux with the predicted contamination of the aperture by background stars, and $df_i/d\var{bg}$ is the correlation of flux with the estimated background level in the image. The quantities  \var{contam} and \var{bg} are normalized so that the coefficients give the amplitude of the trend in each light curve. This analysis uses implicit roll-angle decorrelation with $N_{\rm roll} = 1$.}
\label{tab:gj436_pars}
\begin{tabular}{lrl}
\hline\hline
\multicolumn{1}{@{}l}{Parameter}   & \multicolumn{1}{l}{Value} & Notes \\
\hline
\multicolumn{3}{@{}l}{Input parameters} \\
\noalign{\smallskip}
T$_{\rm eff}$~~[K]         & $   3505 \pm 51 $ & 1 \\
$\log g$~~(cgs)            & $  4.91 \pm 0.07 $ & 1 \\
{[Fe/H]}                   & $ -0.04 \pm 0.16 $ & 1 \\
$M_\star$~~[$M_{\odot}$]   & $0.445 \pm 0.018 $ &  \\
P [d]                      &            2.643898 & 2 \\
$K$~~[m\,s$^{-1}$]         & $  17.38 \pm 0.17 $ & 3 \\

\noalign{\smallskip}
\multicolumn{3}{@{}l}{Model parameters} \\
\noalign{\smallskip}
$D$                               &$    0.00700 \pm    0.00018 $ & \\  
$W$                               &$    0.01593 \pm    0.00015 $ & \\
$b$                               &$      0.802 \pm      0.012 $ & \\
$T_0$                             &$ 1947.26212 \pm    0.00012 $ & $\mathcal{N}(1947.262, 0.01)$, 4  \\
$h_1$                             &$      0.733 \pm      0.051 $ & $\mathcal{N}(0.73, 0.1)$  \\
$h_2$                             &$                    =0.633 $ & \\
$\ln{\sigma_w}$                   &$      -12.1 \pm        3.3 $ & $\mathcal{N}(-10, 5)$  \\
$c_1$ [10$^6$ e-/s]               &$   \bmath{15.36531 \pm   0.00045} $ & \\   
$df_1/dt$~~[d$^{-1}$]             &$    0.00059 \pm    0.00017 $ & \\
$c_2$ [10$^6$ e-/s]               &$   \bmath{15.40448 \pm   0.00082} $ & \\ 
$df_2/dt$~~[d$^{-1}$]             &$    0.00076 \pm    0.00016 $ & \\
$df_2/d{\tt bg}$                  &$   -0.00036 \pm    0.00016 $ & \\
$df_2/d{\tt contam}$              &$   0.000373 \pm   0.000094 $ & \\
$c_3$ [10$^6$ e-/s]               &$   \bmath{15.34905 \pm   0.00054} $ & \\ 
$df_3/dt$~~[d$^{-1}$]             &$    0.00034 \pm    0.00020 $ & \\

\noalign{\smallskip}
\multicolumn{3}{@{}l}{Derived parameters} \\
\noalign{\smallskip}
$M_p$~~[$M_{\oplus}$]             &$      21.72 \pm       0.63 $ & \\
$R_{\rm p}$~~[$R_{\oplus}$]             &$\bmath{3.85 \pm       0.10} $ & \\  
$R_{\star}$~~[$R_{\odot}$]        &$\bmath{0.422 \pm      0.010} $ & \\   
$R_{\rm p}/R_{\star}$                   &$     0.0837 \pm     0.0011 $ & \\
$a/R_{\star}$                     &$      14.56 \pm       0.30 $ & \\
$i$~~[$\degr$]                    &$      86.84 \pm       0.11 $ & \\
$\log(\rho_{\star}/\rho_{\odot})$ &$      0.773 \pm      0.027 $ & $\mathcal{N}(0.724, 0.032)$  \\
$g_p$~~[m\,s$^{-2}$]              &$      14.35 \pm       0.67 $ & \\
$\rho_p$~~[g\,cm$^{-3}$]          &$       2.09 \pm       0.15 $ & \\
$\sigma_w$~~[ppm]                 &$          6 \pm         29 $ & \\
\noalign{\smallskip}
\hline
\end{tabular}

\medskip
\parbox{\columnwidth}{
1: \citet{2019A+A...625A..68S}. 
2: \citet{2014A+A...572A..73L}. 
3: \citet{2018A+A...609A.117T}.
\mbox{4: BJD$_{\rm TDB}-2457000$}. 
}
\end{table}

\begin{figure*} 
\resizebox{0.8\textwidth}{!}{\includegraphics{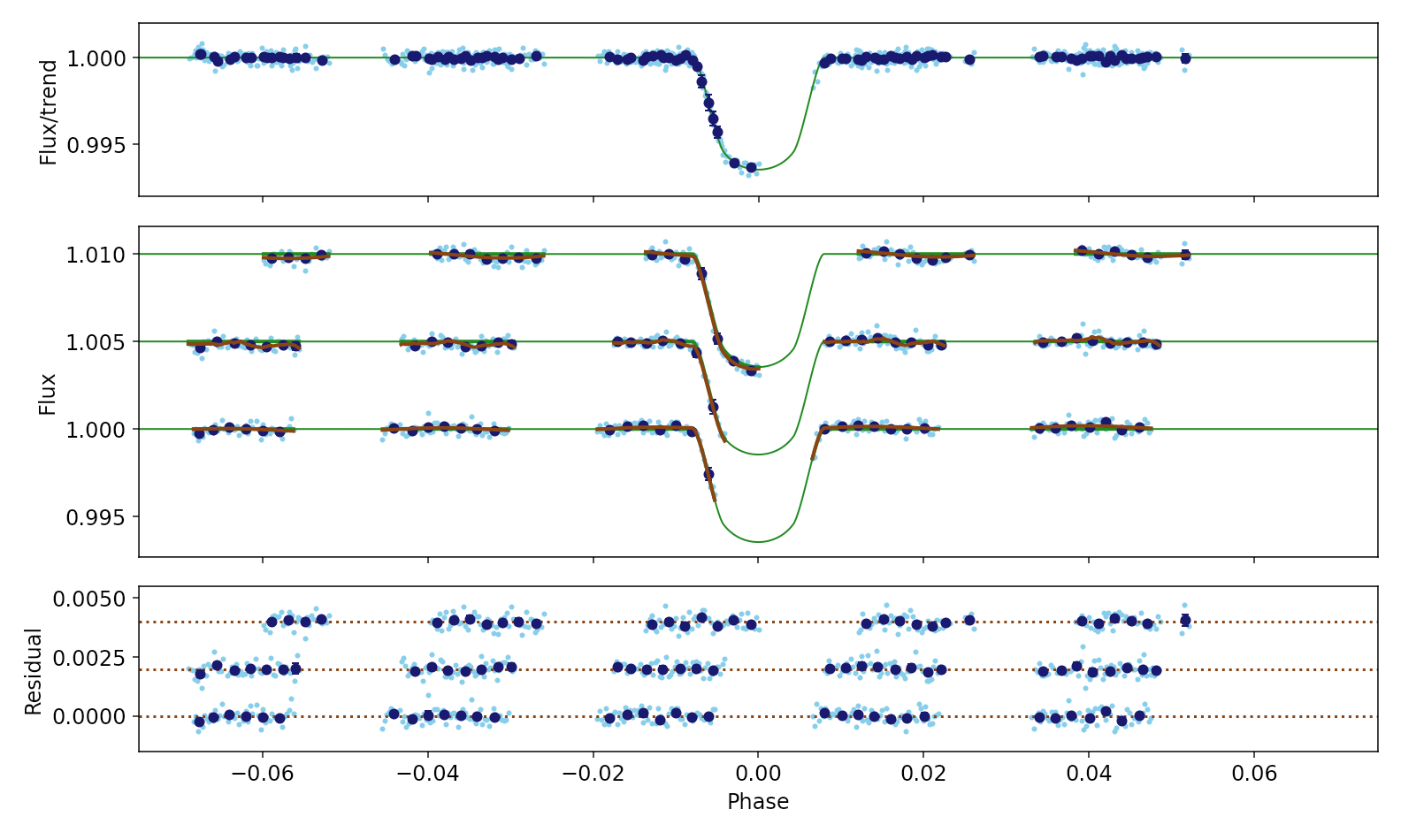}}
\caption{\label{fig:gj436_lcfit} \CHEOPS\ transit light curves of GJ~436\,b. \textit{Upper panel}: All data after removing trends. Observed light curves are displayed in cyan. The dark blue points are the data points binned over 0.002 phase units. The best-fit transit model is shown in green. \textit{Middle panel}: Observed light curves are displayed in cyan offset by multiples of 0.005 units. The dark blue points are the data points binned over 0.002 phase units. The full model including instrumental trends is shown in brown and the transit model without trends is shown in green.  \textit{Lower panel}: Residuals obtained after subtraction of the best-fit model in the same order as the upper plot offset by multiples of 0.002 units.}
\end{figure*}

\begin{figure*} 
\resizebox{0.8\textwidth}{!}{\includegraphics{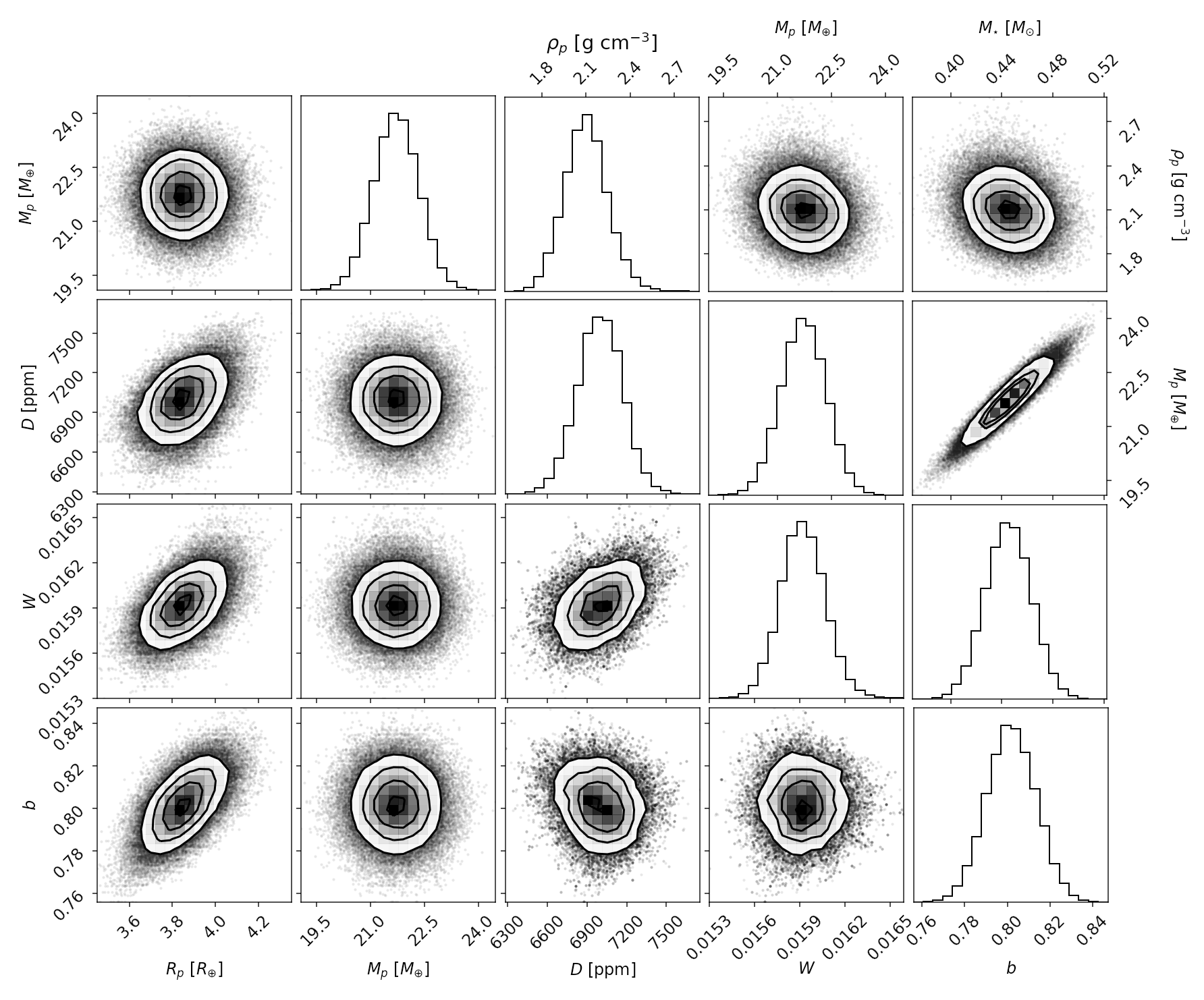}}
\caption{\label{fig:GJ436_corner} Correlation plot for selected parameters from our analysis of GJ~436.}
\end{figure*}


\subsubsection{GJ 436 b}

The warm-Neptune GJ~436\,b orbits a moderately-bright M2.5\,V star ($V=10.6, G=9.6$) with an orbital period of 2.64~days \citep{Butler04}. It was the first Neptune-mass exoplanet found to transit its host star \citep{Gillon2007}. Several studies have scrutinised the evaporating atmosphere of this planet using observations from ultraviolet \citep{2014ApJ...786..132K, Ehrenreich2015, Lavie2017, DosSantos2019} to infrared wavelengths \citep{Pont2009, 2011ApJ...735...27K, 2014Natur.505...66K, 2014A+A...572A..73L}.  A second planet has been posited to explain the significant orbital eccentricity of GJ~436\,b \citep[$e\approx 0.15$;][]{2008ApJ...677L..59R, 2007PASP..119...90M} but recent studies based on extensive radial velocity data have not confirmed previous claims for the existence of this second planet  \citep{2014A+A...572A..73L, 2018A+A...609A.117T}. The orbit of GJ~436\,b is significantly misaligned with the rotation axis of its host star  \citep{Bourrier2018}.

To estimate the mass and mean stellar density of GJ\,436 we used the empirical calibrations implemented in the software \software{kmdwarfparam} \citep{2015AJ....149..166H}.  These empirical relations are well-determined for stars with masses and radii similar to GJ~436. For the input to \software{kmdwarfparam} we used the apparent magnitudes in the V, J, H and K bands listed on SIMBAD and the parallax from {\it Gaia} EDR3. The results are summarised in Table~\ref{tab:gj436_pars}. The mass and radius obtained from \software{kmdwarfparam} agree very well with our values obtained using the methods described in Section~\ref{sec:irfm} ($M=0.444 \pm 0.034M_{\odot}$, $R=0.444 \pm 0.059R_{\odot}$) but are more precise. These radius estimates also agree well with the value $R=0.455 \pm 0.018R_{\odot}$ measured directly using interferometry by \citet{2012ApJ...753..171V}.

We observed three transits of GJ~436\,b (Table~\ref{tab:obslog}). The transit ingress was observed on all three visits but only the final visit covers the point of mid-transit and the egress was only partly observed during the first visit.  We first analysed the transits individually using \code{Dataset.lmfit\_transit} in order to identify which decorrelation parameters are needed for each visit. We fixed the orbital period at the value $P=2.6438980$\,d \citep{2014A+A...572A..73L}. We also fixed the limb darkening parameters at the values inferred from the tables provided by \cite{2019RNAAS...3...17C}. The results are summarised in Table~\ref{tab:individual}. Between 1 and 3 useful decorrelation parameters were identified per visit, with the highest-order term needed for decorrelation against roll angle being $\sin(\phi)$.  GJ\,436 is moderately bright and there is little contamination of the photometric aperture from other stars. As a result, the instrumental noise trends in the light curves have very low amplitudes ($\loa 300$\,ppm). A small but significant linear trend with time is seen for all three visits which we ascribe to stellar variability on time scales longer than the visit duration. The power spectral density (PSD) of the residuals from these initial fits are shown in Fig.~\ref{fig:fft_GJ436}. The small amount of power near orbital frequency of the CHEOPS spacecraft and its first harmonic for data set 1 is not statistically significant, i.e. the PSDs of the residuals are consistent with the white-noise level expected based on the typical error bar per datum. The trends in the data with spacecraft roll angle and our fit to this trend for data set 3 are shown in Fig.~\ref{fig:roll_GJ436}.
 
For the combined analysis of the visits using \code{MultiVisit} we set priors on $f_c$ and $f_s$ based on the values of $e=0.152\pm 0.009$ and $\omega = 325.8{\degr} \pm 5.7{\degr}$ from \citet{2018A+A...609A.117T}. The limb-darkening parameter $h_2$ has only a subtle effect on the light curve during the ingress and egress phases of the transit so we decided to fix this parameter at the value inferred from the tables provided by \cite{2019RNAAS...3...17C}. We include $h_1$ as a free parameter in the analysis  with a Gaussian prior centred on the value obtained from the same tables with an arbitrary choice of 0.1 for the standard error. We also imposed a prior on the mean stellar density based on the values obtained from \software{kmdwarfparam} \citep{2015AJ....149..166H}. Based on the results of the analysis for the individual visits we decided to use $N_{\rm roll}=1$.  Increasing this value by 1 or 2 has a negligible effect on the results. The results from this analysis are given in Table~\ref{tab:gj436_pars} and the fits to the light curves are shown in Fig.~\ref{fig:gj436_lcfit}. Correlations between selected parameters from this analysis are shown in Fig.~\ref{fig:GJ436_corner}. These results are discussed in the context of previous studies of GJ\,436\,b in Section~\ref{sec:gj436_discuss}.


\begin{figure*}
\resizebox{0.8\textwidth}{!}{\includegraphics{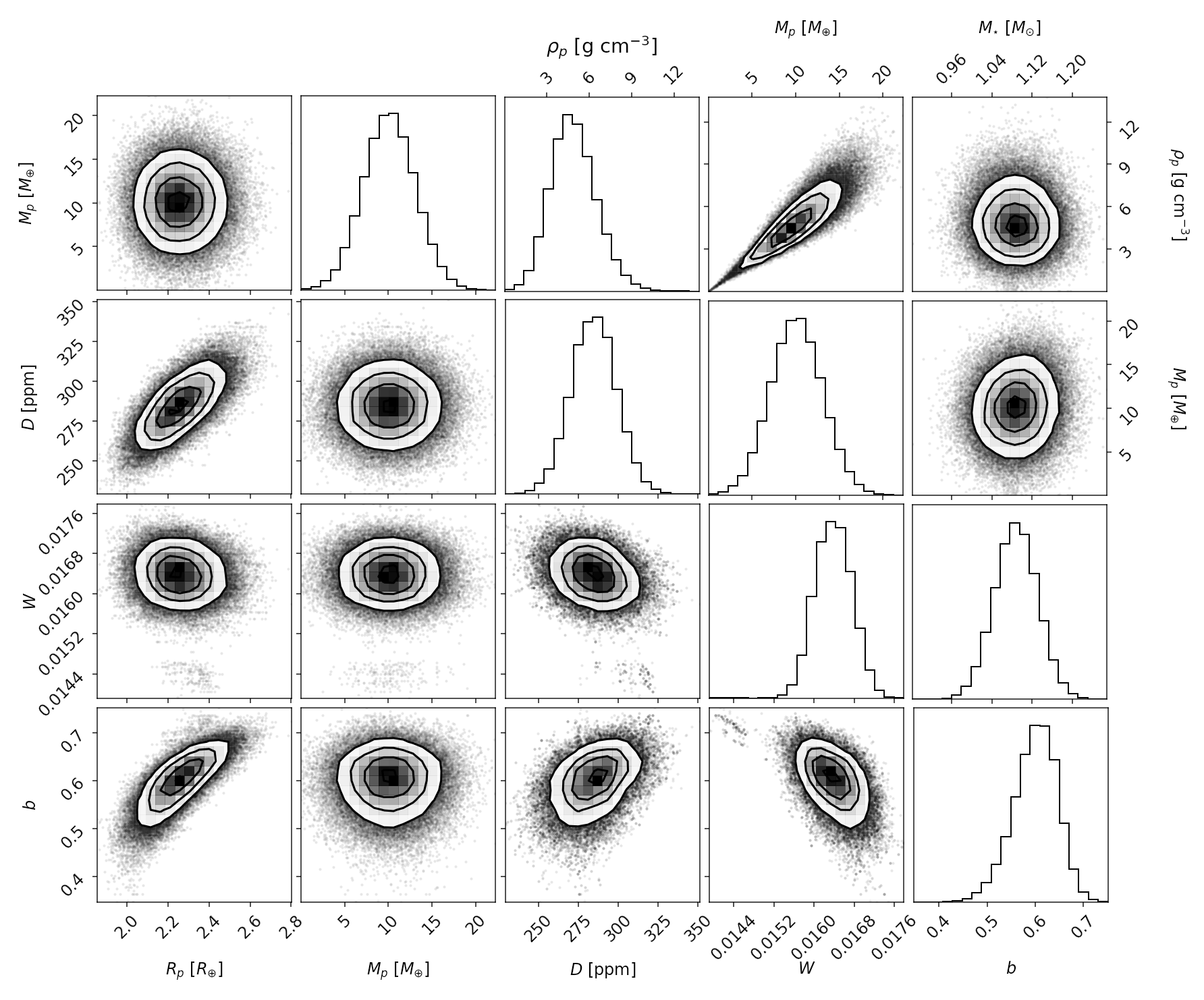}}  
\caption{\label{fig:HD106315_corner} Correlation plot for selected parameters from our analysis of HD~106315.}
\end{figure*}

\begin{table}
\caption{Mass, radius and mean stellar density estimates for HD~106315. 
\label{tab:hd106315_rhostar}}
\begin{center}
\begin{tabular}{@{}cccr}
\hline
\multicolumn{1}{@{}l}{$M_{\star}$ [$M_{\odot}$]} & 
\multicolumn{1}{l}{$R_{\star}$ [$R_{\odot}]$} & 
\multicolumn{1}{l}{$\log(\rho_{\star}/\rho_{\odot})$}  & Ref. \\
\hline
$1.227 \pm 0.064 $&$ 1.257 \pm 0.014 $&$ -0.209 \pm 0.027$ & 1 \\
$1.154 \pm 0.042 $&$ 1.269 \pm 0.024 $&$ -0.248 \pm 0.029$ & 2 \\
$1.091 \pm 0.036 $&$ 1.296 \pm 0.058 $&$ -0.300 \pm 0.060$ & 3 \\
$1.027 \pm 0.034 $&$ 1.281 \pm 0.058 $&$ -0.311 \pm 0.061$ & 4 \\
$ 1.07 \pm  0.03 $&$  1.18 \pm  0.03 $&$ -0.186 \pm 0.035$ & 5 \\
\hline\noalign{\smallskip}
$\bmath{1.088} \pm 0.043 $&$ 1.252 \pm 0.041 $&$ -0.229 \pm 0.045$ & Mean \\
\hline
\end{tabular}
\end{center}
\parbox{\columnwidth}{
1. This work. 
2. \citet{2021AJ....161...47K} 
3. \citet{2017A+A...608A..25B} 
4. \citet{2017AJ....153..256R}
5. \citet{2017AJ....153..255C}
}
\end{table}

\begin{table}
\centering
\caption{
Results from our analysis of the transits for HD 106315\,b.
Gaussian priors on parameters with mean $\mu$ and standard deviation $\sigma$ are
noted using the notation $\mathcal{N}(\mu,\sigma)$. For each data set $i$, $c_i$ is the mean 
count rate out of eclipse, $df_i/dt$ is the linear trend with time and 
$df_i/d\var{smear}$ is the correlation of flux with the smear correction. The quantity \var{smear} is normalized so that the coefficient gives the amplitude of the trend in the light curve. This analysis uses implicit roll-angle decorrelation with $N_{\rm roll} = 1$.}
\label{tab:hd106315_pars}
\begin{tabular}{lrl}
\hline\hline
\multicolumn{1}{@{}l}{Parameter}   & \multicolumn{1}{l}{Value} & Notes \\
\hline
\multicolumn{3}{@{}l}{Input parameters} \\
\noalign{\smallskip}

T$_{\rm eff}$~~[K]         & $   6450 \pm 105 $ & \\
$\log g$~~(cgs)            & $  4.28 \pm 0.10 $ & \\
{[Fe/H]}                   & $ -0.09 \pm 0.05 $ & \\
{[Mg/H]}                   & $ -0.09 \pm 0.12 $ & \\
{[Si/H]}                   & $ -0.05 \pm 0.06 $ & \\
$M_\star$~~[$M_{\odot}$]   & $1.091 \pm 0.029 $ & \\
P [d]                      &           9.552105 & 1 \\
$K$~~[m\,s$^{-1}$]           & $  2.88 \pm 0.85 $ & 1 \\
\noalign{\smallskip}
\multicolumn{3}{@{}l}{Model parameters} \\
\noalign{\smallskip}
$D$                               &$   0.000284 \pm   0.000014 $ & \\  
$W$                               &$    0.01637 \pm    0.00038 $ & \\
$b$                               &$      0.601 \pm      0.045 $ & \\
$T_0$                             &$  1952.4979 \pm     0.0017 $ & 2 \\
$h_1$                             &$      0.777 \pm      0.012 $ & $\mathcal{N}(0.777, 0.012)$  \\
$h_2$                             &$      0.419 \pm      0.055 $ & $\mathcal{N}(0.421, 0.055)$  \\
$\ln\sigma_w$                     &$      -9.34 \pm       0.10 $ & $\mathcal{N}(-9.3, 1.0)$  \\
$c_1$ [10$^6$ e-/s]               &$   20.05254 \pm    0.00028 $ & \\
$df_1/dt$~~[d$^{-1}$]             &$  -0.000154 \pm   0.000020 $ & \\
$df_1/d{\tt bg}$                  &$   0.000029 \pm   0.000037 $ & \\
$df_1/d{\tt smear}$               &$   0.000089 \pm   0.000030 $ & \\
$c_2$ [10$^6$ e-/s]               &$   20.02291 \pm    0.00024 $ & \\
\noalign{\smallskip}
\multicolumn{3}{@{}l}{Derived parameters} \\
\noalign{\smallskip}
$M_p$~~[$M_{\oplus}$]             &$       10.1 \pm        3.0 $ & \\ 
$R_{\rm p}$~~[$R_{\oplus}$]       &$       2.25 \pm       0.10 $ & \\
$R_{\star}$~~[$R_{\odot}$]        &$      1.222 \pm      0.045 $ & \\
$R_{\rm p}/R_{\star}$             &$    0.01686 \pm    0.00041 $ & \\
$a/R_{\star}$                     &$      15.95 \pm       0.55 $ & \\
$i$~~[$\degr$]                    &$      87.84 \pm       0.23 $ & \\
$\log(\rho_{\star}/\rho_{\odot})$ &$     -0.224 \pm      0.045 $ & $\mathcal{N}(-0.229, 0.045)$  \\
$g_p$~~[m\,s$^{-2}$]              &$       19.5 \pm        6.0 $ & \\
$\rho_p$~~[g\,cm$^{-3}$]          &$        4.8 \pm        1.6 $ & \\
$\sigma_w$~~[ppm]                 &$         87 \pm          9 $ & \\
\noalign{\smallskip}
\hline
\noalign{\smallskip}
\multicolumn{3}{@{}l}{{\it K2} light curve analysis} \\
\noalign{\smallskip}
$D$                               &$   0.000277 \pm   0.000016 $ & \\
$W$                               &$    0.01662 \pm    0.00050 $ & \\
$b$                               &$      0.586 \pm      0.054 $ & \\
$T_0$                             &$   615.2030 \pm     0.0020 $ & 2 \\
$h_1$                             &$      0.778 \pm      0.012 $ & $\mathcal{N}(0.78, 0.012)$  \\
$h_2$                             &$      0.422 \pm      0.054 $ & $\mathcal{N}(0.419, 0.055)$  \\
$\ln\sigma_{\rm w}$             &$     -9.942 \pm      0.037 $ & \\
$R_{\rm p}/R_{\star}$                   &$    0.01663 \pm    0.00048 $ & \\
$a/R_{\star}$                     &$      15.92 \pm       0.56 $ & \\
$i$~~[$\degr$]                    &$      87.89 \pm       0.25 $ & \\
$\log(\rho_{\star}/\rho_{\odot})$ &$     -0.227 \pm      0.046 $ & $\mathcal{N}(-0.229, 0.045)$   \\
$\sigma_{\rm w}$~~[ppm]           &$         48 \pm          2 $ & \\

\noalign{\smallskip}
\hline
\end{tabular}
\medskip
\parbox{\columnwidth}{
1: \citet{2021AJ....161...47K}.
\mbox{2: BJD$_{\rm TDB}-2457000$}. 
}
\end{table}

\begin{figure*}
\resizebox{0.9\textwidth}{!}{\includegraphics{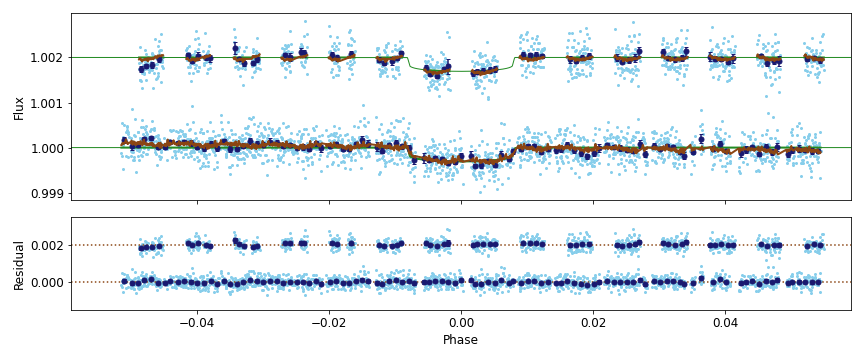}}  
\caption{\label{fig:hd106315_lcfit} \CHEOPS\ transit light curves of HD~106315\,b. \textit{Upper panel}: Observed light curves are displayed in cyan offset by multiples of 0.002 units. The dark blue points are the data points binned over 0.001 phase units. The full model including instrumental trends is shown in brown and the transit model without trends is shown in green.  \textit{Lower panel}: Residuals obtained after subtraction of the best-fit model in the same order as the upper plot offset by multiples of 0.002 units.}
\end{figure*}


\subsubsection{HD 106315 b}
\label{sec:HD106315}

HD~106315 is a F5\,V star with a $V$-band magnitude of $8.95$ that is known to host at least two planets \citep{2017AJ....153..255C, Rodriguez17}. The inner planet (b) is a super-Earth with a radius of 2.44~\Rearth\ and an orbital period of 9.55~days; the outer planet (c) is a Neptune-size planet with a radius of 4.35~\Rearth\ and a period of 21.06~days \citep{2017A+A...608A..25B}. \citet{2021AJ....161...47K} have measured accurate masses for these planets based on extensive multi-year radial velocity measurements for these planets together with transits observed with {\it Spitzer}. That study was motivated by on-going and planned observing programmes with Hubble Space Telescope ({\it HST}) and James Webb Space Telescope ({\it JWST}) to characterise the atmospheres of these planets. These authors find that the orbital eccentricity of these planets is close to $e=0$ based on their extensive radial velocity data and on stability arguments.

The rotation of HD~106315 measured from spectral line broadening is moderately fast ($v_{\rm rot}\sin i_{\star} \approx 13$\,km\,s$^{-1}$) but the {\it K2} light curve and ground-based photometry show that the intrinsic variability of this star is $\loa 0.2$\% at optical wavelengths \citep{2017AJ....153..255C,2021AJ....161...47K}. There are several published estimates for the mass and radius of this star based on a variety of methods -- these are summarised in Table~\ref{tab:hd106315_rhostar} together with our own estimates based on the methods described in Section~\ref{sec:irfm}. We have used these results to estimate the mass of this star and to set a prior on the mean stellar density for the analysis of the light curve. In both cases we have used the weighted mean value and the weighted sample standard deviation to set the value and its error. We use the sample standard error rather than the standard error in the mean because the values in Table~\ref{tab:hd106315_rhostar} are not completely independent and the differences between these estimates may reflect systematic sources of uncertainty e.g. the unknown helium abundance for this star.

To derive the stellar atmospheric parameters for HD~106315 in Table~\ref{tab:hd106315_rhostar} we used version 5.22 of the Spectroscopy Made Easy \software{sme} package \citep{2017A+A...597A..16P} to analyse the spectrum of this star observed with the High Accuracy Radial velocity Planet Searcher (HARPS) spectrograph on the European Southern Observatory (ESO) 3.6-m telescope. All available HARPS spectra were downloaded from the ESO science archive and co-added prior to analysis. In this package synthetic spectra are calculated starting from a first guess of individual stellar parameters and utilizing a grid of stellar models, in this case taken from the ATLAS-12 set \citep{2013ascl.soft03024K}. Atomic parameters were downloaded from the VALD data base \citep{1995A+AS..112..525P}. Keeping all but one parameter fixed and iterating and minimizing until no further improvement is realized one arrives eventually at a set of stellar parameters  \citep{2017A+A...604A..16F}.

We observed two transits of HD~106315\,b with \CHEOPS\ (Table~\ref{tab:obslog}). The first transit was observed when the target was close to the anti-Sun direction so the observing efficiency is very high. The data set for the second visit shows spurious jumps in values of the spacecraft roll angle versus time due to a software bug that was fixed in DRP version 13.0. These spurious roll angle values were corrected prior to the analysis presented here.  We first analysed both transits individually using \code{Dataset.lmfit\_transit} in order to identify which decorrelation parameters are needed for each visit. We fixed the orbital period at the value $P=9.552105$\,d and assumed that the orbital eccentricity is $e=0$ \citep{2021AJ....161...47K}. We also fixed the limb darkening parameters at the values inferred from the tables provided by \cite{2018A+A...616A..39M}. The second data set does not cover the ingress or egress to the transit so the impact parameter is unconstrained by these data. We fixed the impact parameter to the value determined from the analysis of the first data set for the analysis of the second data set. The results are summarised in Table~\ref{tab:individual}. Between 2 and 4 useful decorrelation parameters were identified per visit, with the highest-order term needed for decorrelation against roll angle being $\sin(\phi)$. HD\,106315 is  bright and there is little contamination of the photometric aperture from other stars. As a result, the instrumental noise trends in the light curves have very low amplitudes ($\loa 120$\,ppm). A small but significant linear trend with time is seen for the first visit which we ascribe to stellar variability on time scales longer than the visit duration. The power spectral density (PSD) of the residuals from these initial fits are shown in Fig.~\ref{fig:fft_HD106315}. There is a small excess in power at low frequencies for the second data set that we assume is related to rapid changes in the scattered light level towards the start and end of each visit. This can lead to a gradients in the background level in some images that is not (yet) accounted for in the data reduction pipeline. The trends in the data with spacecraft roll angle and our fit to this trend for data set 2 are shown in Fig.~\ref{fig:roll_HD106315}.
 
 We used the same fixed values of $e$ and $P$ for the combined analysis of the two visits using \code{MultiVisit}. We set priors on the limb-darkening parameters $h_1$ and $h_2$ based on the results from \cite{2018A+A...616A..39M}. We included the small correction to the tabulated values recommended by \citet{2018A+A...616A..39M}  based on the observed offset between these values and the observed values of $h_1$ and $h_2$ for stars similar to HD~106315. Based on the results of the analysis for the individual visits we decided to use $N_{\rm roll}=1$. Changing this value by $\pm 1$ has a negligible effect on the results. The results from this analysis are given in Table~\ref{tab:hd106315_pars} and the fits to the light curves are shown in Fig.~\ref{fig:hd106315_lcfit}. Correlations between selected parameters from this analysis are shown in Fig.~\ref{fig:HD106315_corner}.   

 We also attempted a similar analysis without the prior on the stellar density. The results from that analysis are consistent with the results presented here but with increased uncertainties, particularly for the impact parameter, $b$ ($D= 0.000283 \pm 0.000028$, $W= 0.01647 \pm 0.00043 $, $b= 0.54 \pm 0.31$). The mean stellar density obtained from this analysis of the light curve with no prior on $\rho_{\star}$  is $\log(\rho_{\star}/\rho_{\odot}) = -0.16\pm0.26$.

These results are discussed in the context of previous studies of HD~106315\,b in Section~\ref{sec:hd106315_discuss}. To aid this discussion, we also performed an analysis of the {\bf 6} transits of HD~106315\,b in the {\it K2} light curve of HD\,106315 using very similar assumptions to those used in our analysis of the \CHEOPS\ light curve. We used the light curve corrected for instrumental effects using the \software{ks2c} algorithm \citep{2015MNRAS.447.2880A} downloaded from the  Mikulski Archive for Space Telescopes\footnote{\url{https://archive.stsci.edu/}} (MAST). There are clear offsets in the mean flux level either side of each transit in this light curve so we used a smooth function generated with a Gaussian process fit to the data between the transits to put the flux level onto a consistent scale for every transit. We used the same light curve model from \pycheops\ used for the analysis of the \CHEOPS\ light curve and set the same priors on the transit parameters and mean stellar density. The priors on the limb-darkening parameters were similar to those used for the analysis of the \CHEOPS\ light curve although the values differ due to the different instrument response functions. We did account for the finite integration time of the {\it K2} observations but did not include any additional parameters for decorrelation of instrumental noise sources. The results from this analysis are also given in Table~\ref{tab:hd106315_pars}. These results and the results from previous studies \citep{2017AJ....153..255C,Rodriguez17,2017A+A...608A..25B} are consistent with one another but the errors on the transit parameters vary by a factor $\approx$2 because of the different assumptions made in each study, e.g. the error on $a/R_{\star}$ is sensitive to the prior used for $\rho_{\star}$.


\begin{table}
\caption{Mass, radius and mean stellar density estimates for HD~97658. The error quoted on the mean value is the standard deviation of the sample.
\label{tab:hd97658_rhostar}}
\begin{center}
\begin{tabular}{@{}rrrr}
\hline
\multicolumn{1}{@{}l}{$M_{\star}$ [$M_{\odot}$]} & 
\multicolumn{1}{l}{$R_{\star}$ [$R_{\odot}$]} & 
\multicolumn{1}{l}{$\log(\rho_{\star}/(\rho_{\odot})$}  & Ref. \\
\hline
$0.758 \pm 0.044 $&$ 0.761 \pm 0.009 $&$ 0.236 \pm 0.030$ & 1 \\
$0.74  \pm 0.02  $&$ 0.74  \pm 0.02  $&$  0.26 \pm  0.04$ & 2 \\
$0.74  \pm 0.01  $&$ 0.73  \pm 0.01  $&$ 0.279 \pm 0.019$ & 3 \\
$0.77  \pm 0.05  $&$ 0.741 \pm 0.024 $&$ 0.276 \pm 0.053$ & 4 \\
\hline\noalign{\smallskip}
$0.752 \pm 0.035 $&$ 0.743 \pm 0.017 $&$ 0.263 \pm 0.037$ & Mean \\
\hline
\end{tabular}
\end{center}
\parbox{\columnwidth}{
1. This work. 
2. \citet{2016ApJS..225...32B}.
3. \citet{bonfanti16}.
4. \citet{2014ApJ...786....2V}.
}
\end{table}

\begin{table}
\centering
\caption{
Results from our analysis of HD~97658.
Gaussian priors on parameters with mean $\mu$ and standard deviation $\sigma$ are
noted using the notation $\mathcal{N}(\mu,\sigma)$. RMS is the standard deviation of the residuals from the best fit.}
\label{tab:hd97658_pars}
\begin{tabular}{lrl}
\hline\hline
\multicolumn{1}{@{}l}{Parameter}   & \multicolumn{1}{l}{Value} & Notes \\
\hline
\multicolumn{3}{@{}l}{Input parameters} \\
\noalign{\smallskip}

T$_{\rm eff}$~~[K]         & $   5137 \pm 36 $ & 1 \\
$\log g$~~(cgs)            & $  4.47 \pm 0.09 $ & 1 \\
{[Fe/H]}                   & $ -0.35 \pm 0.02 $ & 1 \\
{[Mg/H]}                   & $ -0.25 \pm 0.03 $ & 1 \\
{[Si/H]}                   & $ -0.31 \pm 0.04 $ & 1 \\
$M_\star$~~[$M_{\odot}$]   & $0.752 \pm 0.035 $ & \\
P [d]                      &           9.489295 & 2 \\
$K$~~[m\,s$^{-1}$]         & $  2.81 \pm 0.15 $ & 2 \\
\noalign{\smallskip}
\multicolumn{3}{@{}l}{Model parameters} \\
\noalign{\smallskip}
$D$                        &$   0.000825 \pm   0.000017 $ & \\  
$W$                        &$   0.012440 \pm   0.000051 $ & \\
$b$                        &$      0.475 \pm      0.037 $ & \\
$T_0$                      &$ 1961.87639 \pm    0.00023 $ & 3 \\
$h_1$                      &$      0.715 \pm      0.011 $ & $\mathcal{N}(0.72, 0.012)$  \\
$h_2$                      &$      0.406 \pm      0.054 $ & $\mathcal{N}(0.397, 0.055)$  \\
$\ln{\sigma_w}$            &$     -10.70 \pm       0.64 $ & \\
$c$ [10$^6$ e-/s]          &$   56.55066 \pm    0.00082 $ & \\
$df/d\sin(\phi)$           &$   0.000110 \pm   0.000013 $ & $\mathcal{N}(0.0, 0.00015)$  \\
$df/d{\tt bg}$             &$  -0.000101 \pm   0.000032 $ & $\mathcal{N}(0.0, 0.00015)$  \\
\noalign{\smallskip}
\multicolumn{3}{@{}l}{Derived parameters} \\
\noalign{\smallskip}
$M_p$~~[$M_{\oplus}$]      &$       7.62 \pm       0.42 $ & \\   
$R_{\rm p}$~~[$R_{\oplus}$]      &$      2.293 \pm      0.070 $ & \\
$M_{\star}$~~[$M_{\odot}$]  &$      0.741 \pm      0.018 $ & \\
$R_{\rm p}/R_{\star}$            &$    0.02872 \pm    0.00030 $ & \\
$a/R_{\star}$              &$      23.35 \pm       0.51 $ & \\
$i$~~[$\degr$]             &$      88.83 \pm       0.12 $ & \\
$\log(\rho_{\star}/\rho_{\odot})$ &$      0.278 \pm      0.029 $ & $\mathcal{N}(0.267, 0.029)$  \\
$g_p$~~[m\,s$^{-2}$]       &$       14.2 \pm        1.1 $ & \\
$\rho_p$~~[g\,cm$^{-3}$]   &$       3.48 \pm       0.36 $ & \\
$\sigma_w$~~[ppm]          &$         23 \pm         14 $ & \\
RMS                 [ppm]  &    137 & \\
\noalign{\smallskip}
\hline
\noalign{\smallskip}
\multicolumn{3}{@{}l}{{\it TESS} analysis} \\
\noalign{\smallskip}
$T_0$                               &$  1904.9407 \pm     0.0010 $ & 3 \\
$D$                                 &$   0.000805 \pm   0.000039 $ & \\
$W$                                 &$    0.01235 \pm    0.00020 $ & \\
$b$                                 &$      0.498 \pm      0.046 $ & \\
$h_1$                               &$      0.771 \pm      0.012 $ & $\mathcal{N}(0.773, 0.012)$  \\
$h_2$                               &$      0.391 \pm      0.056 $ & $\mathcal{N}(0.39, 0.055)$  \\
$\ln{\sigma_w}$                     &$     -7.905 \pm      0.018 $ & \\
$R_{\rm p}/R_{\star}$                     &$    0.02838 \pm    0.00068 $ & \\
$a/R_{\star}$                       &$      23.21 \pm       0.52 $ & \\
$i$~~[$\degr$]                      &$      88.77 \pm       0.14 $ & \\
$\log(\rho_{\star}/\rho_{\odot})$   &$      0.270 \pm      0.029 $ & $\mathcal{N}(0.267, 0.029)$  \\
$\sigma_w$~~[ppm]                   &$        369 \pm          7 $ & \\
\noalign{\smallskip}
\hline
\end{tabular}
\medskip
\parbox{\columnwidth}{
1. \citet{Sousa2018}. 
2. \citep{2020AJ....159..239G}.
\mbox{3: BJD$_{\rm TDB}-2457000$}. 
}
\end{table}

\begin{figure*}
\resizebox{0.8\textwidth}{!}{\includegraphics{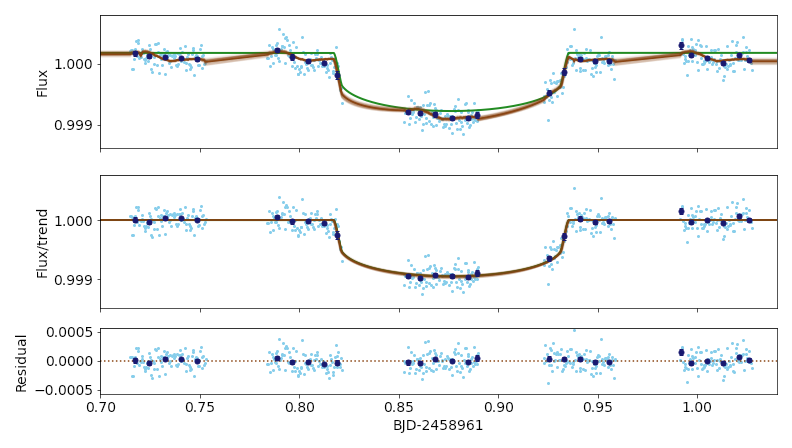}}
\caption{\label{fig:hd97658_lcfit} \CHEOPS\ transit light curve of HD~97658\,b. \textit{Upper panel}: Observed light curve  displayed as cyan points. The dark blue points are the data points binned over 11.5 minutes. The full model including instrumental trends is shown in brown and the transit model without trends is shown in green. Multiple versions of the full model sampled from the PPD are also shown in light brown. \textit{Middle panel}: Same as the upper panel after dividing-out the instrumental trends in the data. \textit{Lower panel}:  Residuals from the best-fit model.}
\end{figure*} 

\begin{figure*}
\resizebox{0.8\textwidth}{!}{\includegraphics{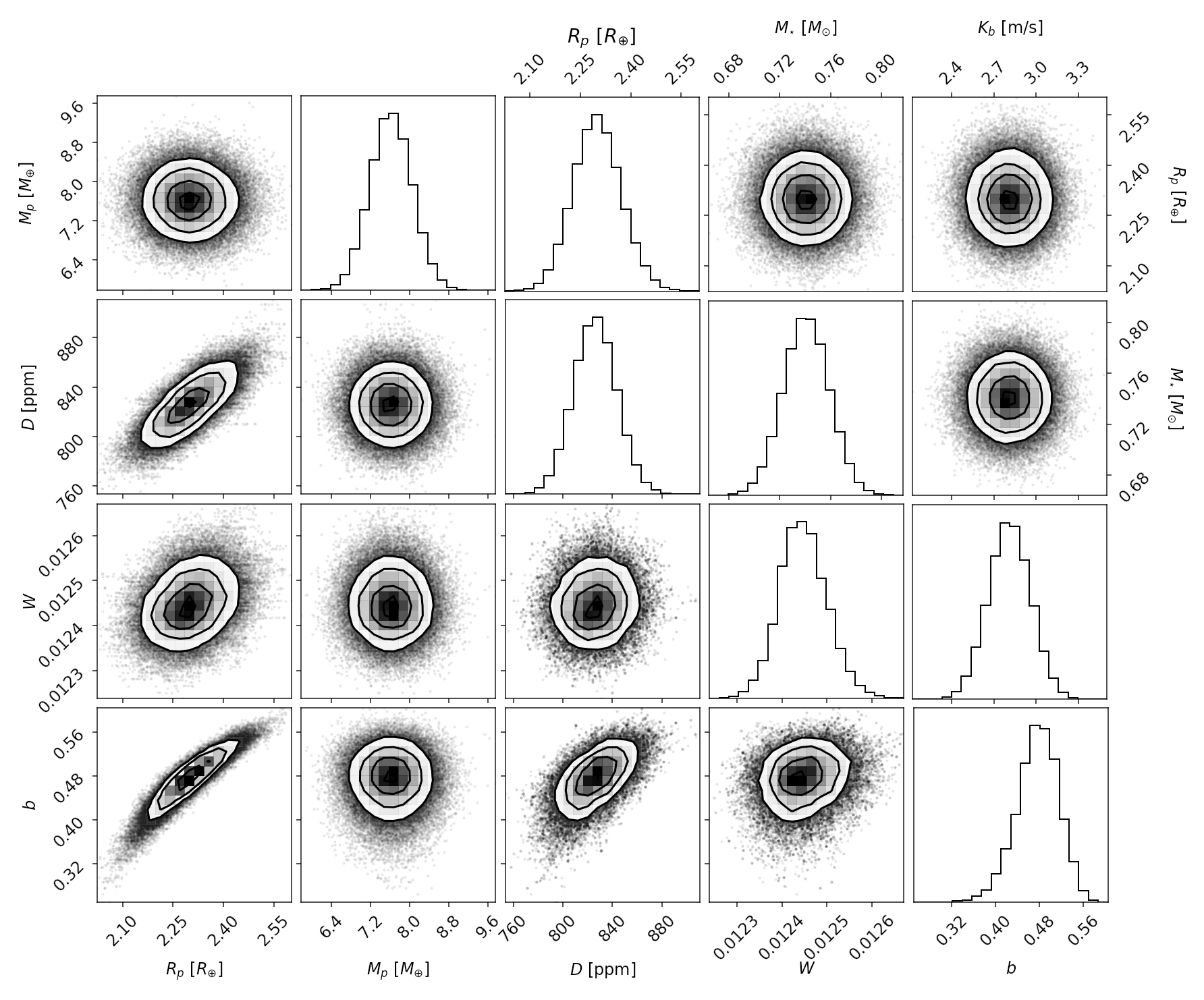}}
\caption{\label{fig:HD97658_corner} Correlation plot for selected parameters from our analysis of HD~97658.}
\end{figure*}

\begin{table}
    \centering
    \caption{Low mass stars with precise $\log g$ measurements.}
    \label{tab:logg-mdwarf}
    \begin{tabular}{@{}lrrrr}
\hline\hline
Star    & 
\multicolumn{1}{l}{Mass/M$_{\odot}$} & 
\multicolumn{1}{l}{$\log g$ [cgs]} & 
\multicolumn{1}{l}{[Fe/H]} & Ref. \\ 
\noalign{\smallskip}
\hline
J0543$-$56 B &$ 0.1641 \pm 0.0059 $ &$   5.09 \pm  0.04 $&$ 0.23 $& 1 \\
J1038$-$37 B &$ 0.1735 \pm 0.0067 $ &$   5.04 \pm  0.04 $&$ 0.31 $& 1 \\
J1013$+$01 B &$ 0.1773 \pm 0.0077 $ &$   5.02 \pm  0.02 $&$ 0.29 $& 1 \\
J1115$-$36 B &$ 0.1789 \pm 0.0061 $ &$   5.12 \pm  0.03 $&$ 0.30 $& 1 \\
J0339$+$03 B &$ 0.2061 \pm 0.0095 $ &$   5.12 \pm  0.05 $&$-0.25 $& 1 \\
J2349$-$32 B &$  0.174 \pm  0.006 $ &$  5.104 \pm 0.014 $&$-0.28 $& 2 \\
SAO 106989 B &$  0.256 \pm  0.005 $ &$  4.818 \pm 0.128 $&$ -0.2 $& 3 \\
HD 24465 B   &$  0.233 \pm  0.002 $ &$  5.029 \pm 0.007 $&$  0.3 $& 3 \\
CM Dra A     &$ 0.2310 \pm 0.0009 $ &$  4.994 \pm 0.007 $&$-0.30 $& 4, 5 \\
CM Dra B     &$ 0.2396 \pm 0.0009 $ &$  5.010 \pm 0.006 $&$-0.30 $& 4, 5 \\
J0522$-$25 A &$ 0.1739 \pm 0.0013 $ &$  5.057 \pm 0.021 $& --     & 6 \\
J0522$-$25 B &$ 0.2168 \pm 0.0048 $ &$  5.007 \pm 0.020 $& --     & 6 \\
J1934$-$42 B &$ 0.1864 \pm 0.0055 $ &$  5.045 \pm 0.012 $&$ 0.29 $& 7 \\
J2046$+$06 B &$ 0.1974 \pm 0.0062 $ &$  5.074 \pm 0.008 $&$  0.00$& 7 \\
\noalign{\smallskip}
\hline
\end{tabular}
\medskip
\parbox{\columnwidth}{
1.~\citet{2019A&A...625A.150V}
2.~\citet{2019A+A...626A.119G}
3.~\citet{2018AJ....156...27C}
4.~\citet{2009ApJ...691.1400M}
5.~\citet{2012ApJ...760L...9T}
6.~\citet{2018MNRAS.481.1897C}
7.~\citet{2021MNRAS.506..306S}
}
\end{table}


\subsubsection{HD 97658 b}

The super-Earth HD~97658\,b orbits a moderately bright K1\,V star ($V=7.7$, $G=7.5$) with a period of $P=9.43$\,d \citep{2011ApJ...730...10H}. Transits of the host star by this planet were found using ground-based observations \citep{2011arXiv1109.2549H} and confirmed using follow-up observations with {\it Spitzer} \citep{2014ApJ...786....2V} and the Microvariability and Oscillations in STars  ({\it MOST}) telescope \citep{2013ApJ...772L...2D}.  \citet{2020AJ....159..239G} analysed near-infrared spectra of HD~97658\,b observed during four transits with the WFC3 instrument on {\it HST}, together with extensive observations of the transit from the STIS instrument on {\it HST}, {\it Spitzer} and  {\it MOST}.  Despite this wealth of data their atmospheric modeling results were inconclusive. \citeauthor{2020AJ....159..239G} were able to rule out previous claims of additional planets in the HD~97658 system based on a large set of radial velocity observations obtained over two decades. Their analysis of these radial velocities also shows that the orbit of HD~97658\,b is circular or nearly so ($e\loa 0.03$). Variability of the activity indicators in the same spectroscopic data set lead to an estimate of $P_{\rm rot} \approx 35$\,d for the rotation period of this star. They conclude that HD~97658\,b is a favourable target for atmospheric characterisation through transmission spectroscopy with {\it JWST}.

The {\it TESS} light curve of HD~97658 shows very little intrinsic variability in this star ($\loa 0.02$\%), as is expected for a very slowly rotating K-dwarf. The results from recent studies of the host star properties are summarised in Table~\ref{tab:hd97658_rhostar} together with the results from our own analysis. We have used the weighted mean of these results to calculate the values of the stellar mass and mean density used in this analysis, and the weighted sample standard deviation to estimate the errors on these parameters. We use the sample standard deviation rather than the standard error in the mean because the values in Table~\ref{tab:hd97658_rhostar} are not completely independent and the differences between these estimates may reflect systematic sources of uncertainty, e.g. the unknown helium abundance for this star. 

We observed a single transit of HD~97658\,b with \CHEOPS\ (Table~\ref{tab:obslog}). Although the observing efficiency is quite high (72\%) the coverage of the ingress to the transit is poor. HD~97658 is a moderately bright and isolated star so the level of instrumental noise in the light curve is very low. 

We used an initial analysis of this transit with \code{Dataset.lmfit\_transit} to determine which decorrelation parameters should be used in our final analysis. We fixed the orbital period at the value $P=9.489295$\,d and assumed a circular orbit \citep{2020AJ....159..239G}. The stellar atmospheric parameters are taken from the SWEET-Cat catalogue \citep{Santos2013, Sousa2018}. These are a homogeneous set of parameters derived using the \software{ares+moog} methodology \citep{Sousa2014} which were originally presented in \citet{Mortier2013}. The limb darkening parameters $h_1$ and $h_2$ were included as free parameters in this initial fit. The mean stellar density with its error from Table~\ref{tab:hd97658_rhostar} was included as a constraint in the least-squares analysis. This initial analysis shows that there are weak trends in the data with amplitudes $\approx 100$ ppm correlated with $\sin(\phi)$ and the background level in the images. There are no other significant instrumental trends in the light curve. If we include a linear trend with time in the least-squares analysis we find that it has an amplitude $\loa 40$\,ppm\,d$^{-1}$. Based on these results we used \code{Dataset.emcee\_sampler} to sample the joint PPD for the transit model parameters, the two decorrelation parameters, and the hyper-parameter $\ln\sigma_{\rm w}$ for our noise model. The results are given in Table~\ref{tab:hd97658_pars}. We set priors on the limb-darkening parameters $h_1$ and $h_2$ based on the results from \citet{2018A+A...616A..39M}. We included the small correction to the tabulated values recommended in \citet{2018A+A...616A..39M} based on the observed offset between these values and the observed values of $h_1$ and $h_2$ for stars similar to HD~97658. The fit to the light curve is shown in Fig.~\ref{fig:hd97658_lcfit} and correlation plots for selected parameters are shown in Fig.~\ref{fig:HD97658_corner}. The power spectral density (PSD) of the residuals shown in Fig.~\ref{fig:fft_HD97658} is consistent with the expected white-noise level based on the median error bar per datum. The trends in the data with spacecraft roll angle and our fit to this trend are shown in Fig.~\ref{fig:roll_HD97658}.

These results are discussed in the context of previous studies of HD~97658\,b in Section~\ref{sec:hd97658_discuss}. To aid this discussion, we also performed an analysis of the 2 transits of HD~97658\,b in the {\it TESS} light curve of HD\,97658 using very similar assumptions to those used in our analysis of the \CHEOPS\ light curve. We used the light curve PDCSAP\_FLUX values provided in the data file downloaded from MAST. Although the variability between the transits in this light curve is very small ($\loa 0.02\%$) we used a smooth function generated with a Gaussian process fit to the data between the transits to ensure that the flux level is on a consistent scale for both transits. We used the same light curve model from \pycheops\ used for the analysis of the \CHEOPS\ light curve and set the same priors on the transit parameters and mean stellar density. The priors on the limb-darkening parameters were similar to those used for the analysis of the \CHEOPS\ light curve although the values differ due to the different instrument response functions. The results from this analysis are also given in Table~\ref{tab:hd97658_pars}.


\begin{table}
\centering
\caption{
Results from our analysis of GJ 1132.
Gaussian priors on parameters with mean $\mu$ and standard deviation $\sigma$ are
noted using the notation $\mathcal{N}(\mu,\sigma)$. For each data set $i$, $c_i$ is the mean 
count rate out of eclipse, $df_i/dt$ is the linear trend with time,  $df_i/d\var{contam}$ is
the correlation of flux with the predicted contamination of the aperture by background stars,
$df_i/d\var{smear}$ is the correlation of flux with the smear correction, and $df_i/d\var{bg}$
is the correlation of flux with the estimated background level in the image. The quantities  
\var{contam}, \var{smear} and \var{bg} are normalized so that the coefficients give the 
amplitude of the trend in each light curve. These results were obtained using implicit roll-angle decorrelation with $N_{\rm roll}=2$.}
\label{tab:gj1132_pars}
\begin{tabular}{lrl}
\hline\hline
\multicolumn{1}{@{}l}{Parameter}   & \multicolumn{1}{l}{Value} & Notes \\
\hline
\multicolumn{3}{@{}l}{Input parameters} \\
\noalign{\smallskip}
T$_{\rm eff}$~~[K]         & $     3090 \pm 65 $ & \\
$\log g$~~(cgs)            & $   5.07 \pm 0.06 $ & \\
{[Fe/H]}                   & $  -0.31 \pm 0.10 $ & \\
$M_\star$~~[$M_{\odot}$]   & $ 0.192 \pm 0.022 $ & \\
$R_\star$~~[$R_{\odot}$]   & $0.207 \pm 0.0124 $ &  \\
P [d]                      &       $=1.6289287 $ & 1 \\
$K$~~[m\,s$^{-1}$]         & $   2.85 \pm 0.34 $ & 2 \\
\noalign{\smallskip}
\multicolumn{3}{@{}l}{Model parameters} \\
\noalign{\smallskip}
$D$                               &$    0.00244 \pm    0.00020 $ & \\
$W$                               &$    0.01876 \pm    0.00054 $ & \\
$b$                               &$       0.43 \pm       0.16 $ & \\
$T_0$                             &$ 1938.91419 \pm    0.00044 $ & $\mathcal{N}(1938.9138, 0.002)$, 3 \\
$h_1$                             &$      0.861 \pm      0.069 $ & $\mathcal{N}(0.75, 0.1)$  \\
$h_2$                             &$                   =0.753  $ &  \\
$\ln{\sigma_w}$                   &$     -7.034 \pm      0.056 $ & $\mathcal{N}(-7.0, 0.5)$  \\
$c_1$ [10$^6$ e-/s]               &$    1.32092 \pm    0.00074 $ & \\
$df_1/dt$~~[d$^{-1}$]             &$    0.00421 \pm    0.00091 $ & \\
$df_1/d{\tt smear}$               &$    0.00117 \pm    0.00063 $ & \\
$df_1/d{\tt contam}$              &$   -0.00149 \pm    0.00052 $ & \\
$c_2$ [10$^6$ e-/s]               &$     1.2976 \pm     0.0016 $ & \\
$df_2/dt$~~[d$^{-1}$]             &$     0.0041 \pm     0.0014 $ & \\
$df_2/d{\tt bg}$                  &$    -0.0022 \pm     0.0011 $ & \\
$df_2/d{\tt contam}$              &$   -0.00158 \pm    0.00052 $ & \\
$c_3$ [10$^6$ e-/s]               &$     1.3038 \pm     0.0023 $ & \\
$df_3/dt$~~[d$^{-1}$]             &$    0.00398 \pm    0.00077 $ & \\
$df_3/d{\tt bg}$                  &$    -0.0022 \pm     0.0010 $ & \\
$df_3/d{\tt contam}$              &$    -0.0060 \pm     0.0011 $ & \\

\noalign{\smallskip}
\multicolumn{3}{@{}l}{Derived parameters} \\
\noalign{\smallskip}
$M_p$~~[$M_{\oplus}$]             &$       1.74 \pm       0.25 $ & \\
$R_{\rm p}$~~[$R_{\oplus}$]             &$       1.11 \pm       0.10 $ & \\
$R_{\star}$~~[$R_{\odot}$]        &$      0.207 \pm      0.016 $ & \\
$R_{\rm p}/R_{\star}$                   &$     0.0494 \pm     0.0021 $ & \\
$a/R_{\star}$                     &$       16.3 \pm        1.1 $ & \\
$i$~~[$\degr$]                    &$      88.50 \pm       0.66 $ & \\
$\log(\rho_{\star}/\rho_{\odot})$ &$      1.338 \pm      0.086 $ & $\mathcal{N}(1.307, 0.089)$  \\
$g_p$~~[m\,s$^{-2}$]              &$       13.7 \pm        2.8 $ & \\
$\rho_p$~~[g\,cm$^{-3}$]          &$        7.0 \pm        1.9 $ & \\
$\sigma_w$~~[ppm]                 &$        881 \pm         50 $ & \\
\noalign{\smallskip}
\hline
\end{tabular}
\medskip
\parbox{\columnwidth}{
1. \citet{2017AJ....153..191S}.
2. \citet{2018A+A...618A.142B}.
\mbox{3: BJD$_{\rm TDB}-2457000$}. 

}
\end{table}

\begin{table}
\centering
\caption{
Results from our reanalysis of the MEarth light curves for GJ~1132.
Gaussian priors on parameters with mean $\mu$ and standard deviation $\sigma$ are
noted using the notation $\mathcal{N}(\mu,\sigma)$. The parameters $z_0\dots z_5$ 
are the coefficients of the polynomial used to model the trend of tabulated flux
with air mass. The tabulated flux values are assumed to all have the same standard 
error, $\sigma_f$.}
\label{tab:gj1132_mearth}
\begin{tabular}{lrl}
\hline\hline
\multicolumn{1}{@{}l}{Parameter}   & \multicolumn{1}{l}{Value} & Notes \\
\hline
\noalign{\smallskip}
\multicolumn{3}{@{}l}{Model parameters} \\
\noalign{\smallskip}
$D$           &$    0.00237 \pm    0.00010 $ & \\
$W$           &$    0.01903 \pm    0.00033 $ & \\
$b$           &$       0.41 \pm       0.23 $ & \\
$T_0$         &$ 2457184.55855 \pm    0.00069 $ & BJD$_{\rm TDB}$ \\
$P$ [days]    &$  1.6289227 \pm  0.0000041 $ & \\
$h_1$         &$      0.805 \pm      0.037 $ & $\mathcal{N}(0.769, 0.15)$ \\
$h_2$         &$                     =0.76 $ & \\
$z_0$         &$   0.000070 \pm   0.000079 $ & \\
$z_1$         &$     0.0082 \pm     0.0012 $ & \\
$z_2$         &$    -0.0423 \pm     0.0055 $ & \\
$z_3$         &$     0.0723 \pm     0.0095 $ & \\
$z_4$         &$    -0.0504 \pm     0.0069 $ & \\
$z_5$         &$     0.0125 \pm     0.0018 $ & \\
$a_{\rm rot}$ &$  -0.000108 \pm   0.000034 $ & \\
$b_{\rm rot}$ &$  -0.000034 \pm   0.000025 $ & \\
$\ln{\sigma_f}$ &$    -5.6410 \pm     0.0039 $ & \\
\noalign{\smallskip}
\multicolumn{3}{@{}l}{Derived parameters} \\
\noalign{\smallskip}
$R_{\rm p}/R_{\star}$ &$     0.0487 \pm     0.0010 $ & \\
$a/R_{\star}$ &$       16.2 \pm        1.8 $ & \\
$i$~~[$\degr$] &$       88.6 \pm        1.0 $ & \\
$\log(\rho_{\star}/\rho_{\odot})$ &$       1.33 \pm       0.15 $ & \\
$\sigma_f$~~[ppm]                 &$       3549 \pm         14 $ & \\
\noalign{\smallskip}
\hline
\end{tabular}
\end{table}

\begin{figure}
\begin{center}
\resizebox{0.7\columnwidth}{!}{\includegraphics{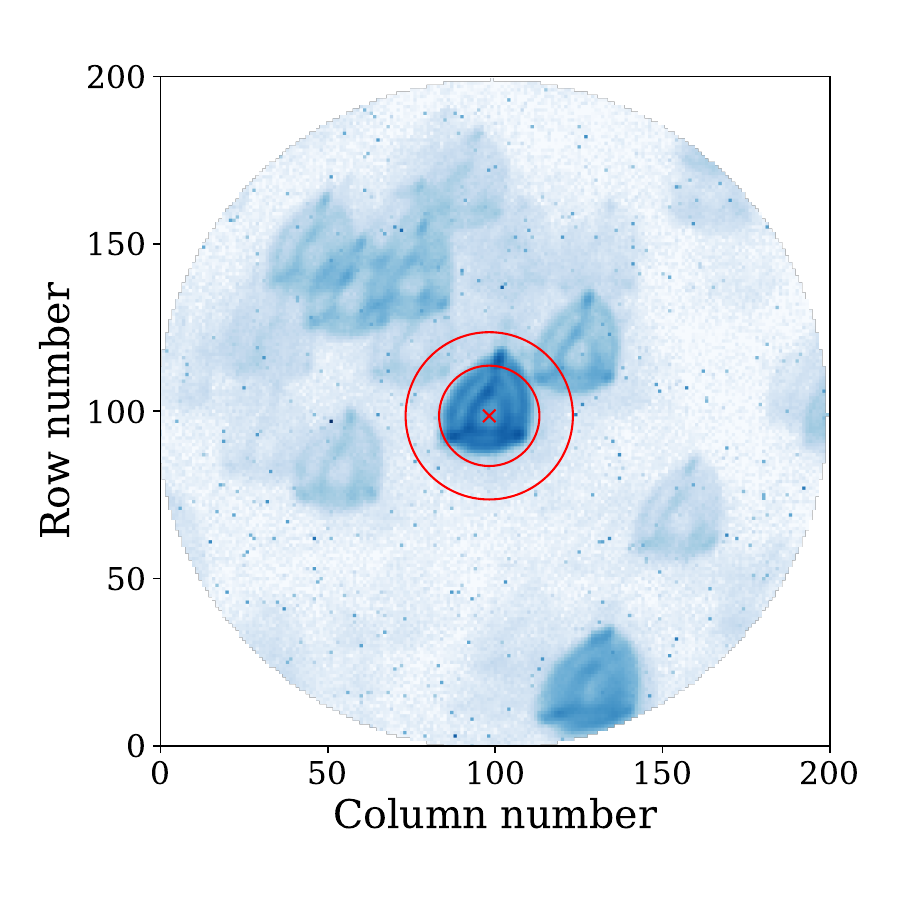}}
\end{center}
\caption{\label{fig:GJ1132_FOV} A typical image of GJ~1132 obtained with \CHEOPS\ prior to calibration and cosmic ray removal. The blue circles indicate photometric apertures with radii of 15.0 and 22.5 pixels.  }
\end{figure}

\begin{figure*}
\resizebox{0.8\textwidth}{!}{\includegraphics{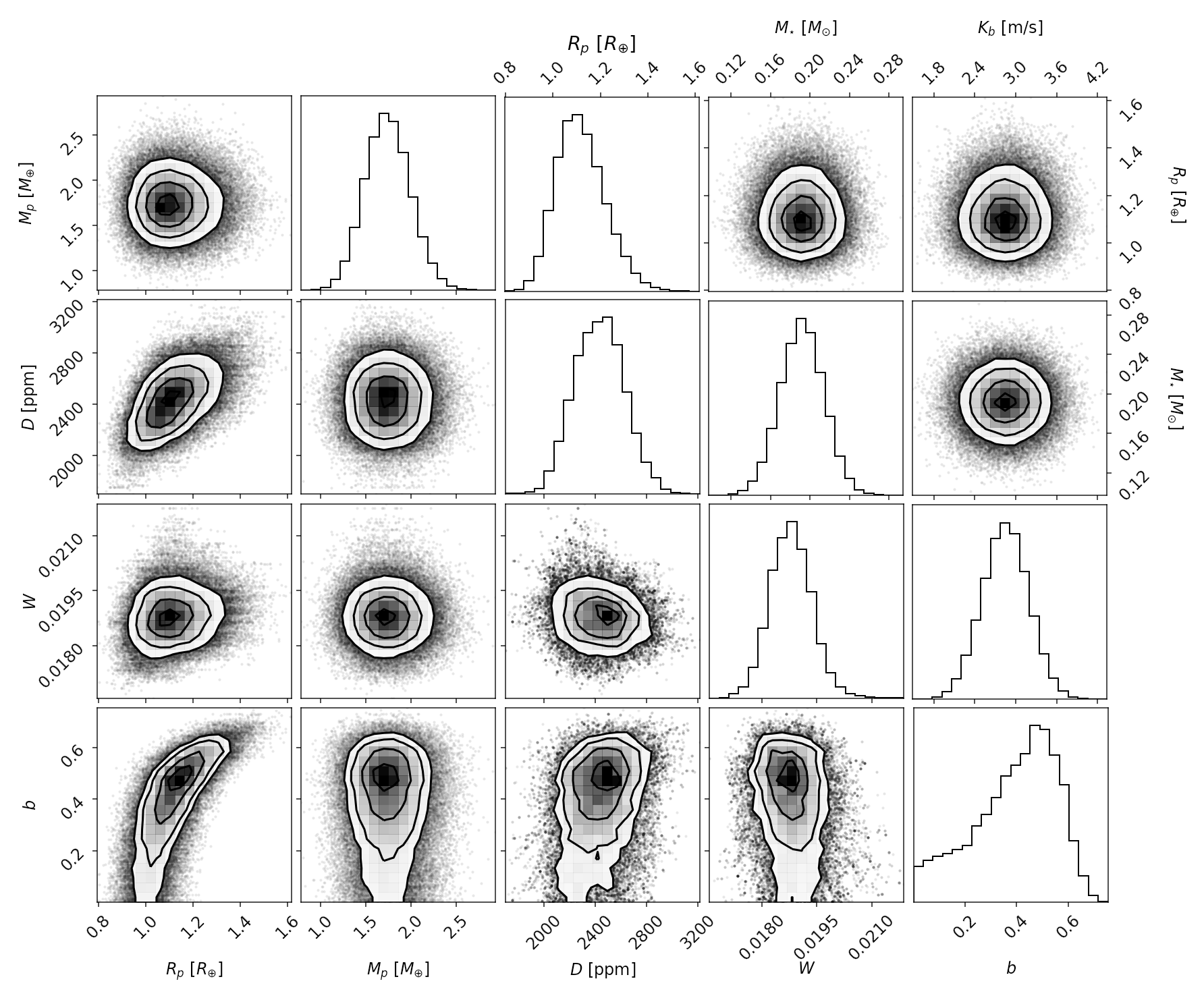}}
\caption{\label{fig:GJ_1132_corner} Correlation plot for selected parameters from our analysis of GJ~1132.
\label{fig:gj1132_corner}}
\end{figure*}

\begin{figure*}
\resizebox{0.8\textwidth}{!}{\includegraphics{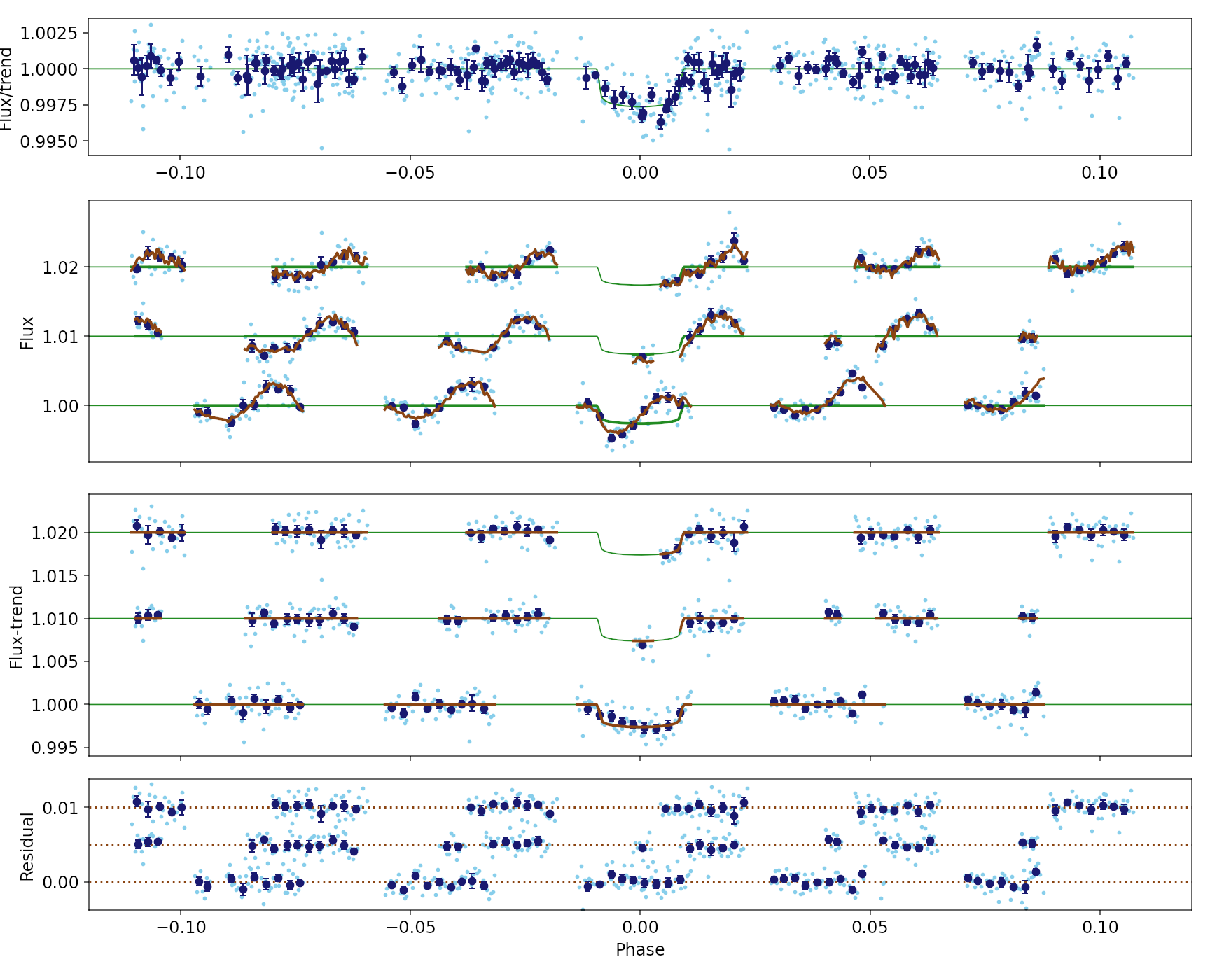}}
\caption{\label{fig:GJ_1132_lcfit-nroll3} \textit{Top}:  \CHEOPS\ observations of 3 transits of GJ~1132\,b. {\it Upper panel:} All data after removing trends. Observed light curves are displayed in cyan. The dark blue points are the data points binned over 0.0025 phase units.  The best-fit transit model is shown in green. {\it Middle-upper plot} Observed light curves are displayed in cyan offset by multiples of 0.01 units. The full model including instrumental trends is shown in brown and the transit model without trends is shown in green. {\it Middle-lower panel:} Same as the middle-upper panel after removing trends correlated with space-craft roll angle. {\it Lower panel:} Residuals obtained after subtraction of the best-fit model in the same order as the upper plot offset by multiples of 0.005 units.
\label{fig:gj1132_lcfit}
}
\end{figure*}

\subsubsection{GJ 1132 b}

GJ~1132 is a nearby  M4.5\,V star ($d\approx 12$\,pc) that was found to host a transiting exoplanet using ground-based photometry from the MEarth project \citep{2015Natur.527..204B}. GJ~1132\,b is a small rocky planet with a radius of $\sim$2.4~$R_\oplus$, a mass of $\sim$1.7~$M_\oplus$, and an orbital period of $P=1.63$ days.  Additional photometry from the MEarth-South telescopes and over 100 hours of observations with {\it Spitzer} by \citet{2017AJ....154..142D} did not reveal any additional transiting exoplanets in this system. Nevertheless, \citet{2018A+A...618A.142B} found evidence for a second non-transiting planet in this system (GJ~1132\,c) with an orbital period $P\approx8.83$\,d from extensive radial velocity observations. \citet{2017AJ....153..191S} claimed the detection of an extended atmosphere on  GJ~1132\,b based on an increased transit depth in the z$^{\prime}$ and K bands relative to other wavelengths. Subsequent spectrophotometric observations with the LDSS3C multi-object spectrograph on the Magellan Clay Telescope by \citet{2018AJ....156...42D} failed confirm the anomalous transit depth around wavelengths of 1\,$\mu$m and are consistent with a featureless spectrum, implying that GJ~1132\,b has a high mean molecular weight atmosphere or no atmosphere at all. More recently, \citet{2021AJ....161..213S} have claimed the detection of atmospheric absorption features in the transmission spectrum of GJ~1132\,b obtained with the WFC3 instrument on {\it HST} over the wavelength range 1.13\,--\,1.64\,$\mu$m, but at a much lower level than the broad-band features claimed by \citeauthor{2017AJ....153..191S} ($\sim 250$\,ppm cf. $\sim$1500\,ppm). \citet{2021arXiv210401873M} found no evidence for molecular absorption in the transmission spectrum of GL~1132\,b from their analysis of the same WFC3 data analysed by \citet{2021AJ....161..213S}. 

Based on its V-band magnitude \citep[V$\approx 14.9$,][]{2011AJ....142...15G}, GJ~1132 lies beyond the faint magnitude limit of \CHEOPS\ (V=12-13). However, the high scientific interest of small planets transiting M dwarfs, which are favourable for atmospheric characterisation, motivated us to assess the precision that \CHEOPS\ can achieve for such faint targets. \CHEOPS\ has a very broad spectral response which is very similar to the {\it Gaia} $G$-band, so the count rate for cool stars like GJ~1132 is equivalent to a Sun-like star with the same $G$-band magnitude but approximately 1 magnitude brighter in the V-band. Nevertheless, GJ~1132 is a faint star ($G=12.1$) in a crowded part of the sky (Fig.~\ref{fig:GJ1132_FOV}) and the transits due to GJ~1132\,b are shallow, so this is a challenging target for observations with \CHEOPS.  

The three transits of GJ~1132\,b we observed with \CHEOPS\ have an observing efficiency from 58\% to 70\%. The duration of the transit is approximately half that of a CHEOPS orbit but we were unfortunate that the majority of the transit falls in a gap for two of the visits. The light curves are dominated by instrumental noise due to contamination of the aperture by nearby stars. For this reason, the OPTIMAL photometric aperture has a radius $\approx$ 15 pixels, much smaller than the aperture size typically used for \CHEOPS\ observations. In addition to the problems with contamination and unfortunate scheduling, it was found that using the science images to track the star during the visits gives worse performance than using the off-axis star trackers. This mode of operation (``payload in the loop'') was disabled for the final visit. The RMS pointing residual was reduced from 2.7\arcsec\ and 3.8\arcsec\ for the first two visits to 0.36\arcsec\ for the final visit. 

GJ\,1132 shows little intrinsic variability. MEarth photometry of GJ~1132 shows rotational modulation with a period $P_{\rm rot}\approx 125$ days and an amplitude $\approx 0.1$\% \citep{2015Natur.527..204B}. To estimate the mass of GJ~1132 we used the mass\,--\,M$_{\rm K}$ relation from \citet{2016AJ....152..141B}. The absolute K-band magnitude of GJ~1132 based on the parallax from {\it Gaia} EDR3 ($\pi = 79.321 \pm 0.018$~mas) and the K$_{\rm s}$-band magnitude from 2MASS (K$_{\rm s} =8.322\pm 0.027$) is M$_{\rm K} = 7.819\pm0.027$. To estimate the error in this value we used the standard deviation of the residuals from this relation for the 9 stars in Benedict et~al. with M$_{\rm K}$ in the range 7.62 to 8.02. Including the small additional uncertainty inherited from the error in M$_{\rm K}$ we estimate that the mass of GJ~1132 is $0.192 \pm 0.022\,M_{\odot}$.
 
To estimate the mean stellar density of GJ~1132 we compiled a sample of stars with accurate and precise surface gravity measurements. We use surface gravity rather than mean stellar density directly because this parameter can be determined independently of any assumptions about the primary star mass for eclipsing binaries where an M-dwarf transits a solar-type star. The properties of these stars are given in Table~\ref{tab:logg-mdwarf}. Note that the value of $\log g$ quoted in Table~4 of \citet{2018MNRAS.481.1897C} is incorrect so we have re-calculated this value based on the mass and radius values given in the same table.   We found that the 5-Gyr solar-metallicity  isochrones from \citet{2015A+A...577A..42B} gives a good estimate for the mass\,--\,$\log g$ relation in this mass range. There is no clear trend with [Fe/H] in the residuals for these stars so we do not account for [Fe/H] when we estimate $\log g$. Based on this isochrone and the standard error of the residuals, we estimate that the surface gravity of GJ~1132 is $\log g  = 5.070 \pm 0.056$. The mean stellar density and radius implied by these values of the mass and $\log g$ are $R  = 0.212 \pm 0.018\,R_{\odot}$ and $\log (\rho/\rho_{\odot}) = 1.307\pm0.089$, respectively. This radius estimate is in very good agreement with the value $R = 0.202 \pm 0.016 R_{\odot}$ inferred from the absolute $G$-band magnitude using the M$_{G}$\,--\,$R$ relation from \citet{2019MNRAS.484.2674R}. Our mass and radius estimates are in good agreement with the values $M=0.181 \pm 0.019$, $R = 0.207 \pm 0.016\,R_{\odot}$ from \citet{2015Natur.527..204B}. The slight increase in the mass and radius are a consequence of the slightly smaller parallax for GJ~1132 from {\it Gaia} EDR3 compared to the value used by Berta-Thompson et al. ($\pi = 83.07 \pm 1.69$~mas).

The T$_{\rm eff}$ and [Fe/H] estimates for GJ~1132 in Table~\ref{tab:gj1132_pars} were obtained using \software{odusseas}, a machine learning tool to derive effective temperature and metallicity for M dwarf stars based on the measurement of the pseudo equivalent widths of stellar absorption lines in high-resolution optical spectra \citep{2020A+A...636A...9A}. We applied \software{odusseas} to the spectrum obtained by combining the spectra of GJ~1132 observed with the HARPS spectrograph.  This estimate of T$_{\rm eff}$ is in reasonably good agreement with the value T$_{\rm eff}  = 3203 \pm 53$\,K based on the star's absolute $G$-band magnitude and the T$_{\rm eff}$\,--\,M$_{\rm G}$ calibration from \citet{2019MNRAS.484.2674R}.

We used an initial analysis of each transit with \code{Dataset.lmfit\_transit} to determine which decorrelation parameters should be used in the combined analysis of the three light curves.  The correction of the ramp effect has not been calibrated for aperture radii less than 22.5 pixels so we did not apply the ramp correction to the light curves used here calculated with aperture radii $\approx15$ pixels. Extrapolating the ramp correction as a function of aperture radius suggests that this correction is  $<30$\,ppm for these light curves. We fixed the orbital period at the value $P=1.6289287$\,d \citep{2017AJ....153..191S} and assumed a circular orbit for this initial analysis, and the limb darkening parameters $h_1$ and $h_2$ were fixed at the values determined from Table~10 of \citet{2019RNAAS...3...17C}. We find that the individual transits provide no constraint on the impact parameter so we fixed this parameter at a nominal value $b=0.77$. The mean stellar density estimate described above ($\log (\rho/\rho_{\odot}) = 1.307\pm0.089$) was included as a constraint in the least-squares analysis. Contamination by background stars is the dominant source of instrumental noise in the light curves so we included \var{dfdcontam} as a decorrelation parameter in the analysis of all the light curves. Other decorrelation parameters were selected in the usual way based on their Bayes factors using the method described in the introduction to this section. A summary of the results from this initial analysis is given in Table~\ref{tab:individual}. The power spectral density (PSD) of the residuals shown in Fig.~\ref{fig:fft_GJ1132} is consistent with the expected white-noise level based on the median error bar per datum for all three data sets. The trends in the data with spacecraft roll angle and our fit to these trends for each data set are shown in Fig.~\ref{fig:roll_GJ1132}.

\citet{2018A+A...618A.142B} find that the eccentricity of the orbit is $e<0.22$ at the 95\% confidence level so for the combined analysis of the visits using \code{MultiVisit} we assumed that the orbit is circular. The limb-darkening parameter $h_2$ has only a subtle effect on the light curve during the ingress and egress phases of the transit so we decided to fix this parameter at the value inferred from the tables provided by \cite{2019RNAAS...3...17C}. We include $h_1$ as a free parameter in the analysis  with a Gaussian prior centered on the value obtained from the same tables with an arbitrary choice of 0.1 for the standard error. We imposed the same prior on the mean stellar density as used in the analysis of the individual visits. Based on the results of the analysis for the individual visits we decided to use $N_{\rm roll}=2$. The results from this analysis are given in Table~\ref{tab:gj1132_pars} and the fits to the light curves are shown in Fig.~\ref{fig:gj1132_lcfit}. Correlations between selected parameters from this analysis are shown in Fig.~\ref{fig:gj1132_corner}. The results found for an analysis with $N_{\rm roll} =3$  or using the \var{unwrap} option are almost indistinguishable from those presented here. We also tried an analysis with $N_{\rm roll}=1$ but there are clear trends in the residuals related to the roll angle. Even so, the results are consistent with those presented here. Very similar results were also found using the RINF aperture with a radius of 22 pixels. The optimum value of $N_{\rm roll}$ for the RINF aperture data is $N_{\rm roll} =3$; the values of $D$ and $b$ obtained are insensitive to the choice of $N_{\rm roll}$ or whether the \var{unwrap} option is used. 

\citet{2017AJ....154..142D} noted that the value of $R_{\rm p}/R_{\star}$ that they measured using MEarth data is inconsistent with the value obtained using {\it Spitzer} photometry at 4.5\,$\mu$m. We have reanalysed the MEarth photometry provided in their Table~1 because there is a clear non-linear trend in these data when plotted as a function of air mass. To model these data we use the qpower2 transit model implemented in \pycheops\ plus a 5th-order polynomial as a function of $\sec z -1$ to account for trends with air mass (where $z$ is the zenith distance of GJ~1132 at the time of observation) plus a sinusoidal model $a_{\rm rot}\sin(2\pi t/P_{\rm rot}) +b_{\rm rot}\cos(2\pi t/P_{\rm rot}) $ with a period $P_{\rm rot} = 125$\,d to account for stellar variability modulated by the stars rotation period. We did not impose a prior on the mean stellar density for the analysis of the MEarth data and only data within 0.075 phase units of the mid-transit were included in the fit. The results from this reanalysis are also given in Table~\ref{tab:gj1132_mearth}.

These results are discussed in the context of previous studies of GJ\,1132\,b in Section~\ref{sec:gj1132_discuss}. 

\subsubsection{Accuracy of the qpower2 algorithm}

We used the \software{ellc} light curve model \citep{2016A&A...591A.111M} to calculate a transit light curve for each of the 4 planets using direct numerical integration of the power-2 limb darkening law. We then fit these light curves with light curves calculated using the qpower2 algorithm to measure the systematic error in the parameters $R_{p}/R_{\star}$ and $a/R_{\star}$. In all cases, we find that this systematic error is negligible compared to the random error in these quantities.

\subsection{Updated transit ephemerides}

\subsubsection{GJ 436 b}
 We used a linear fit to the time of mid-transit from Table~\ref{tab:gj436_pars}, 8 times of mid-transit from \citet{2014A+A...572A..73L}, and 4 times of mid-transit from \citet{2018AJ....155...66L} to establish the following linear ephemeris for the times of mid-transit:
 \[{\rm BJD}_{\rm TDB}(T_0) = 2455475.82450(3) + 2.64389759(7)\times E.\]
 Values in parentheses give the standard error in the final digit of the preceding quantity. 
 There is no evidence for any change in period greater than $\dot{P}/P \approx 6.0\times10^{-10}$ from these data.

 \subsubsection{HD 106315 b}
 \label{sec:hd106315_ephem}
 We used a linear fit to the two times of mid-transit from Table~\ref{tab:hd106315_pars} to establish the following linear ephemeris for the times of mid-transit for HD~106315\,b:
 \[{\rm BJD}_{\rm TDB}(T_0) = 2458427.132(1) + 9.55211(2)\times E.\]

 \subsubsection{HD 97658 b}
 We used a linear fit to the two times of mid-transit from Table~\ref{tab:hd97658_pars}, one time of mid-transit from \citet{2014ApJ...786....2V}, and 18 times of mid-transit from various instruments from \citet{2020AJ....159..239G} to establish the following linear ephemeris for the times of mid-transit:
 \[{\rm BJD}_{\rm TDB}(T_0) = 2457234.82213(16) + 9.4893072(25) \times E.\]

Values in parentheses are the standard error in the final two digits of the preceding quantity.  This is a slight improvement on the value of the orbital period given by \citet{2020AJ....159..239G} ($P = 9.489295(5)$\,d), partly because of the extended baseline including the observation from \CHEOPS, but also because we choose our reference time of mid-transit (cycle $E=0$) to minimize the covariance between this value and $P$. 

Using the same data set we find the following quadratic ephemeris for the time of mid-transit:
\[
\begin{array}{ll}
{\rm BJD}_{\rm TDB}(T_0) =&  2457234.82195(12) + 9.4892968(38) \times E \\
   & + 0.5\times(1.46\pm0.48)\times10^{-7}\times E^2.\\
\end{array}\]

\begin{figure}
\resizebox{\columnwidth}{!}{\includegraphics{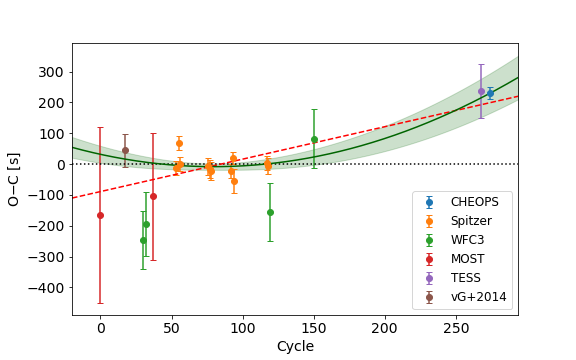}}
\caption{\label{fig:HD97658_tmin} Observed $-$ calculated times of mid-transit for HD~97658\,b based on the linear ephemeris from \citet{2020AJ....159..239G}. The dashed line shows our updated linear ephemeris. The solid line with shaded band shows our updated quadratic ephemeris $\pm$ 1 standard deviation.}
\end{figure}

The Bayesian information criterion for this ephemeris is 37.9 cf. 55.5 for a linear ephemeris, i.e. there is strong evidence from these data that the orbital period of HD~97658\,b is not constant. The observed times of mid-transit and our updated ephemerides are shown as residuals from the linear ephemeris from \citeauthor{2020AJ....159..239G} in Fig.~\ref{fig:HD97658_tmin}.

\subsubsection{GJ 1132 b}

We used a linear fit to the times of mid-transit from Table~\ref{tab:gj1132_pars}, 27 times of mid-transit from \citet{2017AJ....154..142D}, 5 times of mid-transit from \citet{2021arXiv210401873M}, and 9 times of mid-transit from \citet{2017AJ....153..191S} to establish the following linear ephemeris for the times of mid-transit:
 \[{\rm BJD}_{\rm TDB}(T_0) =  2457554.32450(9) + 1.6289292(4)\times E.\]
The errors reported on the times of mid-transit in Table~3 of \citet{2021arXiv210401873M} are clearly too small. We used the root-mean-square residual from a linear fit to these times of mid-transit to assign a more realistic error of 0.00042\,d to these values. There is no evidence for any change in period greater than $\dot{P}/P \approx 3.6\times10^{-9}$ from these data.


\begin{table}
\centering
\caption{Improved planet mass and radius estimates. See Section~\ref{sec:planetradius} for details of the data sources combined to obtain the input values of $R_{\rm p}/R_{\star}$, $a/R_{\star}$  and $\sin i$ used here.}
\label{tab:massradius}
\begin{tabular}{llrrr}
\hline\hline
\multicolumn{1}{@{}l}{Parameter}   & \multicolumn{1}{l}{Units}    & \multicolumn{1}{l}{Value} & \multicolumn{1}{l}{Error}& Notes \\
\hline
\multicolumn{4}{@{}l}{GJ~436\,b} \\
\noalign{\smallskip}
P               &             [d]    &$ 2.64389759 $ &                  &   \\
$M_\star$       & [$M_{\odot}$]      &$      0.445 $ & $\pm\,  0.018 $  &   \\
$K$             & [m\,s$^{-1}$]      &$      17.38 $ & $\pm\,   0.17 $  & 1 \\
$e$             &                    &$      0.152 $ & $\pm\,  0.009 $  & 1 \\
$\sin i$        &                    &$    0.99843 $ & $\pm\,\bmath{0.00004} $  &  \\
$R_{\rm p}/R_{\star}$ &                    &$ \bmath{0.08261} $ & $\pm\,\bmath{0.00022} $  &  \\
$a/R_{\star}$   &                    &$ 14.46 $ & $\pm\, \bmath{0.09} $  &  \\
\noalign{\smallskip}
$R_\star$       & [$R_{\odot}$]      &$      0.425 $ & $\pm\, 0.006 $  &  \\
$M_p$           &   [$M_{\oplus}$]   &$      21.68 $ & $\pm\,  0.63 $  & \\
$R_{\rm p}$           &   [$R_{\oplus}$]   &$\bmath{3.83} $ & $\pm\,  0.06 $  & \\
$g_p$           &   [m\,s$^{-2}$]    &$\bmath{14.50} $ & $\pm\,\bmath{0.24} $  & \\
$\rho_p$        &    [g\,cm$^{-3}$]  &$\bmath{2.12} $ & $\pm\,  0.06 $  & \\
\noalign{\smallskip}
\hline
\noalign{\smallskip}
\multicolumn{3}{@{}l}{HD~106315\,b} \\
\noalign{\smallskip}
P               &             [d]    &$ 9.55211     $ &                 & \\
$M_\star$       & [$M_{\odot}$]      &$\bmath{1.088}  $ & $\pm\,  \bmath{0.043} $ & \\
$K$             & [m\,s$^{-1}$]      &$       2.88  $ & $\pm\,   0.85 $ & 2 \\
$R_{\rm p}/R_{\star}$ &                    &$\bmath{0.01686} $ & $\pm\, \bmath{0.00041}$ & \\
$a/R_{\star}$   &                    &$\bmath{15.95}$ & $\pm\,\bmath{0.55} $ & \\
$\sin i$        &                    &$     0.99931 $ & $\pm\, 0.00013 $ &  \\
\noalign{\smallskip}
$R_\star$       & [$R_{\odot}$]      &$\bmath{1.221}  $ & $\pm\,\bmath{0.045} $  &  \\
$M_p$           &   [$M_{\oplus}$]   &$        10.1 $ & $\pm\,    3.0 $ & \\
$R_{\rm p}$           &   [$R_{\oplus}$]   &$\bmath{2.25} $ & $\pm\, \bmath{0.10}$ & \\
$g_p$           &   [m\,s$^{-2}$]    &$ \bmath{19.6} $ & $\pm\, \bmath{6.0} $ & \\
$\rho_p$        &    [g\,cm$^{-3}$]  &$\bmath{4.9} $ & $\pm\,    1.6 $ & \\
\noalign{\smallskip}
\hline
\noalign{\smallskip}
\multicolumn{3}{@{}l}{HD~97658\,b} \\
\noalign{\smallskip}
P               &             [d]    &$  9.4893072  $ &                 & \\
$M_\star$       & [$M_{\odot}$]      &$      0.752  $ & $\pm\,  0.035 $ & \\
$K$             & [m\,s$^{-1}$]      &$        2.81 $ & $\pm\,   0.15 $ & 3 \\
$R_{\rm p}/R_{\star}$ &                    &$     0.02863 $ & $\pm\, \bmath{0.00030}$ & \\
$a/R_{\star}$   &                    &$ \bmath{23.69}  $ & $\pm\,\bmath{0.49} $ & \\
$\sin i$        &                    &$     0.999816 $ & $\pm\,0.000034 $ &  \\
\noalign{\smallskip}
$R_\star$       & [$R_{\odot}$]      &$       0.724 $ & $\pm\,  0.019 $  &  \\
$M_p$           &   [$M_{\oplus}$]   &$        7.69 $ & $\pm\,   0.47 $ & \\
$R_{\rm p}$           &   [$R_{\oplus}$]   &$        2.26 $ & $\pm\,   0.06 $ & \\
$g_p$           &   [m\,s$^{-2}$]    &$        14.7 $ & $\pm\,    1.0 $ & \\
$\rho_p$        &    [g\,cm$^{-3}$]  &$        3.65 $ & $\pm\,   0.33 $ & \\
\noalign{\smallskip}
\hline
\noalign{\smallskip}
\multicolumn{3}{@{}l}{GJ~1132\,b} \\
\noalign{\smallskip}
P               &             [d]    &$  1.6289289  $ &                 & \\
$M_\star$       & [$M_{\odot}$]      &$       0.192 $ & $\pm\,  0.022 $ & \\
$K$             & [m\,s$^{-1}$]      &$        2.85 $ & $\pm\,   0.34 $ & 4 \\
$R_{\rm p}/R_{\star}$ &                    &$      0.04901 $ & $\pm\, 0.00054 $ & \\
$a/R_{\star}$   &                    &$      16.38  $ & $\pm\,   0.55 $ & \\
$\sin i$        &                    &$      0.9997 $ & $\pm\, 0.0001 $ &  \\
\noalign{\smallskip}
$R_\star$       & [$R_{\odot}$]      &$       0.205 $ & $\pm\,  0.010 $ &  \\
$M_p$           &   [$M_{\oplus}$]   &$        1.74 $ & $\pm\,   0.25 $ & \\
$R_{\rm p}$           &   [$R_{\oplus}$]   &$        1.10 $ & $\pm\,   0.06 $ & \\
$g_p$           &   [m\,s$^{-2}$]    &$        14.2 $ & $\pm\,    2.0 $ & \\
$\rho_p$        &    [g\,cm$^{-3}$]  &$         7.2 $ & $\pm\,    1.2 $ & \\
\noalign{\smallskip}
\hline
\end{tabular}
\medskip
\parbox{\columnwidth}{ 
1:~\citet{2018A+A...609A.117T}.
2:~\citet{2021AJ....161...47K}.
3.~\citet{2020AJ....159..239G}.
4.~\citet{2018A+A...618A.142B}.
}

\end{table}

\begin{figure}
\resizebox{\columnwidth}{!}{\includegraphics{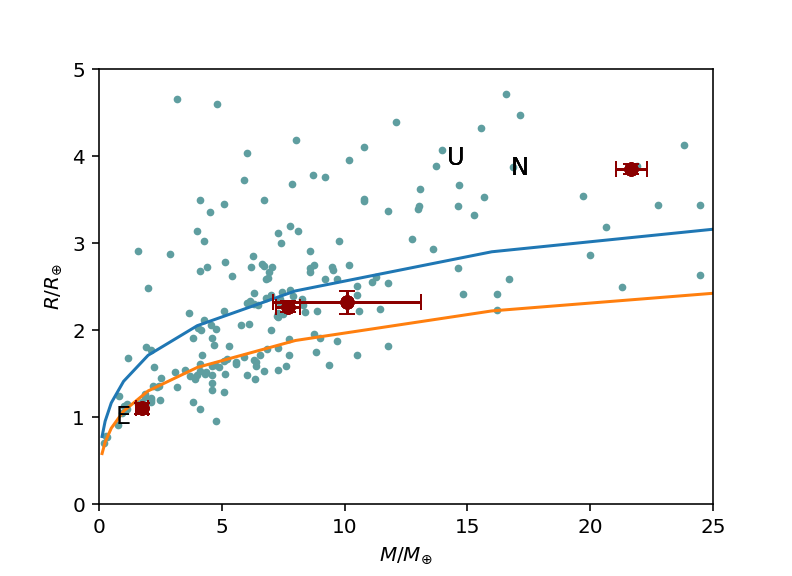}}
\caption{\label{fig:massradius} In order of increasing mass -- GJ~1132\,b, HD~97658\,b, HD~106315\,b, and GJ~436\,b in the mass-radius plane compared to other extrasolar planets with well-determined parameters taken from TEPCat (cyan points) and models from \citet{2016ApJ...819..127Z} for planets composed of 100\%  rock (lower line) or water (upper line). The mass and radius of Earth, Uranus and Neptune are also shown using the initial letters of these planets' names.}
\end{figure}

\subsection{Planet mass and radius estimates}
\label{sec:planetradius}

 The values of the planet mass ($M_{\rm p}$) and radius ($R_{\rm p}$) given in Tables~\ref{tab:gj436_pars}, \ref{tab:hd106315_pars}, \ref{tab:hd97658_pars} and \ref{tab:gj1132_pars} are based on the values of $a/R_{\star}$, $i$ and $k=R_{\rm p}/R_{\star}$ measured from the \CHEOPS\ light curves only. In this section we make improved estimates for $M_{\rm p}$ and $R_{\rm p}$ using all published estimates for these parameters that are of similar precision to the values obtained from the \CHEOPS\ data, or better. For all four planets we have used our best estimate for the stellar mass, $M_{\star}$, together with the mean stellar density, $\rho_{\star}$ derived using Kepler's law from $a/R_{\star}$, to infer a value of $R_{\star}$ and, hence, $R_{\rm p} = k\times R_{\star}$. The masses and radii obtained are shown in Fig.~\ref{fig:massradius}.

\subsubsection{GJ 436 b}
 \citet{2018AJ....155...66L} observed two transits of GJ~436\,b using the STIS spectrograph on {\it HST} with the G750L low-resolution grism covering the wavelength range 0.53\,--\,1.03\,$\mu$m. These observations do not cover the egress of the transit so \citeauthor{2018AJ....155...66L} used fixed values for $a/R_{\star}$ and $i$ from \citet{2015ApJ...802..117M} in their analysis. The weighted mean transit depth from the values at various wavelengths given in their Table~3 using our method described in Appendix~\ref{sec:combine} is $6746 \pm 30$\,ppm. \citeauthor{2018AJ....155...66L} find that using values of $a/R_{\star}$ and $i$ from different sources introduces an additional uncertainty $\approx 130$\,ppm in the transit depth. Taking this into account, we find that the planet-star radius ratio from this study is \mbox{$k=0.08213 \pm 0.00081$}.
 
\citet{2014Natur.505...66K} used the Wide Field Camera 3 (WFC3) instrument on {\it HST} to observe 4 transits of  GJ~436\,b over the wavelength range 1.2\,--\,1.6$\,\mu$m. From their Table~1 we use the values $a/R_{\star} = 14.41\pm0.10$ and $i=86.774\degr \pm 0.030\degr$, and the four values of $R_{\rm p}/R_{\star}$ from each visit which we combine to obtain the weighted average value $k=0.08362 \pm 0.00015$. 

Transits of  GJ~436\,b observed several times with {\it Spitzer} at 3.6\,$\mu$m, 4.5\,$\mu$m and 8.0\,$\mu$m. Some or all of these data have been analysed by \citet{2011ApJ...735...27K}, \citet{2011ApJ...731...16B}, \citet{2015ApJ...802..117M} and \citet{2014A+A...572A..73L}. These studies use a variety of techniques to account for instrumental noise that is comparable to the transit depth in these light curves. Here we use the results from \citeauthor{2014A+A...572A..73L} since this is the only study to use all the available data. From the parameters in their Table~3 we obtain the values $k=0.08258\pm 0.00017$, $i=86.858\degr \pm 0.52$ and $a/R_{\star} = 14.54\pm0.15$. 

None of the studies above find any strong evidence for variations in transit depth with wavelength due to opacity sources in an extended atmosphere on  GJ~436\,b, so we have combined all these estimates of $R_{\rm p}/R_{\star}$ irrespective of wavelength. The values of  $a/R_{\star}$, $\sin i$ and $R_{\rm p}/R_{\star}$ obtained by combining the above estimates with the results from  Table~\ref{tab:gj436_pars} are given in Table~\ref{tab:massradius}, together with the resulting planetary mass and radius estimates. 

\subsubsection{HD 106315 b}
 The values of $\sin i$ and $R_{\rm p}/R_{\star}$ in Table~\ref{tab:massradius} come from combining our results in Table~\ref{tab:hd106315_pars} based on the analysis of the \CHEOPS\ and {\it K2} light curves with those from \citet{2021AJ....161...47K} based on the analysis of two transits of  HD~106315\,b observed with {\it Spitzer} at 4.5\,$\mu$m. We have not used the values of $a/R_{\star}$ from  \citeauthor{2021AJ....161...47K} because they are either inconsistent with the mean stellar density measured independently by several authors shown in Table~\ref{tab:hd106315_rhostar}, or not precise enough to be useful. The values of $a/R_{\star}$ in Table~\ref{tab:hd106315_pars} from the analysis of the \CHEOPS\ and {\it K2} are not independent. They are both strongly constrained by the same prior that we placed on $\rho_{\star}$ for the analysis of both these light curves, so we only used the value of $a/R_{\star}$ from the analysis of the \CHEOPS\ light curve. Where \citeauthor{2021AJ....161...47K} quote asymmetric error bars on a parameter we use the larger of the two error bars. The values from different sources have been combined using the algorithm described in Appendix A. \citeauthor{2021AJ....161...47K} argued that the orbital eccentricity of HD~106315\,b is likely to be small based on the observed radial velocities and on stability arguments for the orbits of the two planets in this system. Based on this analysis we fix $e=0$ for our calculation of the mass and radius of HD~106315\,b.

\subsubsection{HD 97658 b}
Extensive photometry of the transits of HD~97658\,b using {\it Spitzer} and {\it HST} has been presented by \citet{2020AJ....159..239G}. Their Table~2 seems to imply that they were able to establish a value of $a/R_{\star} = 26.7\pm0.4$ from the analysis of their {\it HST} light curves. However, this seems unlikely given that these data have poor coverage of the transit, e.g. the egress was not observed at all, so it is unclear to us where this estimate of $a/R_{\star}$ comes from. It also appears from their Table~2 that they assumed for the analysis of the transits that the orbital eccentricity is $e=0.078$ and that the longitude of periastron is $\omega=90$\degr. Again, it is unclear where these estimates comes from -- previous estimates of $\omega$ have very large uncertainties because the eccentricity of the orbit is low.\footnote{We attempted to contact Xueying Guo via her co-authors but, at the time of writing, we have not obtained clarification of these points.} Unfortunately, this value of $\omega$ maximizes the difference between the value of the mean stellar density inferred from the transit width via Kepler's law assuming either a circular or eccentric orbit. The results of their radial velocity analysis presented in their Table~7 assume that the orbit is circular. Using equation (34) from \citet{2014MNRAS.440.2164K}, the difference is 26\%, with the value derived for $e=0$ being larger than the true value if $e>0$. A full re-analysis of the data in \citet{2020AJ....159..239G} is beyond the scope of this study so we have decided not to use the results from the analysis of the {\it HST} and {\it Spitzer} light curves by \citeauthor{2020AJ....159..239G} in this analysis. However, the results from the radial velocity analysis by \citeauthor{2020AJ....159..239G} are unambiguous so we have followed them in assuming that the orbit is circular and have used the value of $K$ from their Table~7.

Four transits of HD~97658\,b observed by \citet{2013ApJ...772L...2D} with the {\it MOST} satellite provide the following estimates for the transit parameters: $k = 0.0306 \pm 0.0014$, $a/R_{\star}=24.36^{+0.97}_{-1.1}$, \mbox{$i=89\fdg45^{+0.37}_{-0.42}$}. From the analysis of a single transit observed at 4.5$\mu$m with {\it Spitzer} by \citet{2014ApJ...786....2V} we obtain the following values: $k = 0.02780^{+0.00075}_{-0.00077}$, $a/R_{\star}=24.9\pm1.4$, $i=89\fdg14^{+0.52}_{-0.36}$, where the value of $a/R_{\star}$ has been calculated from the values of $D$, $W$ and $b$ in their Table~2. We have combined these estimates with the values of $k$, $a/R_{\star}$ and $i$ from Table~\ref{tab:hd97658_pars} to obtain the values shown in Table~\ref{tab:massradius} using the algorithm described in Appendix A. Where asymmetric error bars are quoted on values we have used the larger value as the standard error estimate. Similarly to the K2 light curve of HD~106315\,b, we have not used the value of $a/R_{\star}$ from the analysis of the {\it TESS} light curve in this calculation because it is not independent of the value from the analysis of the \CHEOPS\ light curve -- both values are strongly constrained by the same prior on $\rho_{\star}$. 
 \citet{2020AJ....159..239G} did not find any strong evidence for features in the transmission spectrum of  HD~97658\,b so we have ignored any possible wavelength dependence in the planetary radius for the calculations summarised in Table~\ref{tab:massradius}.
 
\subsubsection{GJ 1132 b}
Measurements of $R_{\rm p}/R_{\star}$ for GJ~1132\,b from various sources are listed in Table~\ref{tab:gj1132_k_wave}. We have not used the estimates from \citet{2017AJ....153..191S} in our calculations for reasons that will be discussed in Section~\ref{sec:gj1132_discuss}. The values of $a/R_{\star}$ and $\sin i$ in Table~\ref{tab:massradius} are the weighted means of the values from the same sources used to calculate $k$ calculated using the algorithm described in Appendix A. We have ignored any possible wavelength dependence in the planetary radius for the calculations summarised in Table~\ref{tab:massradius}. This point will also be discussed in 
Section~\ref{sec:gj1132_discuss}.

\begin{figure}
\resizebox{\columnwidth}{!}{\includegraphics{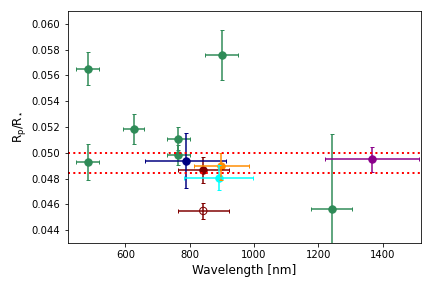}}
\caption{\label{fig:gj1132_k_wave} Planet-star radius ratio as a function of wavelength for GJ~1132\,b. Points are colour-coded as follows: blue --  \CHEOPS, red -- MEarth, orange -- \citet{2018AJ....156...42D}, green -- \citet{2017AJ....153..191S}, magenta -- WFC3, cyan -- TESS. The radius ratio obtained from {\it Spitzer} observations at 4.5\,$\mu$m is indicated with dotted lines. The planet-star radius ratio measured using MEarth data by \citet{2017AJ....154..142D} is plotted with an open circle symbol.}
\end{figure}

\begin{table}
\caption{Planet-star radius ratio as a function of wavelength for GJ~1132\,b. The flux-weighted mean photon wavelength for each observation and its standard deviation are indicated in the column headed $\langle\lambda\rangle$.
\label{tab:gj1132_k_wave}}
\begin{center}
\begin{tabular}{@{}lrrr}
\hline
\multicolumn{1}{@{}l}{Bandpass} & 
\multicolumn{1}{l}{$\langle\lambda\rangle$ [nm]} & 
\multicolumn{1}{l}{$R_{\rm p}/R_{\star}$} & Ref. \\
\hline
\CHEOPS\     & $ 787 \pm 126$ & $0.0494 \pm  0.0021$ & 1 \\
MEarth       & $ 842 \pm ~~79$  & $0.0487 \pm  0.0010 $& 1 \\
Spitzer      & $4442 \pm284$  & $0.0492 \pm  0.0008$ & 2 \\
LDSS3C       & $ 901 \pm ~~90$  & $0.0490\pm 0.0010$ & 3 \\
\noalign{\smallskip}
\hline
Mean         &                & $0.0490 \pm  0.0005 $ &  \\
\hline
MEarth       & $ 842 \pm 79$  & $0.0455 \pm  0.0006 $& 2 \\

g$^{\prime}$ & $ 482 \pm  35$  & $0.0493 \pm  0.0014 $ & 4 \\
r$^{\prime}$ & $ 626 \pm  34$  & $0.0519 \pm  0.0012 $ & 4 \\
i$^{\prime}$ & $ 764 \pm  36$  & $0.0498 \pm  0.0008 $ & 4 \\
z$^{\prime}$ & $ 900 \pm  52$  & $0.0575 \pm  0.0019 $ & 4 \\
J            & $1235 \pm  68$  & $0.0457 \pm  0.0058 $ & 4 \\
H            & $1648 \pm  76$  & $0.0418 \pm  0.0057 $ & 4 \\
K            & $2166 \pm  87$  & $0.0610 \pm  0.0075 $ & 4 \\
g$^{\prime}$ & $ 482 \pm  35$  & $0.0565 \pm  0.0013 $ & 5 \\
i$^{\prime}$ & $ 482 \pm  35$  & $0.0511 \pm  0.0009 $ & 5 \\
WFC3         & $1366 \pm 146$  & $0.0495 \pm  0.0010 $ & 6 \\
{\it TESS}   &  $ 890 \pm 107$  & $0.0481 \pm  0.0010 $ & 6 \\

\hline
\end{tabular}
\end{center}
\parbox{\columnwidth}{
1. This work. 
2. \citet{2017AJ....154..142D}.
3. \citet{2018AJ....156...42D}.
4. \citet{2017AJ....153..191S}, GROND.
5. \citet{2017AJ....153..191S}, PISCO.
6. \citet{2021arXiv210401873M}.
}
\end{table}


\section{Constraints on the internal structure}
\label{sec:structure}
We used the retrieval code already employed in the case of TOI-178 \citep{Leleu21} to constrain the planetary internal structure. Here, we briefly recall the ingredients of the model and apply it to three of the exoplanets observed with \CHEOPS\ during the Early Science observing programme. More details on the code can be found in \citet{Leleu21}.

We use a global Bayesian model to fit the observed properties of the star and planet. The observed properties of the star are its mass, radius, age, effective temperature, and the photospheric abundances  [Si/Fe] and [Mg/Fe].  The  observed properties of the planet are the planet-star radius ratio, the radial-velocity semi-amplitude, and the orbital period. The hidden planetary properties are the mass of solids (where ``solids'' refers to the mass of planet not due to H or He gas), the mass fractions of the core, mantle and water, the mass of the gas envelope, the Si/Fe and Mg/Fe mole ratios in the planetary mantle, the S/Fe mole ratio in the core, and the equilibrium temperature due to irradiation by the star. Then, for any given combination of hidden planetary properties and stellar properties, one can compute the resulting planet-star radius ratio and the radial-velocity semi-amplitude. 

The two important ingredients of such a calculation are the physics included in the forward model that is used to calculate the radius of a planet with a given mass and structure, and the prior distribution on the planetary hidden parameters. We assume in the calculations presented below a fully differentiated planet, consisting of a core composed of Fe and S, a mantle composed of Si, Mg and Fe, a pure water layer, and a gas layer composed of H and He only. The equations of state used for these calculations are taken from \citet{2018Icar..313...61H} and \citet{2016GeoRL..43.6837F} at pressures below 240 GPa, and from \citet{2007Icar..191..337S} and \citet{2020A+A...643A.105H} at higher pressures. The temperature profile is assumed to be adiabatic. For the gas envelope, we use the semi-analytical model of \citet{2014ApJ...792....1L} which provides the thickness of the gas envelope as a function of the gas mass fraction, the equilibrium temperature, the mass and radius of the solid planet, and the age (assumed to be equal to the stellar age).

We assume that the logarithm of the gas-to-solid ratio in the planet has a uniform distribution. The mass of the planet core, the planet mantle and the mass of water have uniform priors except that the mass fraction of water in the solid planet is limited to a maximum value of 0.5. We assume that the bulk Si/Fe and Mg/Fe mole ratios in the planet is equal to the one in the star. This assumption will not be valid for planets that have undergone events such as giant impacts that can strongly affect these mole ratios. From the knowledge of the bulk ratio in the planet as well as the core-to-mantle mass ratio, the Si/Fe and Mg/Fe mole ratios in the mantle can be computed analytically. Importantly, the solid and gas part of the planet are computed independently, which means that we do not include the compression effect of the planetary envelope on its core. Including the feedback from the gas envelope onto the planetary core is left for future work, and is well justified \textit{a posteriori} given the small value of the gas envelope.

\begin{table}
        \caption{List of equation of state (EoS) used in the forward model.}
        \label{Tab:eos_table}
        \centering
        \begin{tabular}{l l l}
                \hline\hline
                Layer & Composition & EoS \\
                \hline
                Core & Fe, FeS & \citet{2018Icar..313...61H},\\ 
                     &         & \citet{2016GeoRL..43.6837F}\\ 
                \noalign{\smallskip}
                Mantle & [Mg,Fe]SiO$_3$, [Mg,Fe]O,  &   \\
                       &  [Mg,Fe]$_2$SiO$_4$,[Mg,Fe]$_2$Si$_2$O$_6$ &  \citet{2007Icar..191..337S}\\
                \noalign{\smallskip}
                Volatile & H$_2$O& \citet{2020A+A...643A.105H} \\
                \hline
        \end{tabular}
\end{table}


\section{Discussion}
\label{sec:discussion}
In this section we compare the results from Section~\ref{sec:analysis}  to the results from previous studies of these planets, and discuss the implication of all these results and the analysis in Section~\ref{sec:structure} for our understanding of these planetary systems and the performance of \CHEOPS\ compared to other instrumentation.


\subsection{GJ 436 b}
\label{sec:gj436_discuss}

The transit depth for GJ~436\,b that we have measured using \CHEOPS\ ($7000 \pm 180$\,ppm) is consistent with the weighted mean value $6800 \pm 30$\,ppm from 8 transits observed with {\it Spitzer} at 3 wavelengths by \citet{2014A+A...572A..73L}. The results in their Table~8 show that this weighted mean is dominated by a single observation at 3.6\,$\mu$m with an uncertainty of 40\,ppm cf. a typical uncertainty of 100\,ppm for the other transits. Thus, the precision in the transit depth measurement we have achieved from 3 visits covering $\sim $half of two transits is about half that achieved with a typical observation of a single visit with {\it Spitzer}. This is a consequence of the larger aperture of the Spitzer Space Telescope cf. CHEOPS, the gaps in the \CHEOPS\ observations, and the red colour of this M-type star favouring observations at infrared wavelengths. 

Although the precision of the transit depth measurement by \citeauthor{2014A+A...572A..73L} is 6 times better than our measurement using \CHEOPS, the precision in the planet radius measurement using all available data in Table~\ref{tab:massradius} ($3.85 \pm 0.06\, R_{\oplus}$) is only a factor of two better than the value based on \CHEOPS\ data only in Table~\ref{tab:gj436_pars} ($4.00\pm 0.13 R_{\oplus}$). This is because the uncertainty in the stellar radius is now the dominant source of uncertainty in the calculation of the planets' radius.  The high cadence of the \CHEOPS\ observations helps to reduce this uncertainty because this allows for an accurate measurement of the transit shape and width, from which we can infer an accurate measurement of the mean stellar density.  

GJ~436 is a slowly-rotating star ($P_{\rm rot}\approx 50$\,d) that shows little intrinsic variability at optical wavelengths \citep[$\loa 0.5$\%,][]{2011ApJ...735...27K, 2018AJ....155...66L}. We might then expected changes in flux at the rate $df/dt\sim 0.0001$\,d$^{-1}$ if this intrinsic variability is due to modulation in the visibility of long-lived star spots by rotation. The observed values of $df/dt$ in Table~\ref{tab:gj436_pars} show variability at a rate several times larger than this estimate over a timescale $\approx 8$ hours.  If there is variability with an amplitude $\approx 0.002$ due to short-lived bright or dark regions in the photosphere of GJ~436 that are not occulted by the planet then there will be a systematic error $\approx 0.1\%$ in $R_{\rm p}/R_{\star}$.   

Our internal structure models suggest that GJ~436b has a significant gas envelope, with a mass between 0.67 $ M_\oplus$ and 1.73 $ M_\oplus$ (all given values are the 5\% or 95 \% quantiles). The mass fraction of water in the planet is essentially unconstrained (comprised between 0.08 and 0.41 of the mass of the core).


\subsection{HD 106315 b}
\label{sec:hd106315_discuss}
The value of the orbital period derived in Section~\ref{sec:hd106315_ephem} is significantly different from the value given by \citet{2021AJ....161...47K} based on their analysis of two transits observed with {\it Spitzer} and the published time of minimum based on {\it K2} data ($9.55287\pm0.00021$\,d). The time of conjunction given by \citeauthor{2021AJ....161...47K} for the transit observed with {\it Spitzer} on the date 2017-09-10 is clearly discrepant by over an hour. This discrepancy introduces a systematic error in the predicted time of mid-transit of almost 7 hours using their linear ephemeris for the observing date 2025 discussed by \citeauthor{2021AJ....161...47K}. The uncertainty on the time of mid-transit for observations in 2025 with our updated linear ephemeris is now less than 10 minutes.

The precision in the planet radius we derive from two transits of HD~106315\,b observed with \CHEOPS\ is very similar to that obtained from about 80 days of observations with {\it K2} covering {\bf 6} transits. Although {\it Kepler} has a larger aperture that {\it CHEOPS} and observed more transits during the {\it K2} mission, 3 of the transits contain only 1 or 2 valid observations and all the transits are affected by missing data points. As a result, there are only 20 valid {\it K2} observations during the transit of HD~106315\,b. These data are also affected by inaccuracies in the correction for spurious flux variations due to the spacecraft motion. There is very good agreement between the transit depth measurements from the two instruments. \CHEOPS\ is very well suited to observations of bright, isolated stars like HD~106315, and the very low levels of instrumental noise for such targets allows for accurate and precise characterisation of broad, shallow transits such as those produced by HD~106315\,b.

In term of internal structure, the internal structure modelling shows that HD~106315b has a gas envelope smaller than $10^{-3} M_\oplus$ (all given values are the 5\% or 95 \% quantiles),  a large mass fraction of water (comprised between 0.04 and 0.47 of the mass of the core, with some preference for large water fraction), and an iron mass fraction in the planet  smaller than for the Earth. Both explains why the density of the planet is smaller than the one of the Earth, and of a pure silicate sphere (see Fig. \ref{fig:massradius}).


\subsection{HD 97658 b}
\label{sec:hd97658_discuss}
HD~97658 is the brightest target observed during the Early Science programme so any systematic noise sources not  removed by the DRP or our decorrelation techniques are most likely to be seen in the light curve of this star. We experimented with using a Gaussian process to model correlated noise in the analysis of this visit using the kernel described in Section~\ref{sec:noise}. This requires some thought about the use of priors on the hyper-parameters of the noise model to avoid the transit signal being modelled as noise with an amplitude $\approx D$ correlated on a timescale $\approx W$. To avoid this problem we use an intermediate step where the transit parameters $D$, $W$, $T_0$, etc. are fixed at the values obtained in the  least-squares fit and we use \software{emcee} to sample the joint PPD of the decorrelation parameters and the hyper-parameters of the GP, $S_0$ and $\omega_0$ ($Q$ is fixed at the value $1/\sqrt{2}$). We find that the convergence of the sampler is improved if we also set a prior on the parameter, $c$, the mean flux level out of transit. We set a Gaussian prior on $c$ with the same mean as the flux values out of transit and a width 4 times the standard error on the mean on these values. Based on the results from this intermediate step, we set Gaussian priors on $S_0$ and $\omega_0$ centred on the mean of the values sampled from the PPD and with standard deviation equal to twice the standard deviation of the sampled PPD. This enables us explore the correlations between the transit parameters and the hyper-parameters of the noise model without exploring unreasonable parts of the parameter space, e.g. solutions where the light curve contains no transit. The results from this analysis are indistinguishable from the results in Table~\ref{tab:hd97658_pars}, e.g. $D= 0.000822 \pm 0.000019 $, $W=0.012442 \pm 0.000054 $.
The amplitude of correlated noise estimated from the standard deviation of the Gaussian process, $\sigma_{\rm GP} = \sqrt{S_0\,\omega_0\,Q}$,  is $25\pm35$\,ppm based on this analysis. The value of the Bayesian information criterion (BIC) for the best fit including a Gaussian process is slightly lower than that without a GP, but the difference is less than 10 so the evidence that the GP is fitting a real signal is not strong. 

The precision of our transit depth measurement from a single visit with \CHEOPS\ improves on the measurement based on two transits observed with {\it TESS} by a factor 2. There is good agreement in the transit depth measured by the two instruments. The precision of the planet-star radius ratio from the combined measurement is less than 1\%. The error in the planet radius, $R_{\rm p}$, is now dominated by the uncertainty in the stellar radius (Table~\ref{tab:massradius}).  

The internal structure of HD~97658b is comparable to the one of HD~106315b, with however a larger mass of the gas envelope (smaller than $\sim 10^{-2} M_\oplus$), a similar water mass fraction (between 0.06 and 0.48 of the mass of the core), and a similar iron mass fraction in the planet. Interestingly, the posterior distribution of the water mass fraction peaks at large values compared to the two planets discussed above (in particular GJ~436\,b).


\subsection{GJ 1132 b}
\label{sec:gj1132_discuss}

Some care is needed when comparing values of $R_{\rm p}/R_{\star}$ as a function of wavelength for observations obtained through broadband filters because GJ~1132 is an M4.5V-type star that has a very red spectrum with strong features due to molecular absorption. These features of the stellar spectral energy distribution should be accounted for when calculating the effective wavelength and bandwidth for observations obtained with different instruments. For the results shown in Fig.~\ref{fig:gj1132_k_wave} and given in Table~\ref{tab:gj1132_k_wave} we  used a synthetic spectrum from the BT-Settl grid of models \citep{2014IAUS..299..271A} to calculate the  effective wavelength and bandwidth for each observation from the flux-weighted mean photon wavelength and its standard error. The MEarth instrument uses a long-pass filter with a cutoff wavelength of 715\,nm. We assumed that MEarth has the same instrument response as \CHEOPS\ for wavelengths redder than this cutoff and 0 response otherwise. For the LDSS3C instrument used by \citet{2018AJ....156...42D} we assume a uniform response over the wavelength range 710\,--\,1030\,nm.

From Fig.~\ref{fig:gj1132_k_wave} it is clear that there is significant disagreement between the value of $R_{\rm p}/R_{\star}$ observed in the  $z^{\prime}$ bandpass by \citet{2017AJ....153..191S} and the values obtained using \CHEOPS, MEarth, {\it TESS} and LDSS3C, despite the substantial overlap in the bandpass for each instrument. Light curves of GJ~1132 from ground-based instruments using broadband filters will be affected by systematic errors because these observations require the use of nearby stars to monitor the atmospheric transparency and extinction. These comparison stars typically have very different spectra to GJ~1132, so they are not affected by changes in observing conditions in the same way as GJ~1132. This is particularly true for observations at infrared wavelengths that are affected by variable water absorption bands. We conclude that the large radius for GJ~1132\,b observed by \citeauthor{2017AJ....153..191S} in the $z^{\prime}$ is not strong evidence for an extended atmosphere on this planet.
 
In the case of the MEarth data we were able to account for the systematic noise correlated with air mass because there is a large amount of data available for this star obtained over many nights. The data from the LDSS3C instrument are not affected by this effect because the extinction correction was done in multiple narrow pass bands. There is excellent agreement between the values of $R_{\rm p}/R_{\star}$ and other transit parameters measured using these instruments and with the values derived using extensive data from {\it Spitzer} at $4.5\,\mu$m. This gives us some reassurance that \CHEOPS\ data analysed using \pycheops\ can provide accurate and precise measurements for the properties of transiting planets, even in cases such as this where there is poor coverage of the individual transits, the field of observation is crowded, and the target is fainter than the design specification of the instrument. 
It should be noted that this is partly due to a well-determined mass--density and mass--absolute-magnitude relations for stars with masses $\approx 0.18$\,M$_{\odot}$. This demonstrates the importance of having a good understanding the host star for accurate characterisation of exoplanet systems.

The focus of the study by \citet{2021AJ....161..213S} using observations of GJ~1132\,b with the WFC3 instrument on {\it HST} was the detection and interpretation of subtle features in the transmission spectrum over the wavelength range 1.13\--\,1.64\,$\mu$m. That study does not report all the transit parameters derived from their analysis of the ``white light'' light curve produced by combining the data at all observed wavelengths. The time of mid-transit is reported in their Table~1 with ``MJD'' in the units column. The value given has the wrong number of digits for a modified Julian date and is 0.5 days less than the time of mid-transit from \citet{2017AJ....153..191S} quoted as a prior in the same table. We assume that this time of mid-transit is actually given as BJD$_{\rm TDB}-0.5$. In that case, the offset of this time of mid-transit from the value predicted by our updated linear ephemeris is $-118\pm 18$\,s, i.e. significantly earlier than expected. There is some ambiguity here as it is unclear what time scale has been used for the value given in their Table~1. \citeauthor{2021AJ....161..213S} state that they derive the key parameter $R_{\rm p}/R_{\star}$ from the white-light data but do not quote the result. It appears that the value of the parameter $R_{\star}/a$ was fixed in their analysis although the value selected is not given. A fixed value for the orbital semi-major axis, $a$, is provided in their Table~1 but it is unclear why since this parameter has a negligible effect on the shape and depth of the transit, unless it is used indirectly with some other parameter to infer $R_{\star}/a$. The value of the orbital inclination is quoted in their Table~1 as $i=87\fdg3577$ with upper and lower limits of $0\fdg0430105$ and  $0\fdg044457$, respectively. We take this to mean that they have derived a value $i=87\fdg358 \pm 0\fdg044$. This value is marginally consistent with the average value $i=88\fdg62 \pm 0\fdg30$ derived from the four data sets used to determine the mass and radius of the planet in Table~\ref{tab:massradius}.

During the preparation of this manuscript, \citet{2021arXiv210401873M} published an analysis of the same HST data used by \citet{2021AJ....161..213S} using two different methods. In contrast to the results from  \citeauthor{2021AJ....161..213S},  \citeauthor{2021arXiv210401873M} found no evidence for any molecular
signatures in the wavelength range covered by the WFC3 instrument. This study was published subsequent to the analysis presented in Sections~\ref{sec:planetradius} and \ref{sec:structure}. We have not updated the analysis in those sections because the planet radius derived from their ``white-light'' light curves is very close to the value used in our analysis. There is also very good agreement between the planet-star radius ratio obtained by \citeauthor{2021arXiv210401873M} from their analysis of the {\it TESS} light curve and the value we have obtained from the analysis of the \CHEOPS\ and MEarth light curves, as can be seen from the values listed in Table~\ref{tab:gj1132_k_wave} and from Fig.~\ref{fig:gj1132_k_wave}.

The internal structure models (see Fig. \ref{fig:struct_GJ1132}) indicate that the planet is a bare dry core. The gas fraction is negligible and the water mass fraction could be up to $27\%$. The mass fraction of the iron core ranges between 2\% and 35\%, with a small fraction of sulphur in it. All these values are similar to some extent to Earth values, so the planet could be pictured as a very hot (massive) Earth analogue. This scenario is consistent with the lack of any detected spectral features in the transmission spectrum of GJ~1132\,b.

\section{Conclusions}
\label{sec:conclusion}
We have used observations of stars observed during the Early Science programme to demonstrate that \CHEOPS\ data can be analysed straightforwardly using \pycheops\ in order to determine accurate and precise transit parameters for transiting extrasolar planets. The performance of \CHEOPS\ is comparable to or better than other space-based instrumentation despite its modest aperture because of the very low levels of instrumental noise by design for this instrument. Compared to {\it K2}, {\it MOST} and {\it Spitzer}, \CHEOPS\ also has the distinct advantage that it is currently operational. \CHEOPS\ also has the flexibility to schedule observations  to coincide with the transits and eclipses of known exoplanets, or to search for suspected transiting exoplanets in multi-planets systems \citep{2021A+A...646A.157B}. \pycheops\ has already been used for the analysis of \CHEOPS\ data in several studies \citep{2021A+A...646A.157B, 2020A+A...643A..94L, 2021ExA....51..109B, 2021arXiv210109260L, borsato2021,2021arXiv210410462V}. \CHEOPS\ observations are on-going so we can look forward to the publication of many exciting results from the partnership of this unique instrument and the \pycheops\ software.

\section*{Acknowledgements}
CHEOPS is an ESA mission in partnership with Switzerland with important contributions to
the payload and the ground segment from Austria, Belgium, France, Germany, Hungary, Italy,
Portugal, Spain, Sweden, and the United Kingdom. The CHEOPS Consortium would like to
gratefully acknowledge the support received by all the agencies, offices, universities, and
industries involved. Their flexibility and willingness to explore new approaches were essential
to the success of this mission.

PM acknowledges support from STFC research grant number ST/M001040/1.
This project has received funding from the European Research Council (ERC) under the European Union’s Horizon 2020 research and innovation programme (project {\sc Four Aces}; grant agreement No 724427).
This project has been carried out in the frame of the National Centre for Competence in Research PlanetS supported by the Swiss National Science Foundation (SNSF).
ACC and TW acknowledge support from STFC consolidated grant number ST/R000824/1.
 YA and MJH  acknowledge  the  support  of  the  Swiss  National  Fund  under  grant 200020\_172746.
SH gratefully acknowledges CNES funding through the grant 837319.
S.G.S. acknowledge support from FCT through FCT contract nr. CEECIND/00826/2018 and POPH/FSE (EC).
The Belgian participation to CHEOPS has been supported by the Belgian Federal Science Policy Office (BELSPO) in the framework of the PRODEX Program, and by the University of Li{\`e}ge through an ARC grant for Concerted Research Actions financed by the Wallonia-Brussels Federation.
We acknowledge support from the Spanish Ministry of Science and Innovation and the European Regional Development Fund through grants ESP2016-80435-C2-1-R, ESP2016-80435-C2-2-R, PGC2018-098153-B-C33, PGC2018-098153-B-C31, ESP2017-87676-C5-1-R, MDM-2017-0737 Unidad de Excelencia “María de Maeztu”- Centro de Astrobiología (INTA-CSIC), as well as the support of the Generalitat de Catalunya/CERCA programme. The MOC activities have been supported by the ESA contract No. 4000124370.
S.C.C.B. acknowledges support from FCT through FCT contracts nr. IF/01312/2014/CP1215/CT0004.
XB, SC, DG, MF and JL acknowledge their role as ESA-appointed CHEOPS science team members.
ABr was supported by the SNSA.
This project was supported by the CNES.
This work was supported by FCT - Funda\c{c}\~{a}o para a Ci\^{e}ncia e a Tecnologia through national funds and by FEDER through COMPETE2020 - Programa Operacional Competitividade e Internacionalização by these grants: UID/FIS/04434/2019; UIDB/04434/2020; UIDP/04434/2020; PTDC/FIS-AST/32113/2017 \& POCI-01-0145-FEDER- 032113; PTDC/FIS-AST/28953/2017 \& POCI-01-0145-FEDER-028953; PTDC/FIS-AST/28987/2017 \& POCI-01-0145-FEDER-028987.
O.D.S.D. is supported in the form of work contract (DL 57/2016/CP1364/CT0004) funded by national funds through FCT.
B.-O.D. acknowledges support from the Swiss National Science Foundation (PP00P2-190080).
MF and CMP gratefully acknowledge the support of the Swedish National Space Agency (DNR 65/19, 174/18).
DG gratefully acknowledges financial support from the CRT foundation under Grant No. 2018.2323 ``Gaseousor rocky? Unveiling the nature of small worlds''.
M.G. is an F.R.S.-FNRS Senior Research Associate.
This work was granted access to the HPC resources of MesoPSL financed by the Region Ile de France and the project Equip@Meso (reference ANR-10-EQPX-29-01) of the programme Investissements d’Avenir supervised by the Agence Nationale pour la Recherche.
This work was also partially supported by a grant from the Simons Foundation (PI Queloz, grant number 327127).
Acknowledges support from the Spanish Ministry of Science and Innovation and the European Regional Development Fund through grant PGC2018-098153-B- C33, as well as the support of the Generalitat de Catalunya/CERCA programme.
This  project  has  been  supported  by  the  Hungarian National Research, Development and Innovation Office (NKFIH) grants GINOP-2.3.2-15-2016-00003, K-119517,  K-125015, and the City of Szombathely under Agreement No.\ 67.177-21/2016.
V.V.G. is an F.R.S-FNRS Research Associate. This research was supported in part by the National Science Foundation under Grant No. NSF PHY-1748958. MS acknowledges supported from STFC, grant number ST/T506175/1. The authors acknowledge support from the Swiss NCCR PlanetS and the Swiss National Science Foundation. YA acknowledges  the  support  of  the  Swiss  National  Fund  under  grant 200020\_172746. 
GPi, VNa, GSs, IPa, LBo, and RRa acknowledge the funding support from Italian Space Agency (ASI) regulated by ``Accordo ASI-INAF n. 2013-016-R.0 del 9 luglio 2013 e integrazione del 9 luglio 2015 CHEOPS Fasi A/B/C''.
 This publication makes use of data products from the Two Micron All Sky Survey, which is a joint project of the University of Massachusetts and the Infrared Processing and Analysis Center/California Institute of Technology, funded by the National Aeronautics and Space Administration and the National Science Foundation.
 This work has made use of data from the European Space Agency (ESA) mission
{\it Gaia} (\url{https://www.cosmos.esa.int/gaia}), processed by the {\it Gaia}
Data Processing and Analysis Consortium (DPAC,
\url{https://www.cosmos.esa.int/web/gaia/dpac/consortium}). Funding for the DPAC
has been provided by national institutions, in particular the institutions
participating in the {\it Gaia} Multilateral Agreement.
KGI is the ESA CHEOPS Project Scientist and is responsible for the ESA CHEOPS Guest Observers Programme. She does not participate in, or contribute to, the definition of the Guaranteed Time Programme of the CHEOPS mission through which observations described in this paper have been taken, nor to any aspect of target selection for the programme.
 This research has made use of the SIMBAD database, operated at CDS, Strasbourg, France
Based on observations made with ESO Telescopes at the La Silla Paranal Observatory under programme IDs 191.C-0873 (GJ~1132), 198.C-0838 (GJ~1132), 198.C-0169 (HD~106315) and 098.C-0304 (HD~106315).
This publication makes use of The Data \& Analysis Center for Exoplanets (DACE), which is a facility based at the University of Geneva (CH) dedicated to extrasolar planets data visualisation, exchange and analysis. DACE is a platform of the Swiss National Centre of Competence in Research (NCCR) PlanetS, federating the Swiss expertise in Exoplanet research. The DACE platform is available at \url{https://dace.unige.ch}.
LD is an F.R.S.-FNRS Postdoctoral Researcher. The Belgian participation to CHEOPS has been supported by the Belgian Federal Science Policy Office (BELSPO) in the framework of the PRODEX Program, and by the University of Li{\`e}ge through an ARC grant for Concerted Research Actions financed by the Wallonia-Brussels Federation.
A.De. acknowledges support from the European Research Council (ERC) under the European Union's Horizon 2020 research and innovation programme (project {\sc Four Aces}, grant agreement No. 724427), and from the National Centre for Competence in Research ``PlanetS'' supported by the Swiss National Science Foundation (SNSF).

\section*{Data Availability}
The data underlying this article are available in the CHEOPS mission archive (\url{https://cheops.unige.ch/archive_browser/})

%


\bibliographystyle{mnras}
\bibliography{pycheops} 


\appendix

\section{Mean and error estimates for quantities that may be affected by systematic errors}
\label{sec:combine}
Where we have multiple estimates for a stellar or planetary  parameter that may be affected by systematic errors, we assume that the systematic error on all these estimates has the same value, $\sigma_{\rm sys}$. Note that $\sigma_{\rm sys}$ may also be used to characterise the variance due to interesting astrophysical signals, e.g., changes in planet radius with wavelength or transit timing variations. The log-likelihood to obtain the observed measurements $\bmath{y} = \{y_i \pm \sigma_i, i=1,\dots,N\}$ is then
 \[\ln\,p(\bmath{y}\,|\,\mu,\sigma_{\rm sys}) = -\frac{1}{2} \sum_i \left[
    \frac{(y_i-\mu)^2}{s_i^2}
    + \ln \left ( 2\pi\,s_i^2 \right )
\right], \]
where $s_i^2 = \sigma_i^2 + \sigma_{\rm sys}^2$. We assume a broad uniform prior on the mean, $\mu$ and a broad uniform  prior on $\ln \sigma_{\rm sys}$. We then sample the posterior probability distribution using \software{emcee} with 1500 steps and 128 walkers. We  discard the first 500 ``burn-in'' steps of the Markov chain and use the remaining sample to calculate the mean and standard deviation of the posterior probability distribution for $\mu$, i.e. our best estimate for the value of the parameter and its standard error. 


\section{Power spectral density of the residuals}
\label{sec:plot_fft}
The figures in this appendix show power spectral density of the residuals for each of the CHEOPS data sets.

\begin{figure}
  \resizebox{\columnwidth}{!}{\includegraphics[]{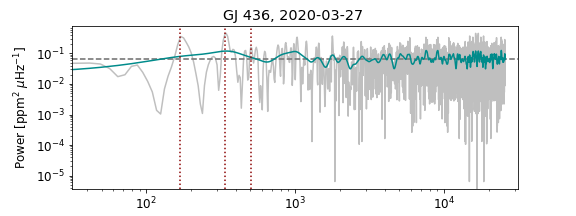}}
  \resizebox{\columnwidth}{!}{\includegraphics[]{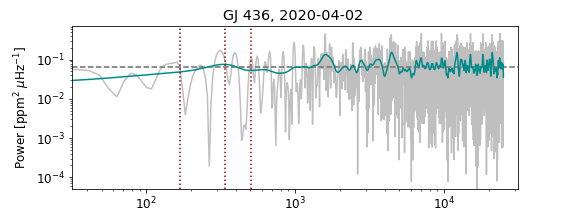}}
  \resizebox{\columnwidth}{!}{\includegraphics[]{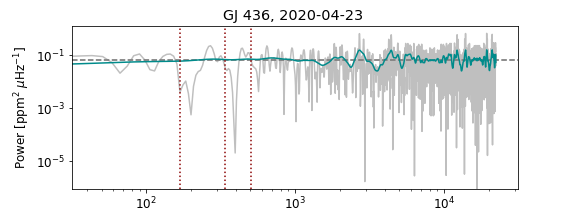}}
  \caption{Power spectral density of the residuals from the best fit to our CHEOPS observations of GJ~436. The cyan line is the PSD smoothed over 10 points with a Gaussian kernel. Each panel is labelled by the date of observation for each data set. Vertical dotted lines show the orbital frequency of the CHEOPS spacecraft and it first two harmonics. The white-noise level expected based on the typical error bar per datum is shown as a dashed horizontal line.}
  \label{fig:fft_GJ436}
\end{figure}

\begin{figure}
  \resizebox{\columnwidth}{!}{\includegraphics[]{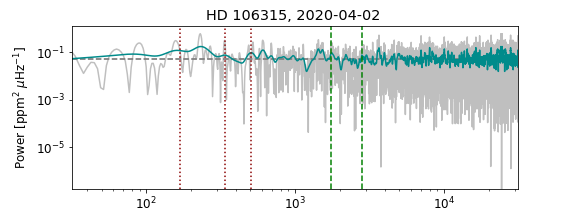}}
  \resizebox{\columnwidth}{!}{\includegraphics[]{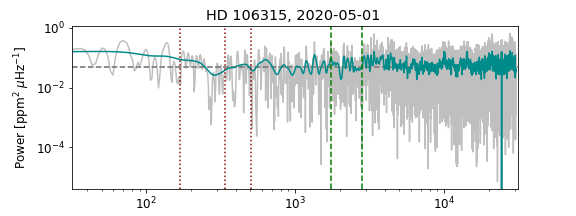}}
  \caption{Power spectral density of the residuals from the best fit to our CHEOPS observations of HD~106315.  The cyan line is the PSD smoothed over 10 points with a Gaussian kernel. Each panel is labelled by the date of observation for each data set. Vertical dotted lines show the orbital frequency of the CHEOPS spacecraft and it first two harmonics. Dashed vertical lines indicate the expected value of $\nu_{\rm max}$ for solar-like oscillations based on equation 20 of \citet{2016ApJ...830..138C}. The white-noise level expected based on the typical error bar per datum is shown as a dashed horizontal line.}
  \label{fig:fft_HD106315}
\end{figure}

\begin{figure}
  \resizebox{\columnwidth}{!}{\includegraphics[]{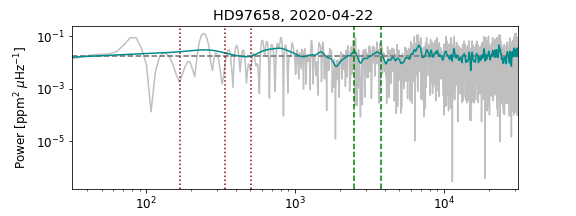}}
  \caption{Power spectral density of the residuals from the best fit to our CHEOPS observations of HD~97658.  The cyan line is the PSD smoothed over 10 points with a Gaussian kernel. Vertical dotted lines show the orbital frequency of the CHEOPS spacecraft and it first two harmonics. Dashed vertical lines indicate the expected value of $\nu_{\rm max}$ for solar-like oscillations based on equation 20 of \citet{2016ApJ...830..138C}. The white-noise level expected based on the typical error bar per datum is shown as a dashed horizontal line.}
  \label{fig:fft_HD97658}
\end{figure}

\begin{figure}
  \resizebox{\columnwidth}{!}{\includegraphics[]{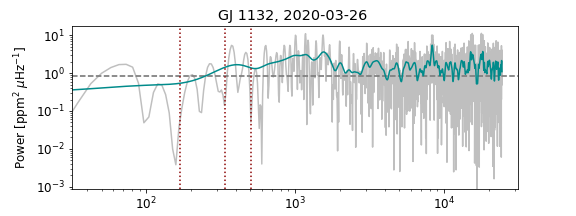}}
  \resizebox{\columnwidth}{!}{\includegraphics[]{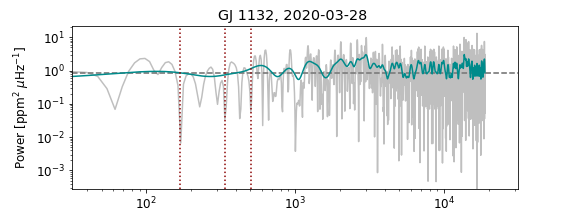}}
  \resizebox{\columnwidth}{!}{\includegraphics[]{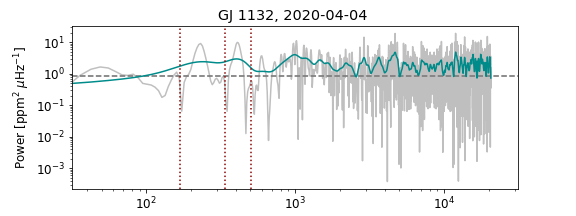}}
  \caption{Power spectral density of the residuals from the best fit to our CHEOPS observations of GJ~1132.  The cyan line is the PSD smoothed over 10 points with a Gaussian kernel. Each panel is labelled by the date of observation for each data set. Vertical dotted lines show the orbital frequency of the CHEOPS spacecraft and it first two harmonics. The white-noise level expected based on the typical error bar per datum is shown as a dashed horizontal line.}
  \label{fig:fft_GJ1132}
\end{figure}


\section{Trends in the CHEOPS data as a function of spacecraft roll angle}
\label{sec:plot_roll}
The figures in this appendix show the trends in the data for each of the CHEOPS data sets as a function of spacecraft roll angle.

\begin{figure}
  \resizebox{\columnwidth}{!}{\includegraphics[]{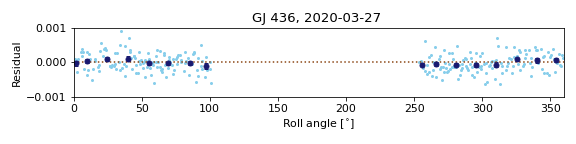}}
  \resizebox{\columnwidth}{!}{\includegraphics[]{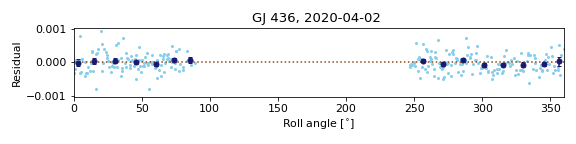}}
  \resizebox{\columnwidth}{!}{\includegraphics[]{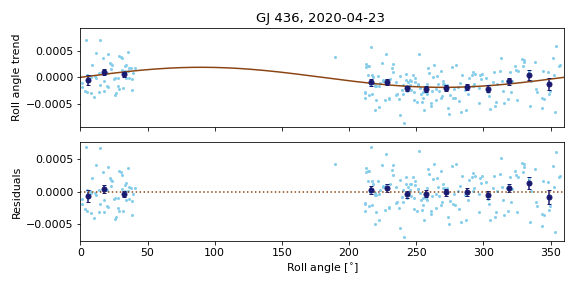}}
  \caption{Trends in our CHEOPS data for GJ~436 as a function of spacecraft roll angle. The upper panel for each visit shows the residuals from the best-fit transit model to each data set together with the best-fit trend as a function of roll angle. The lower panel for each visit shows the residuals from the best-fit model including trends with spacecraft roll angle in the two cases where such trends were included in the fit. The plots for each visit are labelled by the date of observation for each data set. }
  \label{fig:roll_GJ436}
\end{figure}

\begin{figure}
  \resizebox{\columnwidth}{!}{\includegraphics[]{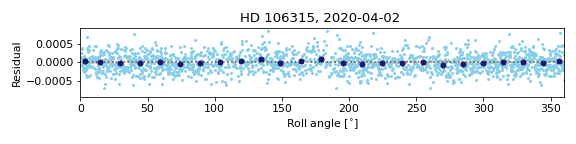}}
  \resizebox{\columnwidth}{!}{\includegraphics[]{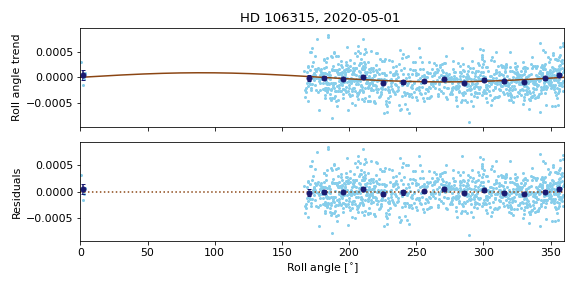}}
  \caption{Trends in our CHEOPS data for HD~106315 as a function of spacecraft roll angle. The upper panel for each visit shows the residuals from the best-fit transit model to each data set together with the best-fit trend as a function of roll angle. The lower panel for each visit shows the residuals from the best-fit model including trends with spacecraft roll angle in the two cases where such trends were included in the fit. The plots for each visit are labelled by the date of observation for each data set. }
  \label{fig:roll_HD106315}
\end{figure}

\begin{figure}
  \resizebox{\columnwidth}{!}{\includegraphics[]{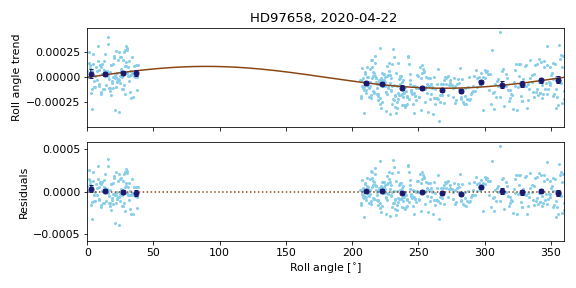}}
  \caption{Trends in our CHEOPS data for HD~97658 as a function of spacecraft roll angle. The upper panel shows the residuals from the best-fit transit model with the best-fit trend as a function of roll angle. The lower panel shows the residuals from the best-fit model including trends with spacecraft roll angle.}
  \label{fig:roll_HD97658}
\end{figure}

\begin{figure}
  \resizebox{\columnwidth}{!}{\includegraphics[]{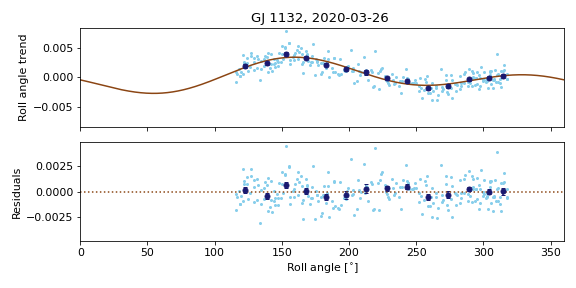}}
  \resizebox{\columnwidth}{!}{\includegraphics[]{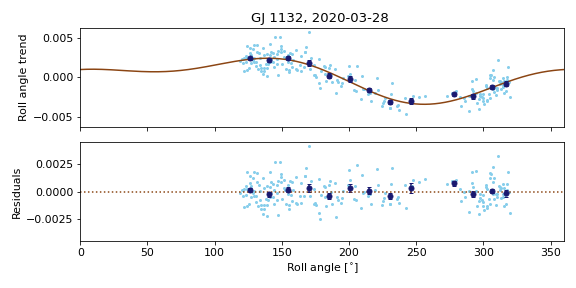}}
  \resizebox{\columnwidth}{!}{\includegraphics[]{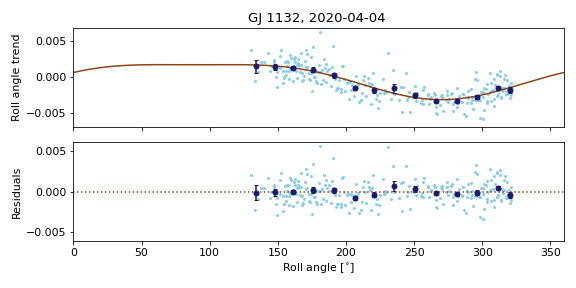}}
  \caption{Trends in our CHEOPS data for GJ~1132 as a function of spacecraft roll angle. The upper panel for each visit shows the residuals from the best-fit transit model to each data set together with the best-fit trend as a function of roll angle. The lower panel for each visit shows the residuals from the best-fit model including trends with spacecraft roll angle in the two cases where such trends were included in the fit. The plots for each visit are labelled by the date of observation for each data set. }
  \label{fig:roll_GJ1132}
\end{figure}


\section{Internal structure parameters}
\label{sec:cornerplots}
The figures in this appendix show the correlations between internal structure parameters derived from the analysis described in section~\ref{sec:structure}. 

\begin{figure*}
  \resizebox{\hsize}{!}{\includegraphics[]{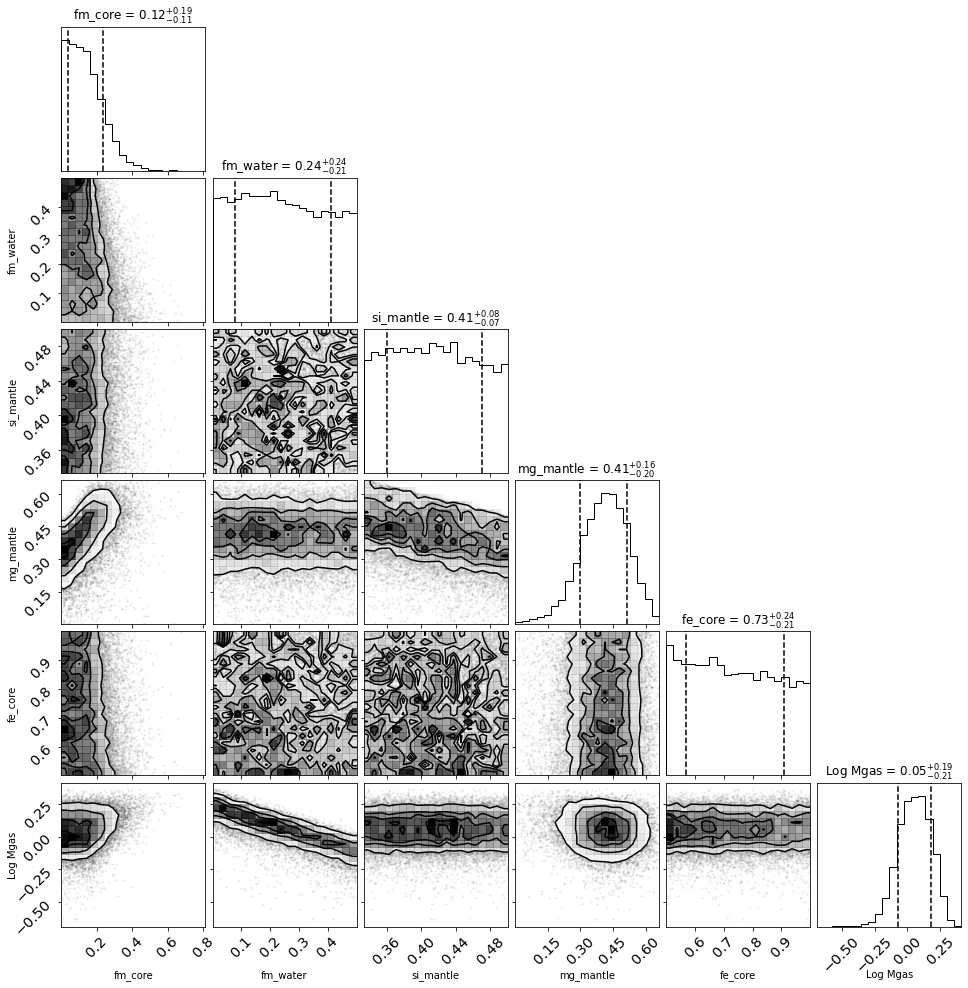}}
  \caption{Corner plot showing the main internal structure parameters of GJ~436b. Shown are the mass fraction of the inner core, the mass fraction of water, the Si and Mg mole fraction in the mantle, the Fe mole fraction in the inner core, and the mass of gas (log scale). The values on top of each column are  the mean and 5\% and 95\% quantiles.}
  \label{fig:struct_GJ436}
\end{figure*}

 \begin{figure*}
  \resizebox{\hsize}{!}{\includegraphics[]{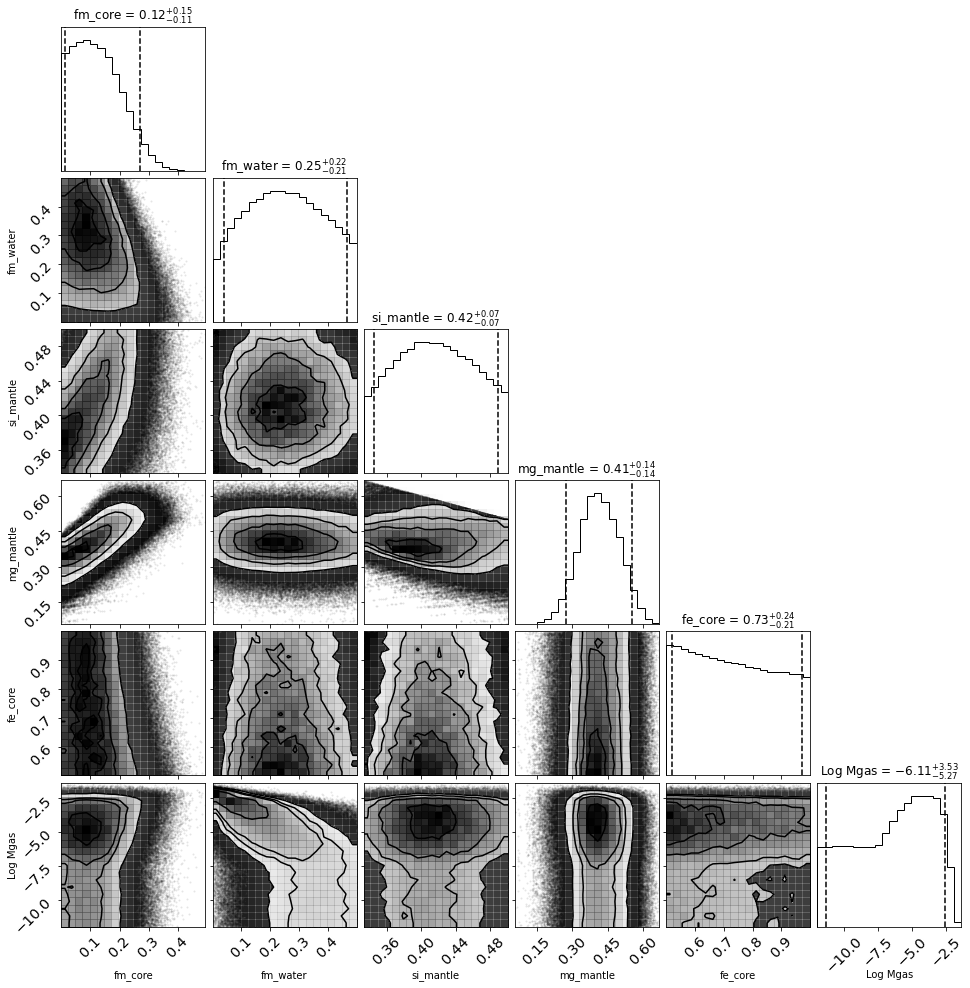}}
  \caption{Corner plot showing the main internal structure parameters of HD~106315b. Shown are the mass fraction of the inner core, the mass fraction of water, the Si and Mg mole fraction in the mantle, the Fe mole fraction in the inner core, and the mass of gas (log scale). The values on top of each column are  the mean and 5\% and 95\% quantiles.}
  \label{fig:struct_HD106315}
\end{figure*}

 \begin{figure*}
  \resizebox{\hsize}{!}{\includegraphics[]{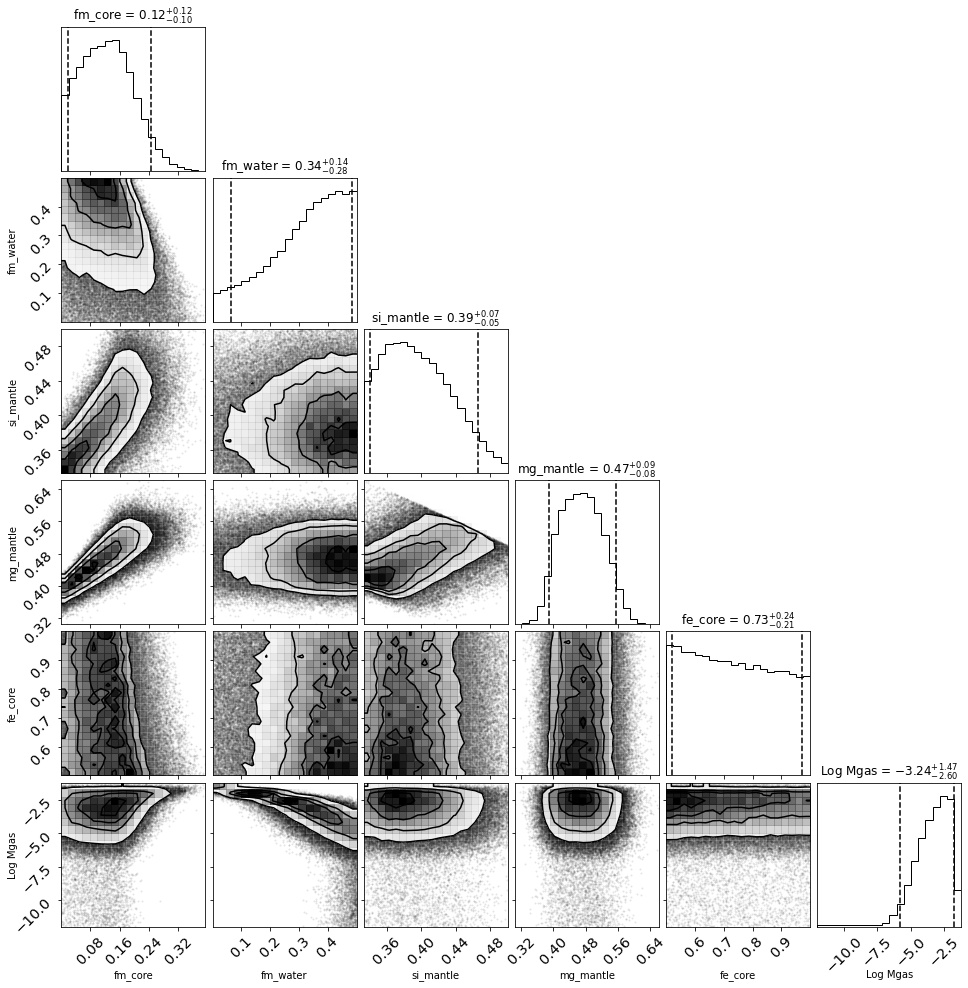}}
  \caption{Corner plot showing the main internal structure parameters of HD~106315b. Shown are the mass fraction of the inner core, the mass fraction of water, the Si and Mg mole fraction in the mantle, the Fe mole fraction in the inner core, and the mass of gas (log scale). The values on top of each column are  the mean and 5\% and 95\% quantiles.}
  \label{fig:struct_HD97658}
\end{figure*}

\begin{figure*}
  \resizebox{\hsize}{!}{\includegraphics[]{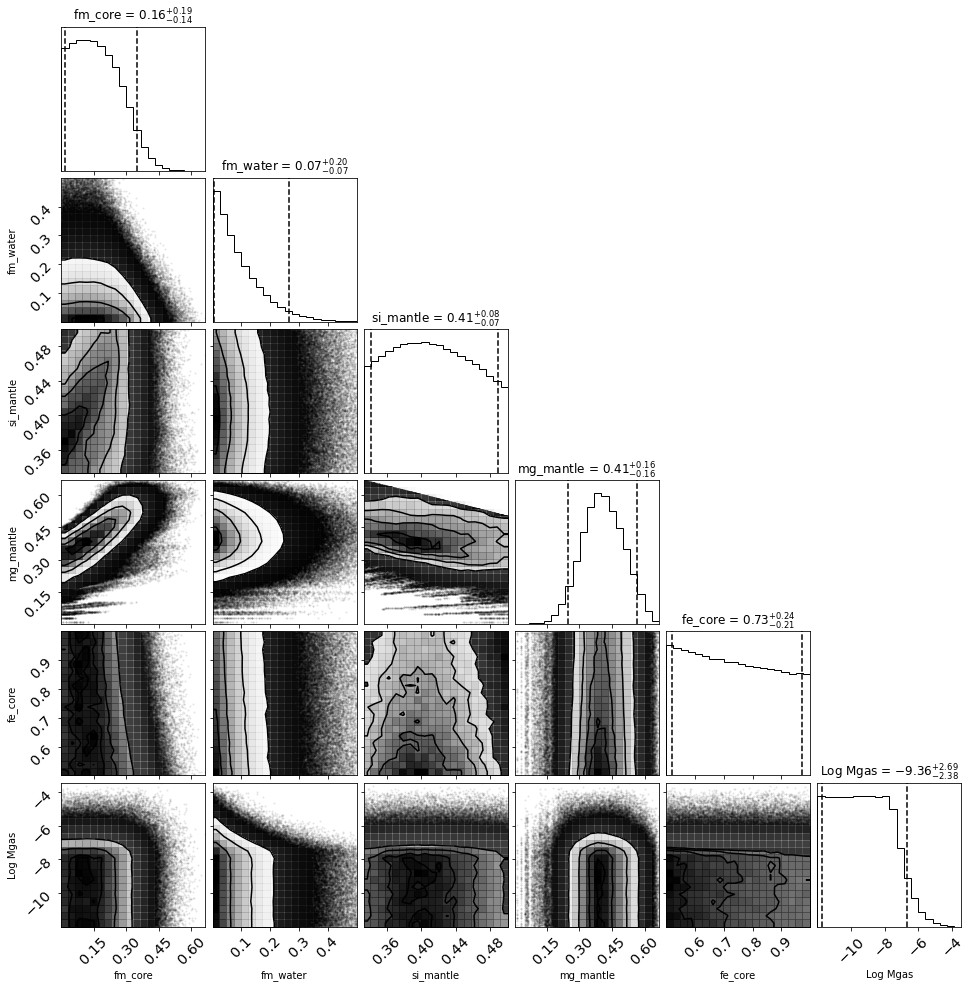}}
  \caption{Corner plot showing the main internal structure parameters of GJ~1132b. Shown are the mass fraction of the inner core, the mass fraction of water, the Si and Mg mole fraction in the mantle, the Fe mole fraction in the inner core, and the mass of gas (log scale). The values on top of each column are  the mean and 5\% and 95\% quantiles.}
  \label{fig:struct_GJ1132}
\end{figure*}

\bigskip
$^{1}$ Astrophysics Group, Keele University, Staffordshire, ST5 5BG, United Kingdom\\
$^{2}$ Department of Astronomy, University of Geneva, Chemin Pegasi 51, Versoix, Switzerland\\
$^{3}$ Centre for Exoplanet Science, SUPA School of Physics and Astronomy, University of St Andrews, North Haugh, St Andrews KY16 9SS, UK\\
$^{4}$ Physikalisches Institut, University of Bern, Gesellsschaftstrasse 6, 3012 Bern, Switzerland\\
$^{5}$ Aix Marseille Univ, CNRS, CNES, LAM, Marseille, France\\
$^{6}$ Instituto de Astrof\'isica e Ci\^encias do Espa\c{c}o, Universidade do Porto, CAUP, Rua das Estrelas, 4150-762 Porto, Portugal\\
$^{7}$ Department of Astronomy, Stockholm University, AlbaNova University Center, 10691 Stockholm, Sweden\\
$^{8}$ Astrobiology Research Unit, Universit\'e de Li\`ege, All\'ee du 6 Ao\^ut 19C, B-4000 Li\`ege, Belgium\\
$^{9}$ Space sciences, Technologies and Astrophysics Research (STAR) Institute, Universit{\'e} de Li{\`e}ge, All{\'e}e du 6 Ao{\^u}t 19C, 4000 Li{\`e}ge, Belgium\\
$^{10}$ Space Research Institute, Austrian Academy of Sciences, Schmiedlstrasse 6, A-8042 Graz, Austria\\
$^{11}$ INAF, Osservatorio Astronomico di Padova, Vicolo dell'Osservatorio 5, 35122 Padova, Italy\\
$^{12}$ Instituto de Astrof\'\i sica de Canarias, 38200 La Laguna, Tenerife, Spain\\
$^{13}$ Departamento de Astrof\'\i sica, Universidad de La Laguna, 38206 La Laguna, Tenerife, Spain\\
$^{14}$ Institut de Ci\`encies de l'Espai (ICE, CSIC), Campus UAB, Can Magrans s/n, 08193 Bellaterra, Spain\\
$^{15}$ Institut d'Estudis Espacials de Catalunya (IEEC), 08034 Barcelona, Spain\\
$^{16}$ Depto. de Astrofísica, Centro de Astrobiologia (CSIC-INTA), ESAC campus, 28692 Villanueva de la Cãda (Madrid), Spain\\
$^{17}$ Departamento de F\'isica e Astronomia, Faculdade de Ci\^encias, Universidade do Porto, Rua do Campo Alegre, 4169-007 Porto, Portugal\\
$^{18}$ Center for Space and Habitability, Gesellsschaftstrasse 6, 3012 Bern, Switzerland\\
$^{19}$ Max Planck Institute for Extraterrestrial Physics, Giessenbachstrasse 1, 85748 Garching, Germany\\
$^{20}$ Université Grenoble Alpes, CNRS, IPAG, 38000 Grenoble, France\\
$^{21}$ Admatis, Miskok, Hungary\\
$^{22}$ Institute of Planetary Research, German Aerospace Center (DLR), Rutherfordstrasse 2, 12489 Berlin, Germany\\
$^{23}$ Université de Paris, Institut de physique du globe de Paris, CNRS, F-75005 Paris, France\\
$^{24}$ ESTEC, European Space Agency, 2201AZ, Noordwijk, NL\\
$^{25}$ Lund Observatory, Dept. of Astronomy and Theoretical Physics, Lund University, Box 43, 22100 Lund, Sweden\\
$^{26}$ Leiden Observatory, University of Leiden, PO Box 9513, 2300 RA Leiden, The Netherlands\\
$^{27}$ Department of Space, Earth and Environment, Chalmers University of Technology, Onsala Space Observatory, 43992 Onsala, Sweden\\
$^{28}$ Dipartimento di Fisica, Universit\`a degli Studi di Torino, via Pietro Giuria 1, I-10125, Torino, Italy\\
$^{29}$ University of Vienna, Department of Astrophysics, Türkenschanzstrasse 17, 1180 Vienna, Austria\\
$^{30}$ Division Technique INSU, BP 330, 83507 La Seyne cedex, France\\
$^{31}$ Department of Physics, University of Warwick, Gibbet Hill Road, Coventry CV4 7AL, United Kingdom\\
$^{32}$ Science and Operations Department - Science Division (SCI-SC), Directorate of Science, European Space Agency (ESA), European Space Research and Technology Centre (ESTEC),
Keplerlaan 1, 2201-AZ Noordwijk, The Netherlands\\
$^{33}$ Konkoly Observatory, Research Centre for Astronomy and Earth Sciences, 1121 Budapest, Konkoly Thege Miklós út 15-17, Hungary\\
$^{34}$ IMCCE, UMR8028 CNRS, Observatoire de Paris, PSL Univ., Sorbonne Univ., 77 av. Denfert-Rochereau, 75014 Paris, France\\
$^{35}$ Institut d'astrophysique de Paris, UMR7095 CNRS, Université Pierre \& Marie Curie, 98bis blvd. Arago, 75014 Paris, France\\
$^{36}$ Department of Astrophysics, University of Vienna, Tuerkenschanzstrasse 17, 1180 Vienna, Austria\\
$^{37}$ INAF, Osservatorio Astrofisico di Catania, Via S. Sofia 78, 95123 Catania, Italy\\
$^{38}$ Institute of Optical Sensor Systems, German Aerospace Center (DLR), Rutherfordstrasse 2, 12489 Berlin, Germany\\
$^{39}$ Dipartimento di Fisica e Astronomia "Galileo Galilei", Università degli Studi di Padova, Vicolo dell'Osservatorio 3, 35122 Padova, Italy\\
$^{40}$ Cavendish Laboratory, JJ Thomson Avenue, Cambridge CB3 0HE, UK\\
$^{41}$ Center for Astronomy and Astrophysics, Technical University Berlin, Hardenberstrasse 36, 10623 Berlin, Germany\\
$^{42}$ Institut für Geologische Wissenschaften, Freie Universität Berlin, 12249 Berlin, Germany\\
$^{43}$ ELTE Eötvös Loránd University, Gothard Astrophysical Observatory, 9700 Szombathely, Szent Imre h. u. 112, Hungary\\
$^{44}$ MTA-ELTE Exoplanet Research Group, 9700 Szombathely, Szent Imre h. u. 112, Hungary\\
$^{45}$ Institute of Astronomy, University of Cambridge, Madingley Road, Cambridge, CB3 0HA, United Kingdom\\

\bsp	
\label{lastpage}
\end{document}